\documentclass[twocolumn,tighten,times,trackchanges]{aastex631}

\newif\ifprintauthors
\printauthorstrue

\usepackage{amsmath}
\let\tablenum\relax

\usepackage{siunitx}
\DeclareSIUnit\parsec{pc}
\DeclareSIUnit\Mpc{\mega\parsec}

\usepackage{txfonts}
\newlength{\capheight}
\usepackage{textgreek}
\usepackage{array}
\usepackage{tabularx}

\usepackage{array}
\usepackage{acronym}
\usepackage{xspace}

\usepackage{amstext}

\makeatletter
\@ifpackageloaded{cleveref}{
  \AtBeginDocument{
    \def\ltx@label#1{\cref@label{#1}}%
    \def\label@in@display@noarg#1{\cref@old@label@in@display{#1}}%
    \def\label@in@mmeasure@noarg#1{%
      \begingroup%
        \measuring@false%
        \cref@old@label@in@display{#1}%
      \endgroup}%
  }
}{}
\makeatother

\protected\def\protectedacused{\acused}

\acrodef{LIGO}[LIGO]{Laser Interferometer Gravitational-Wave Observatory}
\acrodef{LHO}[LHO]{\ac{LIGO} Hanford Observatory}
\acrodef{LLO}[LLO]{\ac{LIGO} Livingston Observatory}
\acrodef{KAGRA}[KAGRA]{KAGRA}\acused{KAGRA}
\acrodef{iKAGRA}[iKAGRA]{initial \ac{KAGRA}}
\acrodef{bKAGRA}[bKAGRA]{baseline \ac{KAGRA}}
\acrodef{GEO}[GEO\,600]{GEO\,600 \ac{GW} detector}
\acrodef{aLIGO}{Advanced \ac{LIGO}}
\acrodef{A+}{Advanced+ \ac{LIGO}}
\acrodef{Asharp}[\ensuremath{\text{A}^\sharp}]{\ac{LIGO} \acs{Asharp}}
\acrodef{AdV}{Advanced \acl{Virgo}}
\acrodef{AdV+}{Advanced \acl{Virgo}+}
\acrodef{Virgo}{Virgo}\acused{Virgo}
\acrodef{VirgoNEXT}[Virgo\_nEXT]{Virgo\_nEXT}\acused{VirgoNEXT}

\acrodef{LSC}[LSC]{\acs{LIGO} Scientific Collaboration}
\acrodef{LV}[LV]{\acs{LIGO}--\acs{Virgo} Collaboration\protect\protectedacused{LVC}}
\acrodef{LVC}[LV]{\acs{LIGO}--\acs{Virgo} Collaboration\protect\protectedacused{LV}}
\acrodef{LVK}[LVK]{\acs{LIGO}--\ac{Virgo}--\ac{KAGRA} Collaboration}
\acrodef{IGWN}[IGWN]{International \ac{GWH} Observatory Network}

\acrodef{O1}[O1]{first observing run}
\acrodef{O2}[O2]{second observing run}
\acrodef{O3}[O3]{third observing run}
\acrodef{O3a}[O3a]{first half of the third observing run}
\acrodef{O3b}[O3b]{second half of the third observing run}
\acrodef{O3GK}[O3GK]{observing run}
\acrodef{O4}[O4]{fourth observing run}
\acrodef{O4a}[O4a]{first part of the fourth observing run}
\acrodef{O4b}[O4b]{second part of the fourth observing run}
\acrodef{O4c}[O4c]{third part of the fourth observing run}
\acrodef{IR1}[IR1]{intermediate run 1}
\acrodef{O5}[O5]{fifth observing run}

\acrodef{BH}[BH]{black hole}
\acrodef{BBH}[BBH]{binary \ac{BH}}
\acrodef{BNS}[BNS]{binary \ac{NS}}
\acrodef{IMBH}[IMBH]{intermediate-mass \ac{BH}}
\acrodef{NS}[NS]{neutron star}
\acrodef{BHNS}[BHNS]{\ac{BH}--\ac{NS} binary}
\acrodefplural{BHNS}[BHNSs]{\ac{BH}--\ac{NS} binaries}
\acrodef{NSBH}[NSBH]{\ac{NS}--\ac{BH} binary}
\acrodefplural{NSBH}[NSBHs]{\ac{NS}--\ac{BH} binaries}
\acrodef{PBH}[PBH]{primordial \ac{BH}}
\acrodef{CBC}[CBC]{compact binary coalescence}

\acrodef{GW}[GW]{gravitational wave\protect\protectedacused{GWH}}
\acrodef{GWH}[GW]{gravitational-wave\protect\protectedacused{GW}}
\acrodef{IFO}[IFO]{interferometer}
\acrodef{SNR}[SNR]{signal-to-noise ratio}
\acrodef{FAR}[FAR]{false-alarm rate}
\acrodef{IFAR}[IFAR]{inverse false-alarm rate}
\acrodef{FAP}[FAP]{false alarm probability}
\acrodef{PSD}[PSD]{power spectral density}
\acrodef{GR}[GR]{general relativity}
\acrodef{NR}[NR]{numerical relativity}
\acrodef{PN}[PN]{post-Newtonian}
\acrodef{EOB}[EOB]{effective-one-body}
\acrodef{ROM}[ROM]{reduced-order model}
\acrodef{IMR}[IMR]{inspiral--merger--ringdown}
\acrodef{PDF}[pdf]{probability density function}
\acrodef{PE}[PE]{parameter estimation}
\acrodef{CI}[CI]{credible interval}
\acrodef{CL}[CL]{credible level}
\acrodef{EOS}[EoS]{equation of state}
\acrodef{KLD}[KLD]{Kullback--Leibler divergence}
\acrodef{JSD}[JSD]{Jensen--Shannon divergence}
\acrodef{GCN}[GCN]{General Coordinates Network}
\acrodef{GWTC}[GWTC]{Gravitational-Wave Transient Catalog}
\acrodef{GWOSC}[GWOSC]{Gravitational Wave Open Science Center}
\acrodef{WDM}[WDM]{Wilson--Daubechies--Meyer}
\acrodef{DQR}[DQR]{data-quality report}

\acrodef{CWB}[cWB]{coherent WaveBurst}
\acrodef{LAL}[LAL]{\ac{LIGO} algorithm library}

\acrodef{CHRoCC}{central heating radius of curvature correction}
\acrodef{NonSENS}{non-stationary estimation and noise subtraction}
\acrodef{RF}{radio frequency}
\acrodef{PNC}{phase noise cancellation}
\acrodef{ASC}{alignment sensing and control}
\acrodef{WFS}{wave-front sensing}
\acrodef{BPC}{beam position control}
\acrodef{ADS}{alignment dither systems}
\acrodef{OMC}{output mode cleaner}
\acrodef{LVDTs}{linear variable differential transformers}
\acrodef{GAS}{geometrical anti-spring}

\acrodef{PTA}{Pulsar Timing Array}

\newcommand\gwtc[1][?]{\mbox{GWTC\if#1?\else-#1\fi}}
\newcommand\thisgwtcversionmajor{5}
\newcommand\thisgwtcversionminor{0}
\newcommand\thisgwtcversionfull{\thisgwtcversionmajor.\thisgwtcversionminor}
\newcommand\thisgwtcversion\thisgwtcversionfull

\newcommand{\Msun}{\ensuremath{\mathit{M_\odot}}}

\newcommand\perGpcyr{\ensuremath{\mathrm{Gpc}^{-3}\,\mathrm{yr}^{-1}}}

\newcommand{\chieff}{\ensuremath{\chi_\mathrm{eff}}}
\newcommand{\chip}{\ensuremath{\chi_\mathrm{p}}}

\newcommand{\spinone}{\ensuremath{\chi_1}}

\newcommand\PEpdfp{\ensuremath{p}}

\newcommand{\PEparameter}{\ensuremath{\boldsymbol{\theta}}}%

\newcommand\PEpdf[2][?]{\ensuremath{\PEpdfp({#2}\ifx#1?\else | {#1}\fi)}}

\newcommand\PEpriorpdfpi{\ensuremath{\pi}}
\newcommand\PEpdfprior[1]{\ensuremath{\PEpriorpdfpi({#1})}}
\newcommand\PEprior[1][\PEparameter]{\PEpdfprior{#1}}
\newcommand\PEpriorpe[1][\PEparameter]{{\let\keepPEpriorpdfpi\PEpriorpdfpi\def\PEpriorpdfpi{\keepPEpriorpdfpi_{\text{PE}}}\PEprior[#1]\let\PEpriorpdfpi\keepPEpriorpdfpi}}

\newcommand{\OfourBStartDate}{{{2024~April~10}}}
\newcommand{\OfourBStartTime}{{{15:00:00}}}

\newcommand{\soft}[1]{\textsc{#1}}

\newcommand{\GSTLAL}{\soft{GstLAL}\xspace}

\newcommand{\CWB}{\soft{cWB}\xspace}

\newcommand{\PYCBC}{\soft{PyCBC}\xspace}
\newcommand{\MBTA}{\soft{MBTA}\xspace}

\newcommand{\BILBY}{\soft{Bilby}\xspace}

\newcommand{\IMRPhenomXPHMST}{\soft{IMRPhenomXPHM\_SpinTaylor}\xspace}

\newcommand{\SEOBNRFIVEPHM}{\soft{SEOBNRv5PHM}\xspace}

\newcommand{\IMRPhenomPTWONRTidal}{\soft{IMRPhenomPv2\_NRTidal}\xspace}

\newcommand{\IMRPhenomNSBH}{\soft{IMRPhenomNSBH}\xspace}

\newcommand{\SURSEVENDQFOUR}{\soft{NRSur7dq4}\xspace}

\input{macros_from_json}
\newcommand{\pixelpop}{\textsc{PixelPop}\xspace}
\newcommand{\fullpop}{\textsc{FullPop}\xspace}
\newcommand{\defbbh}{\textsc{Default \ac{BBH}}\xspace}
\newcommand{\bgp}{\textsc{Binned Gaussian Process}\xspace}
\newcommand{\chieffmixture}{\textsc{\chieff{} Mixture}\xspace}
\newcommand{\chieffthree}{\textsc{\chieff{} Three Transitions}\xspace}
\newcommand{\isolatedpeak}{\textsc{Isolated Peak}\xspace}
\newcommand{\bivskewchichi}{\textsc{Bivariate Skewed \chieff/\chip}\xspace}
\newcommand{\lincorr}{\textsc{Linear Correlation}\xspace}
\newcommand{\splcorr}{\textsc{Spline Correlation}\xspace}
\newcommand{\copcorr}{\textsc{Copula Correlation}\xspace}
\newcommand{\bpltp}{\textsc{Broken Power Law + 2 Peaks}\xspace}
\newcommand{\threeSubpopulation}{\textsc{Three Subpopulation}\xspace}
\newcommand{\tgmm}{\textsc{Truncated Gaussian Mixture Model}\xspace}
\newcommand{\tgmms}{TGMM\xspace}
\acrodef{AGN}[AGN]{active galactic nuclei}
\acrodef{ER16}[ER16]{16th engineering run}
\acrodef{FAR}[FAR]{false alarm rate}
\acrodef{PPISN}[PPISN]{pulsational pair-instability supernova}
\acrodefplural{PPISN}[PPISNe]{pulsational pair-instability supernovae}
\acrodef{PISN}[PISN]{pair-instability supernova}
\acrodefplural{PISN}[PISNe]{pair-instability supernovae}
\acrodef{PPD}[PPD]{posterior predictive distribution}
\acrodef{SFR}[SFR]{star-formation rate}

\newcommand{\gwtctwo}[0]{GWTC-2\xspace}
\newcommand{\gwtcthree}[0]{GWTC-3\xspace}
\newcommand{\gwtcfour}[0]{GWTC-4.0\xspace}
\newcommand{\gwtcfive}[0]{GWTC-5.0\xspace}

\newcommand{\gwtcfivendet}[0]{267\xspace}
\newcommand{\gwtcfivendetThresholdOnePerYear}[0]{267\xspace}
\newcommand{\gwtcfivendetThresholdOnePerFourYear}[0]{242\xspace} 

\newcommand{\OfourBndet}[0]{104\xspace}
\newcommand{\OfourBndetThresholdOnePerYear}[0]{104\xspace}
\newcommand{\OfourBndetThresholdOnePerFourYear}[0]{94\xspace}
\newcommand{\OfourAPointOneNdetThresholdOnePerYear}[0]{2\xspace}
\newcommand{\OfourAPointOneNdetThresholdOnePerFourYear}[0]{4\xspace}

\acrodef{ICAR}[ICAR]{intrinsic conditional autoregressive prior}
\newcommand{\CIPlusMinus}[1]{{#1[median]^{+#1[error plus]}_{-#1[error minus]}}}
\newcommand{\CIPlusMinusPercent}[1]{{#1[median]^{+#1[error plus]}_{-#1[error minus]}} \%}

\newcommand{\AtOnePerYear}{\ensuremath{\sim 1.21\times 10^{6}}\xspace}
\newcommand{\AtOnePerFourYears}{\ensuremath{\sim 1.11\times 10^{6}}\xspace}
\newcommand{\AtSNRTen}{\ensuremath{\sim 3.70\times 10^{5}}\xspace}

\newcommand{\deltamuchiq}{\ensuremath{\delta \mu_{\mathrm{\chi_{eff}}|q}}\xspace}
\newcommand{\deltasigmachiq}{\ensuremath{\delta\! \ln \sigma_{\chi_{\mathrm{eff}}|q}}\xspace}
\newcommand{\etachiq}{\ensuremath{\eta_{\mathrm{\chi_{eff}},q}}\xspace}

\newcommand{\deltamuchiz}{\ensuremath{\delta \mu_{\mathrm{\chi_{eff}}|z}}\xspace}
\newcommand{\deltasigmachiz}{\ensuremath{\delta\! \ln \sigma_{\chi_{\mathrm{eff}}|z}}\xspace}
\newcommand{\etachiz}{\ensuremath{\eta_{\mathrm{\chi_{eff}},z}}\xspace}

\newcommand{\Nnoise}{\ensuremath{N_{\mathrm{noise}}}}
\newcommand{\FARthreshold}{\ensuremath{\mathrm{FAR}_\mathrm{threshold}}}
\newcommand{\linbpl}{\textsc{Linear BPL}\xspace}
\newcommand{\linske}{\textsc{Linear Skew}\xspace}

\newcommand{\mlow}{\ensuremath{\tilde{m}_{\mathrm{low}}}\xspace}
\newcommand{\mmid}{\ensuremath{\tilde{m}_{\mathrm{mid}}}\xspace}
\newcommand{\mhigh}{\ensuremath{\tilde{m}_{\mathrm{high}}}\xspace}
\renewcommand{\Msun}{\ensuremath{\mathrm{M}_\odot}\xspace}
\renewcommand{\chieff}{\ensuremath{\chi_\mathrm{eff}}\xspace}
\renewcommand{\chip}{\ensuremath{\chi_\mathrm{p}}\xspace}
\renewcommand{\gwtcthree}{\mbox{GWTC-3.0}\xspace}
\renewcommand{\gwtctwo}{\mbox{GWTC-2.0}\xspace}

\usepackage{xcolor}

\begin{document}

\title{GWTC-5.0: Population Properties of Merging Compact Binaries}
\ifprintauthors
    \suppressAffiliations


\author[0000-0003-4786-2698]{A.~G.~Abac}
\affiliation{Max Planck Institute for Gravitational Physics (Albert Einstein Institute), D-14476 Potsdam, Germany}
\author{A.~Abe}
\affiliation{Department of Physics, Graduate School of Science, Osaka Metropolitan University, 3-3-138 Sugimoto-cho, Sumiyoshi-ku, Osaka City, Osaka 558-8585, Japan  }
\author{I.~Abouelfettouh}
\affiliation{LIGO Hanford Observatory, Richland, WA 99352, USA}
\author{F.~Acernese}
\affiliation{Dipartimento di Fisica ``E.R. Caianiello'', Universit\`a di Salerno, I-84084 Fisciano, Salerno, Italy}
\affiliation{INFN, Sezione di Napoli, I-80126 Napoli, Italy}
\author[0000-0002-8648-0767]{K.~Ackley}
\affiliation{University of Warwick, Coventry CV4 7AL, United Kingdom}
\author{A.~Adam}
\affiliation{OzGrav, University of Western Australia, Crawley, Western Australia 6009, Australia}
\author[0009-0004-2101-5428]{S.~Adhicary}
\affiliation{The Pennsylvania State University, University Park, PA 16802, USA}
\author{D.~Adhikari}
\affiliation{Max Planck Institute for Gravitational Physics (Albert Einstein Institute), D-30167 Hannover, Germany}
\affiliation{Leibniz Universit\"{a}t Hannover, D-30167 Hannover, Germany}
\author[0000-0002-5731-5076]{R.~X.~Adhikari}
\affiliation{LIGO Laboratory, California Institute of Technology, Pasadena, CA 91125, USA}
\author{V.~K.~Adkins}
\affiliation{Louisiana State University, Baton Rouge, LA 70803, USA}
% ...



    \collaboration{0}{The LIGO Scientific Collaboration, the Virgo Collaboration, and the KAGRA Collaboration\\(See the end matter for the full list of authors)}
\else
    \collaboration{0}{The LIGO Scientific Collaboration, the Virgo Collaboration, and the KAGRA Collaboration}
\fi

\correspondingauthor{LSC P\&P Committee, via LVK Publications as proxy}
\email{lvc.publications@ligo.org}

\begin{abstract}
We present the population properties of merging compact binaries inferred using \gwtcfivendet mergers from the cumulative Gravitational-Wave Transient Catalog 5.0. As this data set contains no new sources with a neutron star, we primarily focus on the properties of the binary black hole mergers. We infer the merger rate of binary black holes with component masses between $2.5\,\Msun$ and $200\,\Msun$ to be $\RateEstimates[fp_bbh_z02_limits][5th percentile]\text{--} \RateEstimates[pp_bbh_z02_limits][95th percentile] \, \perGpcyr$ (all intervals at $90\%$ credible levels) at redshift $z = 0.2$. We find evidence for a subpopulation of binary black hole mergers that host a rapidly spinning black hole (dimensionless spins $\chi \sim 0.7$), consistent with signatures of hierarchical mergers. We find that these occur at two mass scales, the first at primary masses $\sim 10$--$20\,\Msun$ and the second above $\sim 45\,\Msun$, and we estimate their total rate at $z=0.2$ to be $\HierarchicalMacros[rate_uniform][5th percentile]\text{--} \HierarchicalMacros[rates_2g1g_total][95th percentile] \, {\rm Gpc}^{-3} {\rm yr}^{-1}$. We infer that, above $40\,\Msun$, the mass distribution of the less massive (secondary) black hole declines more steeply than that of the more massive (primary) one. This is consistent with a flatter mass-ratio distribution and indicates the prevalence of unequal-mass binaries with large primary masses. We find evidence for two features in the black hole mass spectrum: a peak around $10\,\Msun$ and a change of slope at around $35\,\Msun$. Black holes of $\sim 35\,\Msun$ pair preferentially with companions of similar mass. Additionally, we find that the effective inspiral spin distribution of binary black holes is asymmetric about zero, based on which we infer that at least $\OverviewEstimates[bsn_chieff_asymmetry][5th percentile] \%$ of mergers occur in channels with some preference for spin-orbit alignment. We find evidence that the effective inspiral spin distribution is broader for unequal-mass binaries and that it also likely broadens with increasing redshift. Overall, our results support the presence of multiple subpopulations of merging black holes that can potentially arise from different formation pathways.

\end{abstract}

\section{Introduction}
\label{Sec:intro}

With LIGO \citep{2015CQGra..32g4001L}, Virgo \citep{2015CQGra..32b4001A}, and KAGRA \citep{2021PTEP.2021eA101A} {increasing in sensitivity} over the last decade, the number of \ac{GW} detections has rapidly increased and, with it, our ability to study the population of merging \acp{BH} and \acp{NS}. This paper updates our inference on such populations with the latest iteration of the \ac{LVK} transient catalog, \gwtcfive{} \citep{GWTC:Introduction, GWTC:Results}, which includes data through the \ac{O4b}. 

{We analyze sources in \gwtcfive{} with \ac{FAR} $< 1\,\mathrm{yr}^{-1}$ ($< 0.25\,\mathrm{yr}^{-1}$ for systems with \acp{NS}, see Sec.~\ref{Sec:dataset}) using} hierarchical Bayesian inference (Sec.~\ref{Sec:methods}). In \ac{O4b}, \OfourBndet new \ac{BBH} systems were detected with sufficient significance to be included in our population inferences. Furthermore, two new sources with sufficient significance from the \ac{O4a} are also included {following reanalysis} (see App.~A of \citealp{GWTC:Results}). This brings the total number of detected compact binary mergers considered in this paper to \gwtcfivendet, up from 161 {reported in} \cite{2025arXiv250818083T}. No new \acp{NSBH} or \acp{BNS} of sufficient significance were detected. The increased catalog size  enables more robust population inference using both weakly and strongly-parameterized models; we list some key findings below. Throughout this paper, we denote the more massive and less massive compact object in the binary as primary (subscript 1) and secondary (subscript 2), respectively. All results are quoted at a {symmetric} $90\%$ credible level unless stated otherwise. 

\begin{enumerate}
	\item \textbf{The {inferred distribution of \ac{BBH} systems} is broadly consistent with previous results from \cite{2025arXiv250818083T} based on \gwtcfour (Sec.~\ref{Sec:Overview}).}  We find reduced evidence for the previously reported mass-ratio peak at $q \approx 0.7${, where $q=m_2/m_1$,} and instead find a preference for equal masses. Spin measurements continue to indicate that most \ac{BBH} mergers have non-extremal spins, with $\OverviewEstimates[pp_aleq_half_fracs][5th percentile] \text{--}  \OverviewEstimates[bkpl2pk_aleq_half_fracs][95th percentile] \%$ of \acp{BH} inferred to have dimensionless spin magnitudes $\chi \leq 0.5$.  The effective inspiral spin ($\chieff$,~\citealp{2008PhRvD..78d4021R}) distribution peaks {near}
	 zero but is asymmetric, from which we infer that at least $\OverviewEstimates[bsn_chieff_asymmetry][5th percentile] \%$ of \ac{BBH} mergers originate in formation channels with some degree of spin--orbit alignment. On the other hand, $\OverviewEstimates[pp_neg_chieff_frac][5th percentile] \text{--}  \OverviewEstimates[bsn_neg_chieff_frac][95th percentile] \%$ have $\chieff<0$ indicating contributions from formation channels that can produce spins anti-aligned with the orbital angular momentum. {In Sec.~\ref{sec:bbh_z}, we show that the larger dataset yields a more precise estimate of the \ac{BBH} merger rate that remains statistically consistent with \citet{2025arXiv250818083T}.}

	\item \textbf{We continue to find features in the \ac{BBH} mass spectrum, at ${\sim}10~\Msun$ and ${\sim}35\,\Msun$ in primary mass (Secs.~\ref{Sec:10msun} and~\ref{sec:35_msun}).} We find {less-clear, model-dependent} evidence that the former contains a significant subpopulation of mergers with mass ratios peaking around $q \approx 0.7$. We find reduced support for the interpretation that the ${\sim}35\,\Msun$ feature represents a local peak in the distribution. Instead, we find it to be more consistent with a {change in slope} in both primary- and secondary-mass {distributions}. {Binaries with $m_1 \approx 35\,\Msun$ possibly have a preference for equal-mass pairing, and a \chieff distribution that peaks at or {near} zero}.

	\item \textbf{Binaries with $ m_1\gtrsim 40\,\Msun$ likely have a different mass-ratio distribution than the rest of the population (Sec.~\ref{sec:high_mass}).} The distribution of $m_2$ has a significantly sharper drop than $m_1$ at $\gtrsim 40\,\Msun$, consistent with a flatter mass-ratio distribution for \ac{BBH} mergers with $m_1 \gtrsim 40\,\Msun$. {Combined with the observed change in spin properties at {that mass} (Sec.~\ref{sec:hierarchical}), this shift in the mass-ratio distribution might indicate the presence of formation mechanisms like hierarchical mergers.}

	\item \textbf{We find evidence for a subset of \ac{BBH} mergers with at least one rapidly-spinning \ac{BH} (Sec.~\ref{Sec:Overview_spins} and Sec.~\ref{sec:hierarchical})}, using models of varying degrees of flexibility. With models motivated by the physics of hierarchical mergers in dense stellar clusters~\citep{Antonini:2024het}, we locate the rapidly-spinning \ac{BBH} mergers at two mass ranges: the first at {$10\,\Msun \leq m_1 \leq 20\,\Msun$} and the second at $m_1 \gtrsim 45\,\Msun$. We infer a merger rate density of $\HierarchicalMacros[rate_uniform][5th percentile]\text{--} \HierarchicalMacros[rates_2g1g_total][95th percentile] \, {\rm Gpc}^{-3} {\rm yr}^{-1}$ at redshift $z=0.2$ for the rapidly-spinning \ac{BBH} mergers.

	\item \textbf{The width of the $\chieff$ distribution likely exhibits correlations with other \ac{BBH} properties (Sec.~\ref{Sec:Correlations}).} We find increased evidence that the structure of the joint $q \text{--} \chieff$ distribution is driven more by the changing width of the $\chieff$ distribution with mass ratio rather than {by the} changing mean. Concurrently, we still find that the $\chieff$ distribution likely also broadens with redshift. {The details of the broadening exhibit some degree of model dependence, suggesting that the structure of the correlation is complex}. 
\end{enumerate}

Altogether, our analysis of \gwtcfive supports the presence of multiple \ac{BBH} subpopulations, loosely defined as a subset of the population with some properties that are clearly distinct from the rest, and the possibility of multiple formation pathways contributing to the observed  population. The rest of the paper presents details of the analyses and the results, followed by a short forward-looking summary in Sec.~\ref{Sec:conclusion}.

\section{Dataset}
\label{Sec:dataset}

This section describes the selection criteria used to include candidates in the population analyses of this paper, and how samples from their posterior distributions were produced. See also~\citet{2025arXiv250818083T}, which we closely follow.

\subsection{Data Collection}
We analyze \ac{CBC} candidates from the cumulative \gwtcfive~\citep{GWTC:Results}, which includes previously reported \ac{GW} events~\citep{2023PhRvX..13d1039A, 2021PhRvX..11b1053A, 2019PhRvX...9c1040A} and new compact binary candidates from \ac{O4b} and the \ac{ER16} preceding \ac{O4b}~\citep{GWTC:Introduction, GWTC:Results}. We exclude \ac{ER16} data in our analyses to avoid potential human-selection bias; one candidate meeting our significance selection criteria, GW240406\_062847, is thus omitted.

The candidates are statistically significant triggers identified by the template-based matched-filter analyses \GSTLAL~\citep{2017PhRvD..95d2001M, 2019arXiv190108580S, 2020PhRvD.101b2003H, 2021SoftX..1400680C, 2024PhRvD.109d2008E, 2023PhRvD.108d3004T, 2024PhRvD.109d4066S, 2023arXiv230607190R, 2025arXiv250606497J, 2025arXiv250523959J}, \MBTA~\citep{2016CQGra..33q5012A, 2021CQGra..38i5004A, 2022CQGra..39e5002A, 2025CQGra..42j5009A}, and \PYCBC~\citep{2016CQGra..33u5004U, 2017ApJ...849..118N, 2018PhRvD..98b4050N, 2021ApJ...923..254D}, as well as the weakly modeled excess-power search performed by the \CWB pipeline~\citep{2005PhRvD..72l2002K,2008CQGra..25k4029K,2016PhRvD..93d2004K,2025PhRvD.111b3054M, GWTC:Methods}.
Details of the detector instrumentation, network sensitivity, and data quality for \ac{O4b} are described in ~\citet{GWTC:Introduction}. More details of the search algorithms and the their configuration are described in~\citet{GWTC:Methods}.
The new candidates and corresponding search sensitivities are presented in~\citet{GWTC:Results}.

\subsection{Event Selection}
Not all events from \gwtcfive are included here, as its statistical threshold is generally less stringent than that required for population inference~\citep{GWTC:Results}. To limit contamination from noise triggers, we impose a significance threshold for population analyses~\citep{2021ApJ...913L...7A, 2023PhRvX..13a1048A, 2025arXiv250818083T}. Specifically, we require a \ac{FAR} $< 1\,\mathrm{yr}^{-1}$ in at least one search pipeline for \ac{BBH}-only analyses. Analyses that include both \acp{NS} and \acp{BH} are subject to a threshold \ac{FAR} $< 0.25\,\mathrm{yr}^{-1}$ (Sec.~\ref{Sec:Overview}).
These criteria yield \gwtcfivendetThresholdOnePerYear (\gwtcfivendetThresholdOnePerFourYear) \ac{CBC} candidates in \gwtcfive with \ac{FAR} below $1\,\mathrm{yr}^{-1}$ ($0.25\,\mathrm{yr}^{-1}$), of which \OfourBndetThresholdOnePerYear(\OfourBndetThresholdOnePerFourYear) are identified in \ac{O4b}.
Due to catalog updates (App.~A of~\citealt{GWTC:Results}), our \gwtcfivendetThresholdOnePerYear candidates below $1\,\mathrm{yr}^{-1}$ include \OfourAPointOneNdetThresholdOnePerYear additional \ac{O4a} events compared to \citet{2025arXiv250818083T}, while our \gwtcfivendetThresholdOnePerFourYear candidates below $0.25\,\mathrm{yr}^{-1}$ include \OfourAPointOneNdetThresholdOnePerFourYear additional \ac{O4a} events whose FAR estimates were revised from above to below $0.25\,\mathrm{yr}^{-1}$.

{Assuming statistically independent noise triggers, the expected number of triggers with {$\mathrm{\ac{FAR}}<\mathrm{\FARthreshold}$}that are actually noise events can be estimated as \citep{2015PhRvD..91b3005F, 2023arXiv230607190R} $\Nnoise = \sum_k \FARthreshold \times T_k$ where $T_k$ is an estimate of the time examined by the $k$-th search. For a threshold of $1\,\mathrm{yr}^{-1}$, we expect $\Nnoise\approx \SamplePurity[expectedNumFalseTrialsFactor][rounded]$ since the \ac{O1} through \ac{O4b}}.

The \ac{O4b} sample includes several notable \ac{BBH} candidates new to population analyses, namely the high-\ac{SNR} events GW250114\_082203 \citep{2025PhRvL.135k1403A}, GW240925\_005809 and GW250207\_115645 \citep{2026arXiv260511703T}, as well as the {high-spin} systems GW241011\_233834 (henceforth GW241011) and GW241110\_124123 (henceforth GW241110) \citep{2025ApJ...993L..21A}. No new candidates involving \acp{NS} satisfy our selection criteria in \gwtcfive.

Population inference requires consistent modeling of selection effects (see Sec.~\ref{Sec:methods}); we therefore exclude candidates from independent analyses~\citep{2019PhRvD.100b3011V, 2020PhRvD.101h3030V, 2022PhRvD.106d3009O, Mehta:2023zlk, Wadekar:2023gea, 2019PhRvD.100b3007Z, 2019ApJ...872..195N, 2020ApJ...891..123N, 2021ApJ...922...76N, 2023ApJ...946...59N, 2024PhRvD.110d3036K, 2025PhRvD.111b3054M, 2025MLS&T...6a5054K, Niu:2025nha}, {which may alter the conclusions} \citep{Mehta:2025oge}. 

\subsection{Source Inference}\label{sec.SourceInference}
Posterior samples and associated simulation products used in this work are obtained from~\citet{GWTC:Results}, with the analysis configurations described in~\citet{GWTC:Methods}.
We employ Bayesian inference to generate posterior samples for each detection candidate, conditioned on the observed strain data and assumed signal and noise models~\citep{2019PASA...36...10T,GWTC:Methods}.

{The waveform choices employed by our analyses are detailed in App.~\ref{App.waveforms}. All the waveforms we use are }constructed for quasi-circular binary orbits and our analyses neglect orbital eccentricity in both source modeling and selection-function calculations.
Although nonzero eccentricity has been claimed for some \ac{GW} candidates~\citep{2020ApJ...903L...5R, 2022NatAs...6..344G, 2022ApJ...940..171R, Gamba:2021gap, 2024CoTPh..76g5402F, 2025PhRvD.112j4045G, 2025arXiv250722862M, 2025PhRvD.112l3004P, Morras:2025xfu, Jan:2025fps, 2025arXiv251210803P, 2021ApJ...915L...5A}, conclusions about eccentricity in the literature vary~\citep{2019MNRAS.490.5210R, 2020MNRAS.495..466W, 2020MNRAS.497.1966L, 2023PhRvD.108j4018O, 2024ApJ...972...65I, 2023PhRvD.108l4063R, 2023PhRvD.107f4024B, 2025PhRvD.112d4038G, 2025arXiv251219513X, 2026arXiv260329019G,2025PhRvL.135k1403A,2025ApJ...993L..21A} and eccentricity is not incorporated in the \gwtcfive parameter-estimation products. All posterior samples are reweighted to a prior uniform in comoving volume and source-frame time following App.~C of~\citet{2021PhRvX..11b1053A}.

\section{Methods}
\label{Sec:methods}

\begin{deluxetable*}{lll}
\tablecaption{Population models used in this work{, in order of appearance}. Whenever a parameter is not explicitly mentioned, it takes the value given in the ``Default 1D distributions'' section, following the same defaults as in \citet{2025arXiv250818083T}.
Strongly-parametrized models are marked with a dagger ($^\dagger$).  \label{tab:summary_of_models}}
\tablehead{
    \colhead{Name} & \colhead{Description} & \colhead{References}
}
\startdata
\fullpop $^\dagger$ & \parbox[t]{0.5\textwidth}{A model for all compact binary types (\ac{BNS}, \ac{NSBH}, and \ac{BBH}). Uses two distinct secondary-mass-dependent pairing functions to write the joint mass distribution.}  & \parbox[t]{0.25\textwidth}{\centering App.~\ref{sec:fulllpop}, \cite{Farah:2021qom}, \cite{Fishbach:2020ryj}, \cite{Mali:2024wpq}} \\ \noalign{\vskip 4pt}
\hline
\pixelpop & \parbox[t]{0.55\textwidth}{Measures the posterior distribution of 1 or 2 parameters as a binned Gaussian process with a smoothing hyper prior, using default 1D parametric distributions for all other parameters. For example, ``($m_1,q)$-\pixelpop'' flexibly models $m_1$ and $q$ jointly, and use the default parametric 1D distributions for the rest.}
& \parbox[t]{0.25\textwidth}{\centering App.~\ref{Sec:pixel_pop_appendix}, \cite{Heinzel:2024jlc}, \cite{Heinzel:2024hva}, \cite{Alvarez-Lopez:2025ltt}}\\ \noalign{\vskip 4pt}
\hline
\defbbh$^\dagger$ &  \parbox[t]{0.55\textwidth}{Uses the default 1D distributions for all parameters (see below).} & \parbox[t]{0.25\textwidth}{\centering App.~\ref{Sec:default_bbh_corners},\cite{2018ApJ...856..173T,2025arXiv250818083T}} \\
\hline
\parbox[t]{0.20\textwidth}{\raggedright \tgmm (TGMM)$^\dagger$ } &   \parbox[t]{0.55\textwidth}{Models the two component spins separately, while allowing for correlations between them using a mixture of two bivariate truncated Gaussians.} & \parbox[t]{0.25\textwidth}{\centering App.~\ref{sec:TGIMM}, \cite{2026ApJ...996...71H,Hussain:2025llf}}\\ \noalign{\vskip 4pt}
\hline
\bivskewchichi$^\dagger$ & \parbox[t]{0.55\textwidth}{Models the effective spins $\chieff$/$\chip$ as a bivariate skewed normal distribution truncated between $-1$ and 1 for $\chieff$ and $0$ and $1$ for $\chip$}& \parbox[t]{0.25\textwidth}{\centering App.~\ref{sec:bivariate_skewnormal}, \cite{2023PhRvX..13d1039A,Banagiri:2025dxo, 2020ApJ...895..128M}}  \\ 
\hline
\isolatedpeak & \parbox[t]{0.55\textwidth}{Assumes two subpopulations. The primary mass is a normal distribution for one subpopulation (``peak'') and a cubic spline for the other (``continuum''). Mass ratio, spins and redshifts are modeled as independent cubic splines, different for each subpopulation.} &\parbox[t]{0.25\textwidth}{\centering\cite{Godfrey:2023oxb,2025arXiv250818083T}}\\ \noalign{\vskip 4pt}
\hline
\bgp & \parbox[t]{0.55\textwidth}{Measures the joint distribution of 3 parameters  jointly as a binned Gaussian process, coupling bins with a Gaussian process covariance, assuming an exponential quadratic kernel. Other parameters are fixed to their parameter estimation prior, except for the redshift, which is fixed to a power law $\propto (1+z)^{2.7}$.}  & \parbox[t]{0.25\textwidth}{\centering\cite{Ray:2023upk,2025ApJ...991...17R,Sridhar:2025kvi,2025arXiv250818083T}} \\ \noalign{\vskip 4pt}
\hline
\parbox[t]{0.18\textwidth}{\chieffmixture, \\ \chieffthree$^\dagger$} & \parbox[t]{0.55\textwidth}{Parametrize $\chieff$ as a mixture of a broad uniform component and a narrow Gaussian. The branching ratio is either dependent on the primary mass via a Gaussian process prior (\chieffmixture) or transitions sharply at three values of masses, measured from the data (\chieffthree)}.& \parbox[t]{0.25\textwidth}{\centering App.~\ref{app:zeta_model}, \cite{Tong:2025xir,Antonini:2025ilj,Antonini:2025zzw}}\\ \noalign{\vskip 4pt}
\hline
\lincorr$^\dagger$ & \parbox[t]{0.55\textwidth}{Model for $\chieff$ as a truncated normal distribution whose location and standard deviation depends linearly on another parameter.}   &\parbox[t]{0.25\textwidth}{\centering \cite{2021ApJ...922L...5C,Biscoveanu:2022qac,2023PhRvX..13a1048A,2025arXiv250818083T}}\\
\hline
\splcorr & \parbox[t]{0.55\textwidth}{Model $\chieff$ as a truncated normal distribution whose location and standard deviation depends on another parameter via a cubic spline.}   & \parbox[t]{0.25\textwidth}{\centering \cite{Heinzel:2023hlb,2025arXiv250818083T}}\\ \noalign{\vskip 4pt}
\hline
\copcorr$^\dagger$ & \parbox[t]{0.55\textwidth}{Couple two marginal distributions into a joint distribution using a Frank copula, with a single parameter controlling the strength and sign of their correlation.} & \parbox[t]{0.25\textwidth}{\centering\cite{2022MNRAS.517.3928A,2023ApJ...958...13A,2025arXiv250818083T}}\\ \noalign{\vskip 4pt}
\hline
\threeSubpopulation & \parbox[t]{0.55\textwidth}{Treats high-mass mergers as a subpopulation with separate spin and mass-ratio properties.} & \parbox[t]{0.25\textwidth}{\centering\cite{Banagiri:2025dmy}}\\ \noalign{\vskip 4pt}
\hline\hline
\multicolumn{3}{c}{Default 1D distributions} \\
\hline
Primary mass ($m_1$) $^\dagger$  & \parbox[t]{0.55\textwidth}{A broken power law continuum between a minimum and maximum mass, plus two Gaussian peaks.} & \parbox[t]{0.25\textwidth}{\centering\cite{2018ApJ...856..173T,2025arXiv250818083T}} \\ \noalign{\vskip 4pt}
\hline
Mass ratio ($q$) $^\dagger$  & \parbox[t]{0.55\textwidth}{A power law between a mass-dependent minimum value and 1.} & \parbox[t]{0.25\textwidth}{\centering\cite{2018ApJ...856..173T,2025arXiv250818083T}} \\
\hline
Spin magnitude ($\chi$) $^\dagger$ & \parbox[t]{0.55\textwidth}{A normal distribution truncated at 0 and 1 for \acp{BBH}, or 0.4 for neutron stars in \acp{BNS} and \acp{NSBH}.} & \parbox[t]{0.25\textwidth}{\centering\cite{2025arXiv250818083T}} \\ \noalign{\vskip 4pt}
\hline
Cosine spin tilts ($\cos \theta$) $^\dagger$ & \parbox[t]{0.55\textwidth}{A mixture of an isotropic component and a normal component truncated at $\pm1$.} & \parbox[t]{0.25\textwidth}{\centering \cite{2017CQGra..34cLT01V,Vitale:2022dpa,2025arXiv250818083T}}\\
\hline
Redshift ($z$) $^\dagger$  & \parbox[t]{0.55\textwidth}{A simple power law, $(1+z)^{\kappa_z}$.}& \parbox[t]{0.25\textwidth}{\centering \cite{2018ApJ...863L..41F}} 
\enddata
\end{deluxetable*}

Population {inference} of \ac{GW} sources is based on the framework of hierarchical Bayesian inference. Three main ingredients are needed for this method: posterior samples for each of the sources in the catalog, a method for calculating the selection effects of the detector network, and a model for the distribution of population parameters. {With the assumptions of a Poisson process for the source population and that the data from each event are statistically independent, the population likelihood---i.e., the likelihood of all source-containing data---is}
\begin{equation}
p(\{d_i\} {\mid} \Lambda) \propto N(\Lambda)^{N_\mathrm{obs}}  e^{- \xi(\Lambda) N(\Lambda)} \prod_{i=1}^{N_\mathrm{obs}} \int{\mathrm{d}\theta\, p(d_i {\mid} \theta)\, p(\theta {\mid} \Lambda)}\, .
\end{equation}
Details on this likelihood, including a full derivation, can be found in {, e.g.,} \citet{2002PhRvD..65f3002L,2015PhRvD..91b3005F,2019MNRAS.486.1086M, 2022hgwa.bookE..45V}; here we briefly describe its main elements.

Beginning with the integrand, for any given source, $\theta$ are the \ac{GW} source parameters, such as masses, spins, redshift and orientation~\citep{2016PhRvL.116x1102A}. The last term, $p(\theta {\mid} \Lambda)$ is our model for the astrophysical distribution of source parameters, parametrized by a set of hyperparameters $\Lambda$, in whose values we are ultimately interested. For each source $i$, $p(d_i {\mid} \theta)$ is the likelihood of the corresponding stretch of data, usually estimated by Monte Carlo reweighing and importance sampling of precomputed posterior draws (cf.~Sec.~\ref{sec.SourceInference}). 

The prefactor to the integral accounts for selection effects; it depends on $N(\Lambda)$, which is the {Poisson rate parameter controlling the total} number of sources that merged over the data-taking period. Of those, $N_\mathrm{obs}$ were actually observed. Given an astrophysical model for the population, the fraction of sources expected to be detected is $\xi(\Lambda)$, which we numerically estimate by reweighting a large set of detectable simulated sources \citep[App.~\ref{App.Selection} and][]{2018CQGra..35n5009T}.
In general, a higher number of such {simulated} sources is desired to reduce the uncertainty in the estimation of the instruments' selection effects, and hence in the population likelihood~\citep{2018CQGra..35n5009T,2023MNRAS.526.3495T,2019RNAAS...3...66F,2022arXiv220400461E,2025arXiv250907221H}. Unless stated otherwise, throughout this work, we require a maximum variance of 1 on the population likelihood estimator; for some models, we also show results with a more relaxed variance cut in App.~\ref{Sec.AppendixOverview}. This is implemented as a sharp or smoothly-tapered cutoff \citep[e.g.,][]{2024PhRvX..14b1005C}.

Following our previous work~\citep{2025arXiv250818083T}, we employ two {broad classes of population models}. We use \emph{strongly-parameterized} models where $p(\theta {\mid} \Lambda)$ is written as the product of compact generally-analytical distributions, e.g., combinations of power laws and Gaussians. Within this framework, we can allow for correlations between parameters by directly modeling their joint distributions rather than the marginals. {Stronger models make rigid assumptions (e.g., that a distribution is Gaussian) that can yield narrower posteriors. However, when the model is a poor match to the data, they can be prone to systematic errors that are sometimes hard to uncover~\citep{2022PASA...39...25R,Miller:2026buq,Alvarez-Lopez:2025ltt,Cheng:2023ddt}. }

We also rely on \emph{weakly-parametrized models}, also known as agnostic models, for which all or some of the marginal distributions cannot be written in terms of simple probability distributions but are rather combinations of flexible bases {such as} B-splines or Gaussian processes. These allow for non-trivial correlations between parameters, usually in a way that is less prescriptive than what can be achieved with strongly-parameterized models. Typically, the extra flexibility of weakly-parameterized models comes at the expense of larger statistical uncertainties. 

A list of the models used in the paper, each with a short qualitative description, is given in Tab.~\ref{tab:summary_of_models}. More details are given in the works and appendices listed in the last column. We usually present results from both strongly- and weakly-parameterized models, so that modeling tradeoffs and impacts can be assessed.

\textbf{A note on Figures:} {We report our measurements of event-level parameters by drawing posterior samples from the population {hyperparameters} $\Lambda$, to obtain the distribution $p(\theta {{\mid}} \Lambda)$. We refer to this as the \ac{PPD}~\citep{2023PhRvX..13a1048A}}. {Most \ac{PPD} plots of reconstructed populations} (e.g., Fig.~\ref{fig:p_m1q_bbh}) show hatched areas corresponding to regions of the parameter space where we have under $1\%$ of \ac{PE} samples using default priors, equally weighted across all detections. This is intended as a heuristic proxy for parts of parameter space where the models may not be actively constrained by the data, and where modeling and prior choices may become particularly important. For example, the value of a power-law index may be mainly informed by a data-rich region, {while the trend is} extrapolated to a data-poor one. That said, regions with a dearth or absence of detections may be highly informative if the detector network is sensitive to those regions, so the hatching should not be overinterpreted.

\begin{figure*}[ht!]
\centering
    \includegraphics[width=0.95\textwidth]{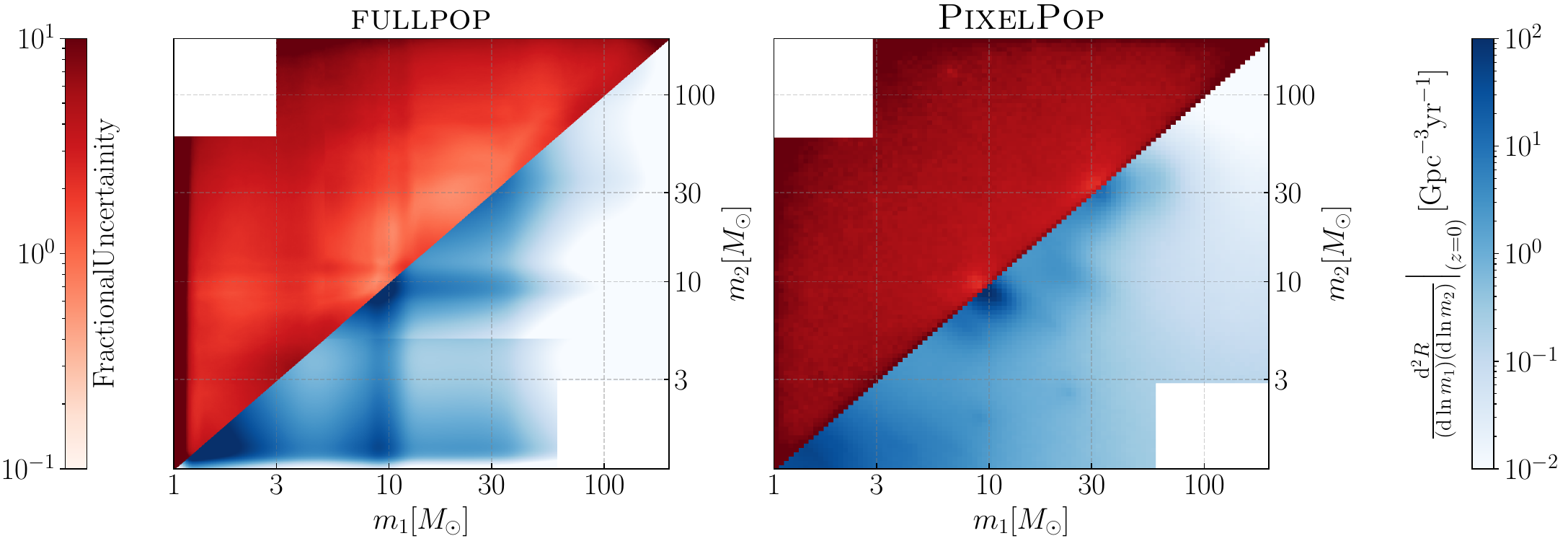}
    \caption{\label{fig:m1m2_fullspectrum} {The differential merger rate distribution as a function of component masses for the full compact binary population inferred using \fullpop (left) and $(m_1, m_2)$-\pixelpop (right) models. The lower triangle in each subplot shows the median mass distribution while the upper triangle shows the corresponding fractional uncertainty, $\Delta R / R$, defined as the difference between the $95$th and $5$th percentile values divided by the median. Since $m_1 \geq m_2$ by definition, the axes are flipped for the upper uncertainty triangles (i.e., vertical axis is $m_1$  for the upper triangles). The overdense vertical stripe at $m_1 \sim 10\,\Msun$ in the \fullpop plot arise because the common pairing function shared by \ac{BNS} and \ac{NSBH} mergers in the model is not well constrained, {resulting in a median mass-ratio distribution that is approximately flat}. The \fullpop model uses different pairing functions for binaries with $m_2$ above and below $5\,\Msun$ explaining the apparently discontinuity there. The regions with $m_1 > 60\,\Msun, m_2 < 3\,\Msun$ have been cut out because of lack of simulated signals (cf.~App.~\ref{App.Selection}) in that region in \gwtcthree which prevents a self-consistent accounting of selection  effects at those masses.}}
\end{figure*}

\begin{table*}[ht]
\centering
 \begin{tabular*}{0.95\textwidth}{@{\extracolsep{\fill}} lccc c @{}}
\hline
\hline
Type & \ac{BNS}$^{\ast}$ & \ac{NSBH}$^{\ast}$ & \ac{BBH}$^{\dagger}$ & {IMBH}$^{\dagger}$\\
\hline
$m_1$ range & $m_1 \in [1, 2.5] \, \Msun$ &  $m_1 \in [2.5, 200] \, \Msun$ & $m_1 \in [2.5, 200] \, \Msun$ & $m_1 \in [100, 200) \, \Msun$  \\
$m_2$ range & $m_2 \in [1, 2.5] \, \Msun$ &  $m_2 \in [1, 2.5] \, \Msun$ & $m_2 \in [2.5, 200] \, \Msun$ & $m_2 \in [100, 200) \, \Msun$   \\
\hline
\pixelpop & $\CIPlusMinus{\RateEstimates[pp_bns_limits]}$ & $\CIPlusMinus{\RateEstimates[pp_nsbh_limits]}$ & $\CIPlusMinus{\RateEstimates[pp_bbh_z02_limits]}$ &$\CIPlusMinus{\RateEstimates[pp_imbh_z02_limits]} \times 10^{-3}$  \\ 
\fullpop & $\CIPlusMinus{\RateEstimates[fp_bns_limits]}$ & $\CIPlusMinus{\RateEstimates[fp_nsbh_limits]}$ & $\CIPlusMinus{\RateEstimates[fp_bbh_z02_limits]}$  &$\CIPlusMinus{\RateEstimates[fp_imbh_z02_limits]} \times 10^{-3}$ \\ 
\hline
Joint & $\RateEstimates[pp_bns_limits][5th percentile] \text{--}  \RateEstimates[fp_bns_limits][95th percentile]$ & $\RateEstimates[fp_nsbh_limits][5th percentile] \text{--}  \RateEstimates[pp_nsbh_limits][95th percentile]$  & $\RateEstimates[fp_bbh_z02_limits][5th percentile] \text{--}  \RateEstimates[pp_bbh_z02_limits][95th percentile]$ & $ (\RateEstimates[pp_imbh_z02_limits][5th percentile] \text{--}  \RateEstimates[pp_imbh_z02_limits][95th percentile] )\times 10^{-3}$  \\ 
\hline
\end{tabular*}

 \caption{Merger rates in units of $\perGpcyr$ of the different categories of \acp{CBC} based on the \fullpop and \pixelpop. The bottom row shows the joint merger rate, calculated as the union of two $90 \%$ credible intervals. {The columns marked with $\ast$ report the merger rates at $z=0$, while those marked by a $\dagger$ report them at $z = 0.2$ as that is approximately the redshift at which we measure the \ac{BBH} rates the best.}}
\label{tab:rates}
\end{table*}

\section{Population Overview}
\label{Sec:Overview}
{This section presents an overview of compact binary merger population properties}, highlights prominent features in their global distribution, and {establishes the basis for a deeper dive in the following sections}.

\subsection{Full Spectrum Mass Distribution}
\label{sec:full_spectrum}

Figure~\ref{fig:m1m2_fullspectrum} shows the median of the joint component-mass distribution of all compact binaries using the strongly-parameterized \fullpop (left) and the weakly-parameterized $(m_1, m_2)$-\pixelpop analyses (right). We direct the reader to Tab.~\ref{tab:summary_of_models} for details of the models.
The upper triangles in Fig.~\ref{fig:m1m2_fullspectrum} plot the fractional uncertainty of the merger rate density ({in }red). Both \fullpop and \pixelpop use the default spin magnitude, tilt and redshift models (Tab.~\ref{tab:summary_of_models}). The two-dimensional {distribution}, $p(m_1, m_2)$  {(in blue)}, displays three notable overdensities: first, an overdensity corresponding to \ac{BNS} systems $(m_1, m_2) \approx (1, 1) \,\Msun$; second, a prominent overdensity centered at $(m_1, m_2) \approx (10, 10) \,\Msun$; and, last, a weaker overdensity at $(m_1, m_2) \approx (35, 30)\,\Msun$. Sections~\ref{Sec:10msun} and \ref{sec:35_msun} provide a detailed investigation of systems into the last two features.

\textbf{With \gwtcfive, our models rule out a completely empty gap between $3 \text{--} 5\,\Msun$}.
Historically, Galactic observations suggested a dearth of compact objects in the ${\sim}3 \text{--} 5\,\Msun$ range, potentially caused by the physics of supernova explosions \citep[e.g.,][]{1998ApJ...499..367B, 2011ApJ...741..103F, 2012ApJ...757...36K}. Recently, this picture has been challenged, both by electromagnetic~\citep{2020Sci...368.3282V,2024NatAs...8.1583W,Barr:2024wwl} and \ac{GW} observations~\citep{2020ApJ...896L..44A,2024ApJ...970L..34A} {of individual sources}. {Our previous study based on \gwtcfour~\citep{2025arXiv250818083T} also found that a completely empty gap is disfavored and this is reinforced by our analyses with \gwtcfive.} While the mass distribution inferred by \fullpop is consistent with a {dip in the} merger rate between the \ac{NS} peak and the ${\sim} 10\,\Msun$ \ac{BH} peak, it rules out a completely empty gap. In this model, a parameter, $A$, controls the gap depth; with $A= 1$ indicating a completely empty gap between $3 \text{--} 5\,\Msun$. We instead find $A \leq \OverviewEstimates[fullpop_gap_depth][99th percentile]$ at $99\%$ credibility. \pixelpop also finds a qualitatively flat mass distribution between the two peaks with no evidence for a pronounced dip. Figure~\ref{fig:lowmass_fullspectrum} {in App.~\ref{sec:fulllpop}} shows the inferred primary mass distribution at the lower end of the mass range.

\begin{figure*}[ht]
	\centering
	    \includegraphics[width=1.0\textwidth]{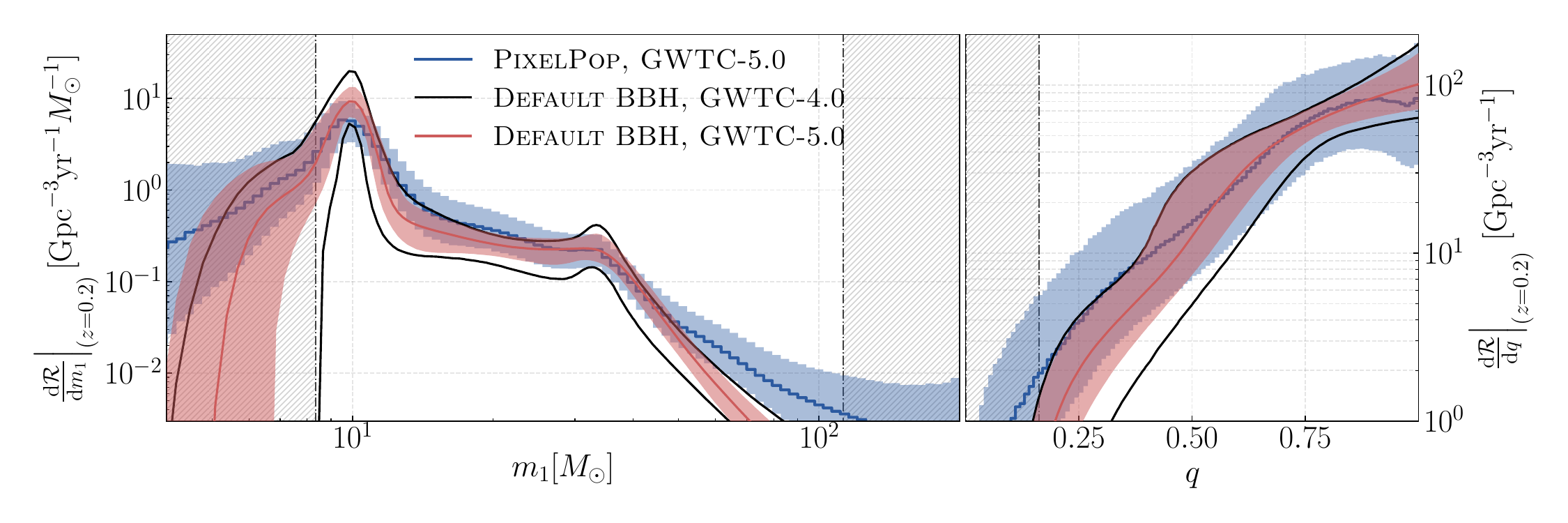}
	    \caption{\label{fig:p_m1q_bbh} The distribution of the primary mass of the \ac{BBH} population (left), and the mass ratio distribution (right). The solid lines show the {median merger-rate density} for each model, with the shaded bands depicting the $90\%$ credible regions. The solid black lines show the $90\%$ credible bounds for the strongly-parameterized model from \gwtcfour. The hatched grey areas behind the dot-dashed lines---in this and all subsequent figures---correspond to regions at the edges of the parameter space with $< 1\%$ of \ac{BBH} parameter estimation samples equally weighted across all detections. {This is intended as a heuristic proxy for parts of parameter space where the models may not be actively constrained by the data, and where modeling and prior choices may become particularly important}. See Sec.~\ref{Sec:methods} for more details.}
\end{figure*}

\subsection{Merger Rates}
\label{sec:full_spectrum_rates}

 The full mass spectrum models described above model the joint $m_1 - m_2$ distribution and allow us to self-consistently estimate the rates of stellar-mass \ac{BBH}, \ac{BNS}, \ac{NSBH}, and \ac{IMBH} mergers. Following \cite{2025arXiv250818083T}, we assume that the threshold between a \ac{NS} and a black hole is  $2.5\,\Msun$, consistent with both astrophysical observations and nuclear physics~\citep{2020PhRvD.101l3007L, 2020Sci...370.1450D, 2021PhRvD.104f3003L, 2022Natur.606..276H, Ai:2023ykc, Dittmann:2024mbo, 2024ApJ...971L..19R, Koehn:2024set, 2017ApJ...850L..19M,2018ApJ...852L..25R,2018PhRvD..97b1501R,2020CQGra..37d5006A,2021ApJ...908L..28N, 2026ApJ...998..272C, 2026PhRvD.113h3013M}. 

Table~\ref{tab:rates} shows the estimated rates for the different category of mergers using \pixelpop and \fullpop. {In order to obtain a conservative merger-rate estimate that is less sensitive to} modeling assumptions, we also calculate a joint merger rate interval estimated as the union of the $90\%$ credible intervals of \fullpop and \pixelpop . Since \gwtcfive includes no new \acp{CBC} with significant support below a component mass of $2.5\,\Msun$, the upper {bounds} of the rates for \ac{NSBH} and \ac{BNS} mergers are lower than in \gwtcfour. 

We also define an \ac{IMBH} source category comprising binaries with both $m_1, m_2 \in [100, 200]\,\Msun$. The most massive \ac{BBH} in \gwtcfive, GW231123\_135430 (henceforth GW231123), has negligible support for source-frame masses above $200\,\Msun$~\citep{2025ApJ...993L..25A} under standard parameter estimation priors; any rate estimate above this mass will therefore be driven either by priors or by model extrapolations from lower masses. While the detector network still has small but non-negligible sensitivity up to source-frame masses of $500\,\Msun$~\citep{GWTC:Methods, 2022A&A...659A..84A}, we find that our upper limits on merger rates in this regime are both highly dependent on modeling assumptions and {astrophysically uninteresting when contrasted with theoretical models~\citep[e.g.,][]{2022ApJ...933..170F}}.

\subsection{BNS and NSBH Properties}
\label{sec:bns_nsbh_props}
{We do not present analyses of \ac{BNS} and \ac{NSBH} systems and refer the readers to see Secs.~4.3 and 5 of \citet{2025arXiv250818083T}, and Sec.~5 of \citet{2024ApJ...970L..34A} for our latest analyses of \ac{NS} population properties inferred using \ac{GW} detections}. Recent claims of eccentricity in the \ac{NSBH} merger, GW200105\_162426~\citep{2024CoTPh..76g5402F,Morras:2025xfu, Jan:2025fps,Planas:2025plq,Tiwari:2025fua}, if confirmed, might warrant revisiting the \ac{NS} mass models used in previous studies~\citep{2021ApJ...921L..25L, Biscoveanu:2022iue}.

\subsection{BBH Properties} 
\label{sec:bbh_props}

{We next focus on \ac{BBH} mergers. Following \gwtcthree and \gwtcfour~\citep{2023PhRvX..13a1048A, 2025arXiv250818083T}, we only include events with \ac{FAR} $\leq 1\,\mathrm{yr}^{-1}$ and for which both $m_1, m_2 \geq 3\,\Msun$ at $99\%$ credibility {with default parameter estimation priors~\citep{GWTC:Methods, GWTC:Results}}. {This yields a cumulative catalog of $259$ \ac{BBH} candidates}. {The \ac{BBH} mass-threshold here is more conservative than what is used in the full-spectrum analyses in Sec.~\ref{sec:full_spectrum} and Sec.~\ref{sec:full_spectrum_rates}. This is because the goal there is to model the whole \ac{CBC} population and self-consistently estimate the rates of each category}. The mass threshold here, instead aims to ensure that sources containing \acp{NS} do not contaminate our inference of \ac{BBH} properties under population priors. Nevertheless, because our formalism for selection effects (see Sec.~\ref{Sec:methods}) does not account for this additional selection criteria, features of the \ac{BBH}-only analyses below ${\sim}8\,\Msun$ should be interpreted with caution.}{ Following our previous papers~\citep{2025arXiv250818083T, 2023PhRvX..13a1048A} we quote all \ac{BBH} differential merger rates at $z=0.2$.}

\subsubsection{Primary Mass Distribution}
\label{sec:pm1_bbh}

The left subplot of Fig.~\ref{fig:p_m1q_bbh} shows the \ac{BBH} merger rate distribution as a function of the primary mass. We use \defbbh as our strongly-parameterized model and a $(m_1, m_2)$-\pixelpop, with the BBH selections described above, as our weakly-parameterized model. We find qualitatively good agreement between the weakly- and strongly-parameterized models for the primary-mass distribution. 

\textbf{We robustly find a global peak at $m_1 \sim 10\,\Msun$ and a feature in the mass spectrum at $m_1 \sim 35\,\Msun$}, consistent with \cite{2025arXiv250818083T}. We find that $\CIPlusMinusPercent{\OverviewEstimates[bkpl2pk_10msun_fracs]}$ ($\CIPlusMinusPercent{\OverviewEstimates[pp_10msun_fracs]}$) {of \ac{BBH} mergers have $m_1 \in (8, 15) \,\Msun$ with the \defbbh (\pixelpop)} model. Both models observe a clear drop in density on both sides of the peak.
 {With the \defbbh model, we find a break in the power law at $m_{\rm break} = \CIPlusMinus{\OverviewEstimates[break_mass]}\,\Msun$}, where the slope changes from $\alpha_1 =  \CIPlusMinus{\OverviewEstimates[bkpl2pk_alpha_1]}$ to $\alpha_2 =  \CIPlusMinus{\OverviewEstimates[bkpl2pk_alpha_2]}$. This implies $\alpha_2 > \alpha_1$ at $\OverviewEstimates[alpha_2_steepness]\%$ credibility, consistently with~\citep{2025arXiv250818083T}. However, it is possible that such steepening is dependent on our assumptions about the maximum mass~\citep{Mould:2026sww}.

Earlier analyses reported hints of additional peaks at $m_1 \sim 20\,\Msun \text{ and } 60\,\Msun$ when using weakly-parameterized models (e.g.~\citealp{2021ApJ...913L..19T, 2023ApJ...946...16E, 2024MNRAS.527..298T, 2024arXiv240702460M, 2025arXiv250818083T, 2025PhRvD.111l3046G, 2026arXiv260103257P}). We do not find evidence for either of them in \gwtcfive based on the \pixelpop primary-mass distribution. Recent works suggest the presence of a subpopulation of mergers with rapidly-spinning \acp{BH}, peaking around $16\,\Msun$ (\citealt{Plunkett:2026pxt, Farah:2026jlc, Tong:2025xir}; also see Sec.~\ref{sec:hierarchical}). More {targeted} joint models of spins and masses might be able to tease out such substructure.

\begin{figure*}[ht]
	\centering
	    \includegraphics[width=1.0\textwidth]{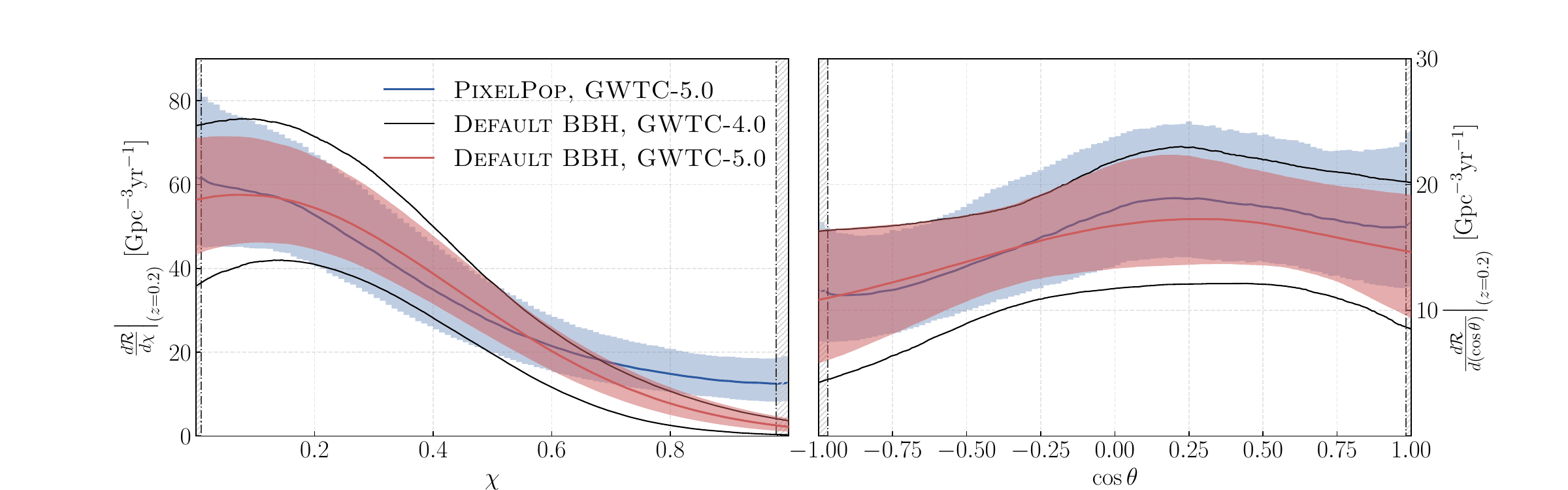}
	    \caption{\label{fig:p_bbh_component_spins} Differential \ac{BBH} merger rate as a function of component spin-magnitudes (left) and tilts (right), assuming that the magnitudes and tilts of both components come from the same underlying distribution. The solid lines show the {median merger-rate density} for each model, with the shaded bands depicting the $90\%$ credible regions. The solid black lines show the $90\%$ credible bounds for the strongly-parameterized model from \gwtcfour}
\end{figure*}

\subsubsection{Mass Ratio Distribution}
\label{sec:pq_bbh}
\textbf{We find that the global \ac{BBH} mass-ratio distribution is consistent with a peak at $q=1$}.
The right subplot of Fig.~\ref{fig:p_m1q_bbh} shows the mass-ratio distribution of the \ac{BBH} population, taking a $(m_1, q)$-\pixelpop analysis as our weakly-parameterized model. This is in-line with the results from \gwtcfour: although the \textsc{B-Spline} model highlighted in~\cite{2025arXiv250818083T} (App.~C) finds evidence for a peak in the global distribution at  $q \approx 0.8$, other flexible models explored there find that the distribution is consistent with a peak at $q = 1$  suggesting some degree of model dependence. {When the mass-ratio distribution is fit with a power law, $p(q) \propto q^{\beta_q}$, we find $\beta_q = \CIPlusMinus{\OverviewEstimates[beta_q]}$. This is a somewhat flatter distribution than the value of $\beta_q = 1.2^{+ 1.2}_{- 1.0}$ from \gwtcfour, but still statistically consistent with it}. 

\subsubsection{Spin Distribution}
\label{Sec:Overview_spins}

We first consider the individual spin magnitudes $(\chi_1, \chi_2)$ and tilts $(\cos \theta_1, \cos \theta_2)$.%
\footnote{Following~\citet{2025arXiv250818083T}, we define tilts at a fixed reference frequency, which is typically 20 Hz but differs for some events. Yet, tilts remain constant over the inspiral to leading order \citep{1994PhRvD..49.6274A} and we have checked that the reference point does not impact our conclusions.}
Figure~\ref{fig:p_bbh_component_spins} shows their distributions inferred using the \defbbh model, and the $(\chi, \cos \theta)$-\pixelpop model. The former models both $\chi_1$ and $\chi_2$ as independently and identically distributed Gaussians truncated over the range of $[0, 1]$.  The spin tilts, $\cos \theta_1$ and $\cos \theta_2$, are assumed to be non-independently distributed as a mixture model consisting of an isotropic and a non-isotropic component (see App.~\ref{Sec:default_bbh_corners}). The $(\chi, \cos \theta)$-\pixelpop model treats both component spins and spin tilts as independent and identically distributed draws from the same underlying distribution. 

Both models infer \ac{BBH} spin distributions that are consistent with our \gwtcfour inference in \citet{2025arXiv250818083T}. The spin tilt distribution is broad, with the possibility of a peak in $\cos \theta$ between $[0, 0.5]$ or a broader plateau (right panel, Fig.~\ref{fig:p_bbh_component_spins}). We find that $\CIPlusMinusPercent{\OverviewEstimates[bkpl2pk_aligned_fracs]}$ ($\CIPlusMinusPercent{\OverviewEstimates[pp_aligned_fracs]}$) of \acp{BH} {are aligned at some degree -- i.e., $\cos \theta > 0$ with} -- the orbital angular momentum when using \defbbh (\pixelpop). The spin-magnitude distribution, on the other hand, peaks at small values but with a tail {extending beyond} $\chi \sim 0.8$ (left panel, Fig.~\ref{fig:p_bbh_component_spins}). Using the \defbbh (\pixelpop) model, we infer that $\CIPlusMinusPercent{\OverviewEstimates[bkpl2pk_aleq_half_fracs]}$ ($\CIPlusMinusPercent{\OverviewEstimates[pp_aleq_half_fracs]}$) of \acp{BH} have $\chi \leq 0.5$.

\begin{figure*}[ht]
	\centering
	    \includegraphics[width=1.0\textwidth]{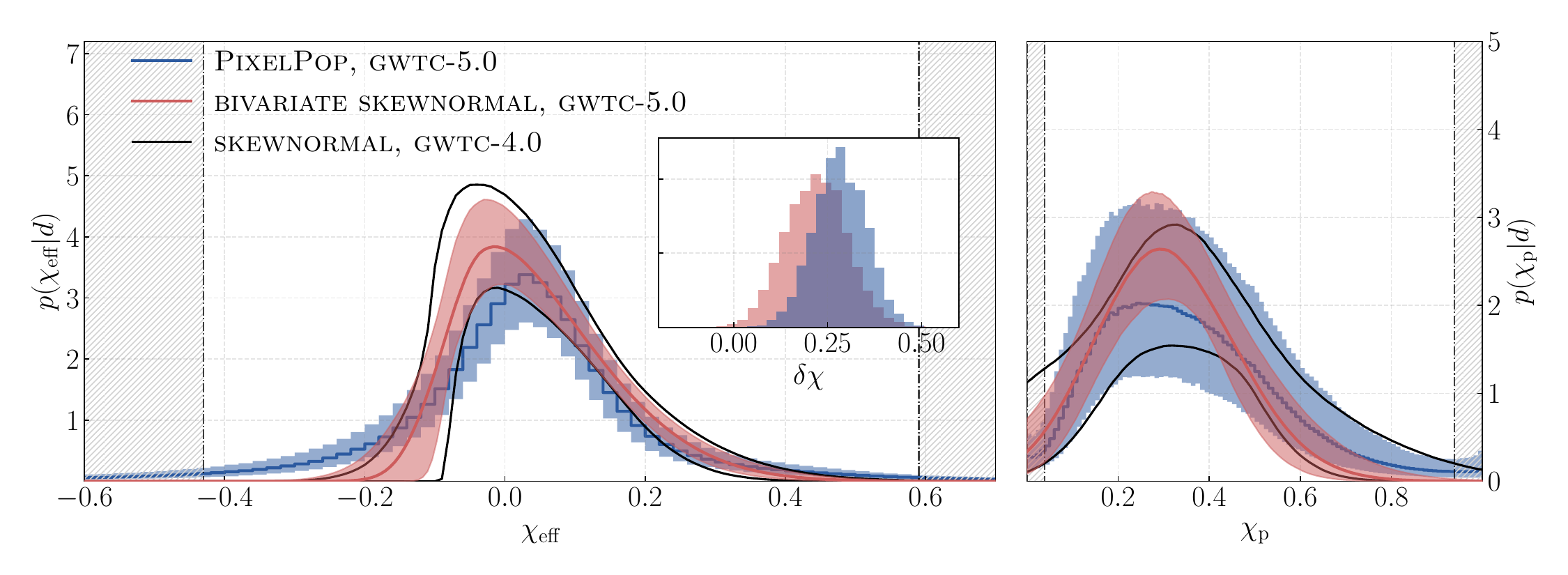}
	    \caption{\label{fig:bbh_efective_spins} Posterior distribution of effective spins in the \ac{BBH} population: $\chieff$ (left) and $\chip$ (right). {The inset shows that the asymmetry about zero, $\delta \chi = p(\chi_{\rm eff} > 0) - p(\chi_{\rm eff} < 0)$ is found to be positive, indicating a preference for aligned binaries}}
\end{figure*}

When modeling the component spin-magnitudes separately and allowing for correlations between them using \tgmms analysis~\citep[App.~\ref{sec:TGIMM};][]{2026ApJ...996...71H,Hussain:2025llf}, we infer that the fraction of \ac{BBH} mergers with small spins $\chi_{1, 2} \lesssim 0.25$ is {$\CIPlusMinus{\OverviewEstimates[TGMM_dominant_mode_fracs]}$}, with the rest belonging to a subpopulation with at least one rapidly-spinning \ac{BH} $\chi \sim 0.7$ (App.~\ref{sec:TGIMM}, Fig.~\ref{Fig:TGMM}). \textbf{We now confidently find this high-spin subpopulation with \gwtcfive} confirming similar indications from previous analyses using \gwtcthree and \gwtcfour~\citep{2026ApJ...996...71H, 2025ApJ...994..261A, 2026ApJ...996..144G, Banagiri:2025dmy, 2026arXiv260505300W}, and consistent with our further findings on \acp{BH} with rapid spins in Sec.~\ref{sec:hierarchical}. {The \tgmms analysis, in particular, also raises the possibility that the secondary \ac{BH} in this subpopulation might also have rapid spins which, if confirmed, has implications for their formation mechanism (see App.~\ref{sec:TGIMM}).}

We obtain a complementary view of the spins by considering the effective inspiral and precessing spin parameters, $\chieff$~\citep{{2008PhRvD..78d4021R,2011PhRvL.106x1101A}} and $\chip$ \citep{{2011PhRvD..84b4046S, 2012PhRvD..86j4063S, 2015PhRvD..91b4043S, Gerosa:2020aiw}}, respectively. These are compressed descriptions of the spin degrees of freedom that respectively encode mass-weighted projections of the component spins along and perpendicular to the orbital angular momentum. {Figure~\ref{fig:bbh_efective_spins} shows their distributions using the \bivskewchichi and the $(\chieff, \chip)$-\pixelpop models. The two analyses disagree about the skewness of the distribution (App.~\ref{sec:bivariate_skewnormal}): while \bivskewchichi finds a right-skewed \chieff distribution (Fig.~\ref{fig:bsn_corners}), \pixelpop infers a \chieff{} distribution with significantly lower skewness. A certain degree of model dependence of the skewness was already highlighted in \gwtcfour (see Fig. 21 of 
\citealt{2025arXiv250818083T}).}

However, both models agree that \ac{BBH} mergers are more likely to have positive (albeit small) rather than negative \chieff values, which could provide robust limits on preferentially-aligned channels such as isolated binaries evolution or AGN disks~\citep{2018PhRvD..98h4036G, 2016ApJ...832L...2R, Banagiri:2025dxo}. We quantify this with the degree of asymmetry about zero, {$\delta \chi = p(\chi_{\rm eff} > 0) - p(\chi_{\rm eff} < 0)$, finding that} $\delta \chi = \CIPlusMinus{\OverviewEstimates[bsn_chieff_delchi]}$ ($\CIPlusMinus{\OverviewEstimates[pp_chieff_delchi]})$ {with} \bivskewchichi (\pixelpop) model  (Fig.~\ref{fig:bbh_efective_spins} inset). {Calculating the union of the $90\%$ credible intervals on $\delta \chi$ inferred by \bivskewchichi and \pixelpop, \textbf{we estimate that at least $\OverviewEstimates[bsn_chieff_asymmetry][5th percentile] \text{--} \OverviewEstimates[pp_chieff_asymmetry][95th percentile] \%$ of mergers must originate from preferentially aligned channels to explain the observed asymmetry.}}

Moreover, consistent with previous catalogs~\citep{2021ApJ...913L...7A, 2023PhRvX..13a1048A, 2025arXiv250818083T}, \textbf{we find that $\CIPlusMinusPercent{\OverviewEstimates[bsn_neg_chieff_frac]}$ ($\CIPlusMinusPercent{\OverviewEstimates[pp_neg_chieff_frac]}$) of binaries have negative $\chieff$ according to the \bivskewchichi (\pixelpop) model; indicating systems with at least one \ac{BH} spin misaligned at more than $90^{\circ}$ with the orbital angular momentum}.  If one assumes that all such binaries come from dynamical formation channels such as dense stellar clusters
this suggests that a majority of \ac{BBH} mergers originate from such channels. {However, it may be possible for anti-aligned binaries to {develop} if the stellar progenitors of \acp{BH} experience large supernova kicks~\citep{Baibhav:2024rkn}. Alternatively, \cite{Kiroglu:2025bbp} posits that the entire \chieff~ distribution might be explained by mergers in dynamical environments where a preference for mild alignment might arise from stellar collisions with compact object binaries.} 

{Finally, we find that the $\chi_{\rm p}$ distribution peaks at a small but non-zero value of $\chi_{p}$ indicating that a non-negligible fraction of the \ac{BBH} population have in-plane spins and experience precession.}

\subsubsection{Redshift Distribution}
\label{sec:bbh_z}

\begin{figure*}[ht!]
	\centering
	    \includegraphics[width=1.0\textwidth]{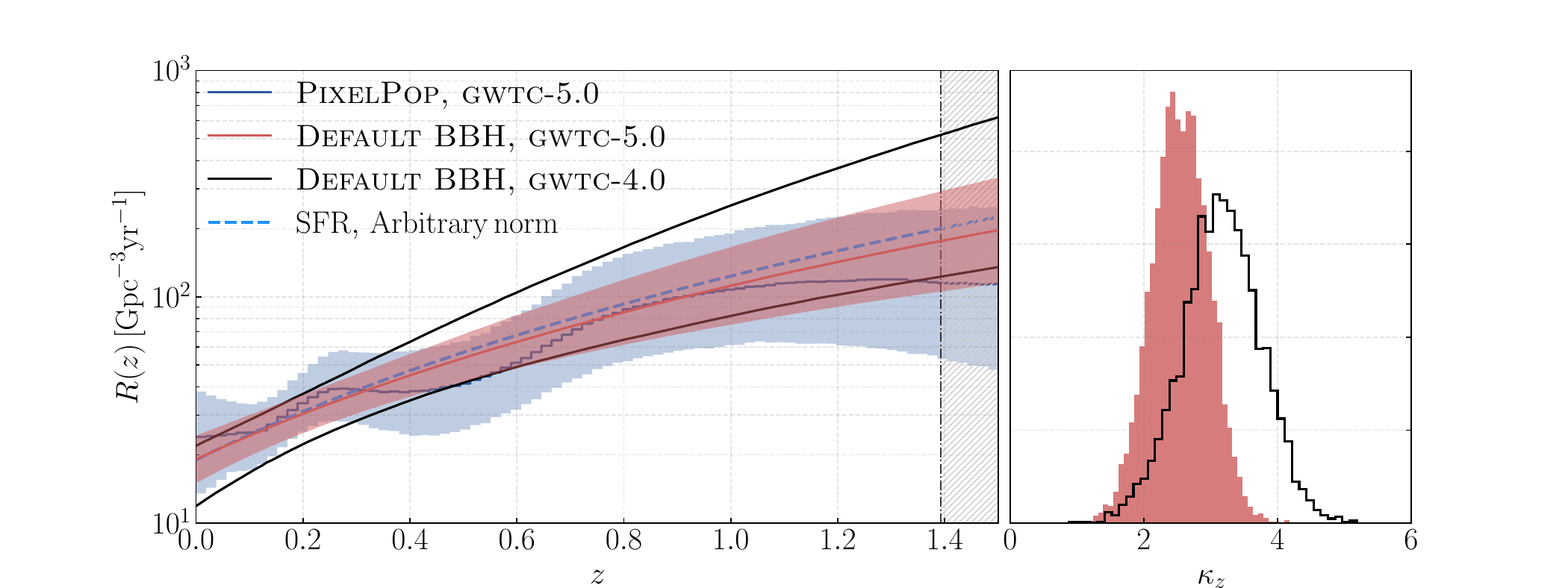}
	    \caption{\label{fig:p_z_bbh} The \ac{BBH} redshift distribution (left), and the inferred value of the redshift spectral index $\kappa_z$ assuming $p(z) \propto (1 + z)^{\kappa_z}$ (right). The dashed blue lines in the left panel show the redshift evolution of the \ac{SFR} (arbitrary normalisation)}
\end{figure*}

{To infer the redshift distribution, we use $z$-\pixelpop as our weakly-parameterized redshift model, and the \defbbh model as our strongly-parameterized one. We assume a $\Lambda$CDM cosmology based on inferred cosmological parameter values from \textsc{planck15}~\citep{2016A&A...594A..13P}}. Figure~\ref{fig:p_z_bbh} shows the comoving source-frame redshift distribution inferred by the two analyses, along with the \gwtcfour bounds for comparison. We estimate the \ac{BBH} merger rate at $z=0.2$ to be
$\CIPlusMinus{\OverviewEstimates[bkpl2pk_bbh_rates]} \, (\CIPlusMinus{\OverviewEstimates[pp_bbh_rates]}) \,  \perGpcyr$ using  the \defbbh (\pixelpop) model for the redshift distribution.\footnote{The rates quoted here and in Tab.~\ref{tab:rates} rest on slightly different assumptions as the lower-limit on black hole mass is different which can lead to a different subset of events contributing to the analysis. Second, all of the binaries in that section are fit with a common redshift distribution, while here we focus only on those that are clearly \ac{BBH} mergers.}. {These rates are consistent with the corresponding \ac{BBH} rates inferred in~~\citet{2025arXiv250818083T} and represents a $\OverviewEstimates[bbh_rate_uncertainity_shrink] \%$ reduction in statistical uncertainty over \gwtcfour when comparing the \defbbh models.}.

With the updated \gwtcfive catalog, we find $\kappa_z = \CIPlusMinus{\OverviewEstimates[bkpl2pk_kappa_z]}$ for  the \defbbh redshift power-law index, which is slightly flatter but  statistically consistent with the estimate using \gwtcfour~\citep{2025arXiv250818083T}, as well as being consistent with the redshift evolution of \ac{SFR}~\citep[]{Madau:2014bja}; see Fig.~\ref{fig:p_z_bbh}. We do not find strong evidence for additional structure in the redshift distribution beyond a power law; a parametric model based on the Madau--Dickinson \ac{SFR} model has a log Bayes factor of $\log_{10}\mathcal{B} = \OverviewEstimates[MD_BBH_def_log10_bayes]$ compared to \defbbh. See App.~\ref{Sec:md_model} for more details. 

It is unclear if the structures in the $z$-\pixelpop inference at $z \leq 0.5$ represent real features or statistical fluctuations. Such features have been noted with past data by non-Bayesian analyses that infer the maximum-likelihood population model~\citep{2023PhRvR...5b3013P, 2026ApJ...996..144G} and with some weakly-parameterized models~\citep{2023ApJ...946...16E}. With present data, these structures are not confidently required in the \pixelpop analysis, and the inference of the model remains statistically consistent with a simple power law.

\newcommand{\TenMsunIsoPeakQFifth}{0.49}
\newcommand{\TenMsunIsoPeakQMedian}{0.80}
\newcommand{\TenMsunIsoPeakQNinetyFifth}{0.97}
\newcommand{\TenMsunIsoPeakQErrorMinus}{0.31}
\newcommand{\TenMsunIsoPeakQErrorPlus}{0.17}
\newcommand{\TenMsunIsoPeakQSummary}{{\TenMsunIsoPeakQMedian}^{+\TenMsunIsoPeakQErrorPlus}_{-\TenMsunIsoPeakQErrorMinus}}
\newcommand{\TenMsunIsoPeakMedianCurveRateRatioQEightToQTen}{1.7}
\newcommand{\TenMsunIsoPeakMedianCurveRateRatioQEightToQSix}{3.2}
\newcommand{\TenMsunIsoPeakRateRatioQEightToQTen}{1.6}
\newcommand{\TenMsunIsoPeakRateRatioQEightToQSix}{3.2}
\newcommand{\TenMsunIsoPeakRateRatioQEightToQTenCredibleFifth}{0.5}
\newcommand{\TenMsunIsoPeakRateRatioQEightToQTenCredibleMedian}{1.6}
\newcommand{\TenMsunIsoPeakRateRatioQEightToQTenCredibleNinetyFifth}{8.2}
\newcommand{\TenMsunIsoPeakRateRatioQEightToQTenCredibleErrorMinus}{1.1}
\newcommand{\TenMsunIsoPeakRateRatioQEightToQTenCredibleErrorPlus}{6.6}
\newcommand{\TenMsunIsoPeakRateRatioQEightToQTenCredibleSummary}{{\TenMsunIsoPeakRateRatioQEightToQTenCredibleMedian}^{+\TenMsunIsoPeakRateRatioQEightToQTenCredibleErrorPlus}_{-\TenMsunIsoPeakRateRatioQEightToQTenCredibleErrorMinus}}
\newcommand{\TenMsunIsoPeakRateRatioQEightToQSixCredibleFifth}{1.1}
\newcommand{\TenMsunIsoPeakRateRatioQEightToQSixCredibleMedian}{3.2}
\newcommand{\TenMsunIsoPeakRateRatioQEightToQSixCredibleNinetyFifth}{14.1}
\newcommand{\TenMsunIsoPeakRateRatioQEightToQSixCredibleErrorMinus}{2.1}
\newcommand{\TenMsunIsoPeakRateRatioQEightToQSixCredibleErrorPlus}{10.9}
\newcommand{\TenMsunIsoPeakRateRatioQEightToQSixCredibleSummary}{{\TenMsunIsoPeakRateRatioQEightToQSixCredibleMedian}^{+\TenMsunIsoPeakRateRatioQEightToQSixCredibleErrorPlus}_{-\TenMsunIsoPeakRateRatioQEightToQSixCredibleErrorMinus}}
\newcommand{\TenMsunIsoPeakContinuumQFifth}{0.22}
\newcommand{\TenMsunIsoPeakContinuumQMedian}{0.73}
\newcommand{\TenMsunIsoPeakContinuumQNinetyFifth}{0.97}
\newcommand{\TenMsunIsoPeakContinuumQErrorMinus}{0.51}
\newcommand{\TenMsunIsoPeakContinuumQErrorPlus}{0.24}
\newcommand{\TenMsunIsoPeakContinuumQSummary}{{\TenMsunIsoPeakContinuumQMedian}^{+\TenMsunIsoPeakContinuumQErrorPlus}_{-\TenMsunIsoPeakContinuumQErrorMinus}}
\newcommand{\TenMsunIsoPeakContinuumMedianCurveRateRatioQEightToQTen}{1.5}
\newcommand{\TenMsunIsoPeakContinuumMedianCurveRateRatioQEightToQSix}{1.9}
\newcommand{\TenMsunIsoPeakContinuumRateRatioQEightToQTen}{1.5}
\newcommand{\TenMsunIsoPeakContinuumRateRatioQEightToQSix}{1.9}
\newcommand{\TenMsunIsoPeakContinuumRateRatioQEightToQTenCredibleFifth}{0.6}
\newcommand{\TenMsunIsoPeakContinuumRateRatioQEightToQTenCredibleMedian}{1.5}
\newcommand{\TenMsunIsoPeakContinuumRateRatioQEightToQTenCredibleNinetyFifth}{5.3}
\newcommand{\TenMsunIsoPeakContinuumRateRatioQEightToQTenCredibleErrorMinus}{1.0}
\newcommand{\TenMsunIsoPeakContinuumRateRatioQEightToQTenCredibleErrorPlus}{3.7}
\newcommand{\TenMsunIsoPeakContinuumRateRatioQEightToQTenCredibleSummary}{{\TenMsunIsoPeakContinuumRateRatioQEightToQTenCredibleMedian}^{+\TenMsunIsoPeakContinuumRateRatioQEightToQTenCredibleErrorPlus}_{-\TenMsunIsoPeakContinuumRateRatioQEightToQTenCredibleErrorMinus}}
\newcommand{\TenMsunIsoPeakContinuumRateRatioQEightToQSixCredibleFifth}{0.9}
\newcommand{\TenMsunIsoPeakContinuumRateRatioQEightToQSixCredibleMedian}{1.9}
\newcommand{\TenMsunIsoPeakContinuumRateRatioQEightToQSixCredibleNinetyFifth}{4.9}
\newcommand{\TenMsunIsoPeakContinuumRateRatioQEightToQSixCredibleErrorMinus}{1.1}
\newcommand{\TenMsunIsoPeakContinuumRateRatioQEightToQSixCredibleErrorPlus}{3.0}
\newcommand{\TenMsunIsoPeakContinuumRateRatioQEightToQSixCredibleSummary}{{\TenMsunIsoPeakContinuumRateRatioQEightToQSixCredibleMedian}^{+\TenMsunIsoPeakContinuumRateRatioQEightToQSixCredibleErrorPlus}_{-\TenMsunIsoPeakContinuumRateRatioQEightToQSixCredibleErrorMinus}}
\newcommand{\TenMsunPixelPopInsideQFifth}{0.48}
\newcommand{\TenMsunPixelPopInsideQMedian}{0.81}
\newcommand{\TenMsunPixelPopInsideQNinetyFifth}{0.98}
\newcommand{\TenMsunPixelPopInsideQErrorMinus}{0.34}
\newcommand{\TenMsunPixelPopInsideQErrorPlus}{0.17}
\newcommand{\TenMsunPixelPopInsideQSummary}{{\TenMsunPixelPopInsideQMedian}^{+\TenMsunPixelPopInsideQErrorPlus}_{-\TenMsunPixelPopInsideQErrorMinus}}
\newcommand{\TenMsunPixelPopInsideMedianCurveRateRatioQEightToQTen}{1.1}
\newcommand{\TenMsunPixelPopInsideMedianCurveRateRatioQEightToQSix}{3.4}
\newcommand{\TenMsunPixelPopInsideRateRatioQEightToQTen}{1.1}
\newcommand{\TenMsunPixelPopInsideRateRatioQEightToQSix}{3.3}
\newcommand{\TenMsunPixelPopInsideRateRatioQEightToQTenCredibleFifth}{0.2}
\newcommand{\TenMsunPixelPopInsideRateRatioQEightToQTenCredibleMedian}{1.1}
\newcommand{\TenMsunPixelPopInsideRateRatioQEightToQTenCredibleNinetyFifth}{6.7}
\newcommand{\TenMsunPixelPopInsideRateRatioQEightToQTenCredibleErrorMinus}{0.9}
\newcommand{\TenMsunPixelPopInsideRateRatioQEightToQTenCredibleErrorPlus}{5.6}
\newcommand{\TenMsunPixelPopInsideRateRatioQEightToQTenCredibleSummary}{{\TenMsunPixelPopInsideRateRatioQEightToQTenCredibleMedian}^{+\TenMsunPixelPopInsideRateRatioQEightToQTenCredibleErrorPlus}_{-\TenMsunPixelPopInsideRateRatioQEightToQTenCredibleErrorMinus}}
\newcommand{\TenMsunPixelPopInsideRateRatioQEightToQSixCredibleFifth}{0.7}
\newcommand{\TenMsunPixelPopInsideRateRatioQEightToQSixCredibleMedian}{3.3}
\newcommand{\TenMsunPixelPopInsideRateRatioQEightToQSixCredibleNinetyFifth}{15.0}
\newcommand{\TenMsunPixelPopInsideRateRatioQEightToQSixCredibleErrorMinus}{2.7}
\newcommand{\TenMsunPixelPopInsideRateRatioQEightToQSixCredibleErrorPlus}{11.7}
\newcommand{\TenMsunPixelPopInsideRateRatioQEightToQSixCredibleSummary}{{\TenMsunPixelPopInsideRateRatioQEightToQSixCredibleMedian}^{+\TenMsunPixelPopInsideRateRatioQEightToQSixCredibleErrorPlus}_{-\TenMsunPixelPopInsideRateRatioQEightToQSixCredibleErrorMinus}}
\newcommand{\TenMsunPixelPopOutsideQFifth}{0.33}
\newcommand{\TenMsunPixelPopOutsideQMedian}{0.80}
\newcommand{\TenMsunPixelPopOutsideQNinetyFifth}{0.98}
\newcommand{\TenMsunPixelPopOutsideQErrorMinus}{0.47}
\newcommand{\TenMsunPixelPopOutsideQErrorPlus}{0.19}
\newcommand{\TenMsunPixelPopOutsideQSummary}{{\TenMsunPixelPopOutsideQMedian}^{+\TenMsunPixelPopOutsideQErrorPlus}_{-\TenMsunPixelPopOutsideQErrorMinus}}
\newcommand{\TenMsunPixelPopOutsideMedianCurveRateRatioQEightToQTen}{0.7}
\newcommand{\TenMsunPixelPopOutsideMedianCurveRateRatioQEightToQSix}{2.1}
\newcommand{\TenMsunPixelPopOutsideRateRatioQEightToQTen}{0.7}
\newcommand{\TenMsunPixelPopOutsideRateRatioQEightToQSix}{2.2}
\newcommand{\TenMsunPixelPopOutsideRateRatioQEightToQTenCredibleFifth}{0.2}
\newcommand{\TenMsunPixelPopOutsideRateRatioQEightToQTenCredibleMedian}{0.7}
\newcommand{\TenMsunPixelPopOutsideRateRatioQEightToQTenCredibleNinetyFifth}{2.1}
\newcommand{\TenMsunPixelPopOutsideRateRatioQEightToQTenCredibleErrorMinus}{0.5}
\newcommand{\TenMsunPixelPopOutsideRateRatioQEightToQTenCredibleErrorPlus}{1.4}
\newcommand{\TenMsunPixelPopOutsideRateRatioQEightToQTenCredibleSummary}{{\TenMsunPixelPopOutsideRateRatioQEightToQTenCredibleMedian}^{+\TenMsunPixelPopOutsideRateRatioQEightToQTenCredibleErrorPlus}_{-\TenMsunPixelPopOutsideRateRatioQEightToQTenCredibleErrorMinus}}
\newcommand{\TenMsunPixelPopOutsideRateRatioQEightToQSixCredibleFifth}{0.8}
\newcommand{\TenMsunPixelPopOutsideRateRatioQEightToQSixCredibleMedian}{2.2}
\newcommand{\TenMsunPixelPopOutsideRateRatioQEightToQSixCredibleNinetyFifth}{5.5}
\newcommand{\TenMsunPixelPopOutsideRateRatioQEightToQSixCredibleErrorMinus}{1.3}
\newcommand{\TenMsunPixelPopOutsideRateRatioQEightToQSixCredibleErrorPlus}{3.4}
\newcommand{\TenMsunPixelPopOutsideRateRatioQEightToQSixCredibleSummary}{{\TenMsunPixelPopOutsideRateRatioQEightToQSixCredibleMedian}^{+\TenMsunPixelPopOutsideRateRatioQEightToQSixCredibleErrorPlus}_{-\TenMsunPixelPopOutsideRateRatioQEightToQSixCredibleErrorMinus}}
\newcommand{\TenMsunBGPInsideQFifth}{0.56}
\newcommand{\TenMsunBGPInsideQMedian}{0.82}
\newcommand{\TenMsunBGPInsideQNinetyFifth}{0.98}
\newcommand{\TenMsunBGPInsideQErrorMinus}{0.26}
\newcommand{\TenMsunBGPInsideQErrorPlus}{0.16}
\newcommand{\TenMsunBGPInsideQSummary}{{\TenMsunBGPInsideQMedian}^{+\TenMsunBGPInsideQErrorPlus}_{-\TenMsunBGPInsideQErrorMinus}}
\newcommand{\TenMsunBGPInsideMedianCurveRateRatioQEightToQTen}{1.3}
\newcommand{\TenMsunBGPInsideMedianCurveRateRatioQEightToQSix}{2.8}
\newcommand{\TenMsunBGPInsideRateRatioQEightToQTen}{1.3}
\newcommand{\TenMsunBGPInsideRateRatioQEightToQSix}{2.8}
\newcommand{\TenMsunBGPInsideRateRatioQEightToQTenCredibleFifth}{0.7}
\newcommand{\TenMsunBGPInsideRateRatioQEightToQTenCredibleMedian}{1.3}
\newcommand{\TenMsunBGPInsideRateRatioQEightToQTenCredibleNinetyFifth}{4.0}
\newcommand{\TenMsunBGPInsideRateRatioQEightToQTenCredibleErrorMinus}{0.6}
\newcommand{\TenMsunBGPInsideRateRatioQEightToQTenCredibleErrorPlus}{2.7}
\newcommand{\TenMsunBGPInsideRateRatioQEightToQTenCredibleSummary}{{\TenMsunBGPInsideRateRatioQEightToQTenCredibleMedian}^{+\TenMsunBGPInsideRateRatioQEightToQTenCredibleErrorPlus}_{-\TenMsunBGPInsideRateRatioQEightToQTenCredibleErrorMinus}}
\newcommand{\TenMsunBGPInsideRateRatioQEightToQSixCredibleFifth}{1.1}
\newcommand{\TenMsunBGPInsideRateRatioQEightToQSixCredibleMedian}{2.8}
\newcommand{\TenMsunBGPInsideRateRatioQEightToQSixCredibleNinetyFifth}{10.8}
\newcommand{\TenMsunBGPInsideRateRatioQEightToQSixCredibleErrorMinus}{1.7}
\newcommand{\TenMsunBGPInsideRateRatioQEightToQSixCredibleErrorPlus}{8.0}
\newcommand{\TenMsunBGPInsideRateRatioQEightToQSixCredibleSummary}{{\TenMsunBGPInsideRateRatioQEightToQSixCredibleMedian}^{+\TenMsunBGPInsideRateRatioQEightToQSixCredibleErrorPlus}_{-\TenMsunBGPInsideRateRatioQEightToQSixCredibleErrorMinus}}
\newcommand{\TenMsunBGPOutsideQFifth}{0.22}
\newcommand{\TenMsunBGPOutsideQMedian}{0.64}
\newcommand{\TenMsunBGPOutsideQNinetyFifth}{0.96}
\newcommand{\TenMsunBGPOutsideQErrorMinus}{0.43}
\newcommand{\TenMsunBGPOutsideQErrorPlus}{0.32}
\newcommand{\TenMsunBGPOutsideQSummary}{{\TenMsunBGPOutsideQMedian}^{+\TenMsunBGPOutsideQErrorPlus}_{-\TenMsunBGPOutsideQErrorMinus}}
\newcommand{\TenMsunBGPOutsideMedianCurveRateRatioQEightToQTen}{1.0}
\newcommand{\TenMsunBGPOutsideMedianCurveRateRatioQEightToQSix}{1.1}
\newcommand{\TenMsunBGPOutsideRateRatioQEightToQTen}{1.0}
\newcommand{\TenMsunBGPOutsideRateRatioQEightToQSix}{1.1}
\newcommand{\TenMsunBGPOutsideRateRatioQEightToQTenCredibleFifth}{0.7}
\newcommand{\TenMsunBGPOutsideRateRatioQEightToQTenCredibleMedian}{1.0}
\newcommand{\TenMsunBGPOutsideRateRatioQEightToQTenCredibleNinetyFifth}{1.6}
\newcommand{\TenMsunBGPOutsideRateRatioQEightToQTenCredibleErrorMinus}{0.3}
\newcommand{\TenMsunBGPOutsideRateRatioQEightToQTenCredibleErrorPlus}{0.6}
\newcommand{\TenMsunBGPOutsideRateRatioQEightToQTenCredibleSummary}{{\TenMsunBGPOutsideRateRatioQEightToQTenCredibleMedian}^{+\TenMsunBGPOutsideRateRatioQEightToQTenCredibleErrorPlus}_{-\TenMsunBGPOutsideRateRatioQEightToQTenCredibleErrorMinus}}
\newcommand{\TenMsunBGPOutsideRateRatioQEightToQSixCredibleFifth}{0.7}
\newcommand{\TenMsunBGPOutsideRateRatioQEightToQSixCredibleMedian}{1.1}
\newcommand{\TenMsunBGPOutsideRateRatioQEightToQSixCredibleNinetyFifth}{1.8}
\newcommand{\TenMsunBGPOutsideRateRatioQEightToQSixCredibleErrorMinus}{0.4}
\newcommand{\TenMsunBGPOutsideRateRatioQEightToQSixCredibleErrorPlus}{0.7}
\newcommand{\TenMsunBGPOutsideRateRatioQEightToQSixCredibleSummary}{{\TenMsunBGPOutsideRateRatioQEightToQSixCredibleMedian}^{+\TenMsunBGPOutsideRateRatioQEightToQSixCredibleErrorPlus}_{-\TenMsunBGPOutsideRateRatioQEightToQSixCredibleErrorMinus}}

\newcommand{\TenMsunIsoPeakContinuumChiNinetyFifth}{0.71}
\newcommand{\TenMsunIsoPeakContinuumChiNinetyMedian}{0.78}
\newcommand{\TenMsunIsoPeakContinuumChiNinetyNinetyFifth}{0.85}
\newcommand{\TenMsunIsoPeakContinuumChiNinetyErrorMinus}{0.07}
\newcommand{\TenMsunIsoPeakContinuumChiNinetyErrorPlus}{0.07}
\newcommand{\TenMsunIsoPeakContinuumChiNinetySummary}{{\TenMsunIsoPeakContinuumChiNinetyMedian}^{+\TenMsunIsoPeakContinuumChiNinetyErrorPlus}_{-\TenMsunIsoPeakContinuumChiNinetyErrorMinus}}
\newcommand{\TenMsunIsoPeakChiNinetyFifth}{0.56}
\newcommand{\TenMsunIsoPeakChiNinetyMedian}{0.68}
\newcommand{\TenMsunIsoPeakChiNinetyNinetyFifth}{0.80}
\newcommand{\TenMsunIsoPeakChiNinetyErrorMinus}{0.12}
\newcommand{\TenMsunIsoPeakChiNinetyErrorPlus}{0.12}
\newcommand{\TenMsunIsoPeakChiNinetySummary}{{\TenMsunIsoPeakChiNinetyMedian}^{+\TenMsunIsoPeakChiNinetyErrorPlus}_{-\TenMsunIsoPeakChiNinetyErrorMinus}}
\newcommand{\TenMsunPixelPopContinuumChiEffFifth}{-0.51}
\newcommand{\TenMsunPixelPopContinuumChiEffMedian}{0.01}
\newcommand{\TenMsunPixelPopContinuumChiEffNinetyFifth}{0.44}
\newcommand{\TenMsunPixelPopContinuumChiEffErrorMinus}{0.52}
\newcommand{\TenMsunPixelPopContinuumChiEffErrorPlus}{0.43}
\newcommand{\TenMsunPixelPopContinuumChiEffSummary}{{\TenMsunPixelPopContinuumChiEffMedian}^{+\TenMsunPixelPopContinuumChiEffErrorPlus}_{-\TenMsunPixelPopContinuumChiEffErrorMinus}}
\newcommand{\TenMsunPixelPopPeakChiEffFifth}{-0.14}
\newcommand{\TenMsunPixelPopPeakChiEffMedian}{0.05}
\newcommand{\TenMsunPixelPopPeakChiEffNinetyFifth}{0.20}
\newcommand{\TenMsunPixelPopPeakChiEffErrorMinus}{0.19}
\newcommand{\TenMsunPixelPopPeakChiEffErrorPlus}{0.15}
\newcommand{\TenMsunPixelPopPeakChiEffSummary}{{\TenMsunPixelPopPeakChiEffMedian}^{+\TenMsunPixelPopPeakChiEffErrorPlus}_{-\TenMsunPixelPopPeakChiEffErrorMinus}}
\newcommand{\TenMsunBGPContinuumChiEffFifth}{-0.20}
\newcommand{\TenMsunBGPContinuumChiEffMedian}{0.02}
\newcommand{\TenMsunBGPContinuumChiEffNinetyFifth}{0.36}
\newcommand{\TenMsunBGPContinuumChiEffErrorMinus}{0.22}
\newcommand{\TenMsunBGPContinuumChiEffErrorPlus}{0.34}
\newcommand{\TenMsunBGPContinuumChiEffSummary}{{\TenMsunBGPContinuumChiEffMedian}^{+\TenMsunBGPContinuumChiEffErrorPlus}_{-\TenMsunBGPContinuumChiEffErrorMinus}}
\newcommand{\TenMsunBGPPeakChiEffFifth}{-0.06}
\newcommand{\TenMsunBGPPeakChiEffMedian}{0.04}
\newcommand{\TenMsunBGPPeakChiEffNinetyFifth}{0.14}
\newcommand{\TenMsunBGPPeakChiEffErrorMinus}{0.10}
\newcommand{\TenMsunBGPPeakChiEffErrorPlus}{0.10}
\newcommand{\TenMsunBGPPeakChiEffSummary}{{\TenMsunBGPPeakChiEffMedian}^{+\TenMsunBGPPeakChiEffErrorPlus}_{-\TenMsunBGPPeakChiEffErrorMinus}}
\newcommand{\TenMsunPixelPopMOneChiEffSpearmanInsideFifth}{-0.07}
\newcommand{\TenMsunPixelPopMOneChiEffSpearmanInsideMedian}{0.13}
\newcommand{\TenMsunPixelPopMOneChiEffSpearmanInsideNinetyFifth}{0.33}
\newcommand{\TenMsunPixelPopMOneChiEffSpearmanInsideErrorMinus}{0.20}
\newcommand{\TenMsunPixelPopMOneChiEffSpearmanInsideErrorPlus}{0.20}
\newcommand{\TenMsunPixelPopMOneChiEffSpearmanInsideSummary}{{\TenMsunPixelPopMOneChiEffSpearmanInsideMedian}^{+\TenMsunPixelPopMOneChiEffSpearmanInsideErrorPlus}_{-\TenMsunPixelPopMOneChiEffSpearmanInsideErrorMinus}}
\newcommand{\TenMsunPixelPopMOneChiEffSpearmanOutsideFifth}{-0.24}
\newcommand{\TenMsunPixelPopMOneChiEffSpearmanOutsideMedian}{-0.01}
\newcommand{\TenMsunPixelPopMOneChiEffSpearmanOutsideNinetyFifth}{0.25}
\newcommand{\TenMsunPixelPopMOneChiEffSpearmanOutsideErrorMinus}{0.23}
\newcommand{\TenMsunPixelPopMOneChiEffSpearmanOutsideErrorPlus}{0.26}
\newcommand{\TenMsunPixelPopMOneChiEffSpearmanOutsideSummary}{{\TenMsunPixelPopMOneChiEffSpearmanOutsideMedian}^{+\TenMsunPixelPopMOneChiEffSpearmanOutsideErrorPlus}_{-\TenMsunPixelPopMOneChiEffSpearmanOutsideErrorMinus}}

\section{Properties of the ${\sim}10\,\Msun$ peak}
\label{Sec:10msun}

The {global maximum of the \ac{BBH} primary-mass distribution at ${\sim}10\,\Msun$} has been extensively studied since it was first identified in \gwtctwo~\citep{Tiwari:2020otp,Edelman:2021zkw,2023PhRvX..13a1048A,GWTC:AstroDist}. The existence of this peak and some of its properties are consistent with expectations from isolated binary evolution \citep{Dominik:2014yma,2020A&A...636A.104B,Giacobbo:2018etu,Wiktorowicz:2019dil,2019MNRAS.490.3740N}, including the stable mass transfer channel \citep[e.g.,][]{vanSon:2022myr, vanSon:2022ylf}.
Those models typically predict that \acp{BH} in the ${\sim}10\,\Msun$ peak should have low spins preferentially aligned with the orbital angular momentum.
Additionally, in the stable mass transfer channel, the combination of mass-ratio reversal, envelope stripping, and the mass-transfer stability cut can cause the mass-ratio distribution to peak away from unity, around $q \sim 0.6\text{--}0.8$ \citep[e.g.,][]{2019MNRAS.490.3740N, 2021ApJ...921L...2O, vanSon:2021zpk, 2021A&A...651A.100O, 2022ApJ...938...45B, 2024MNRAS.530.3706D}.

In \gwtcfive, the peak around $10\, \Msun$ is discernible in both the primary and secondary masses, whether the data are analyzed with strongly- or weakly-parameterized models (see Fig.~\ref{fig:m1m2_fullspectrum} and Fig.~\ref{fig:p_m1q_bbh}).
The overdensity appearing in both components is consistent with the fact that very uneven mass pairings are disfavored by the data (see also Fig.~\ref{fig:highmass_mass_ratio_distribution}). Nevertheless, \cite{Godfrey:2023oxb} found signs in \gwtcthree that systems with primaries $m_1 \sim 10\,\Msun$ are more likely to have a lighter companion, an observation that persisted in \gwtcfour \citep{GWTC:AstroDist}, where we found a preference for mass ratio $q \sim 0.8$ for systems with primaries $m_1 \sim 10\,\Msun$.

\begin{figure*}[ht]
	\centering
	\includegraphics[width=\textwidth]{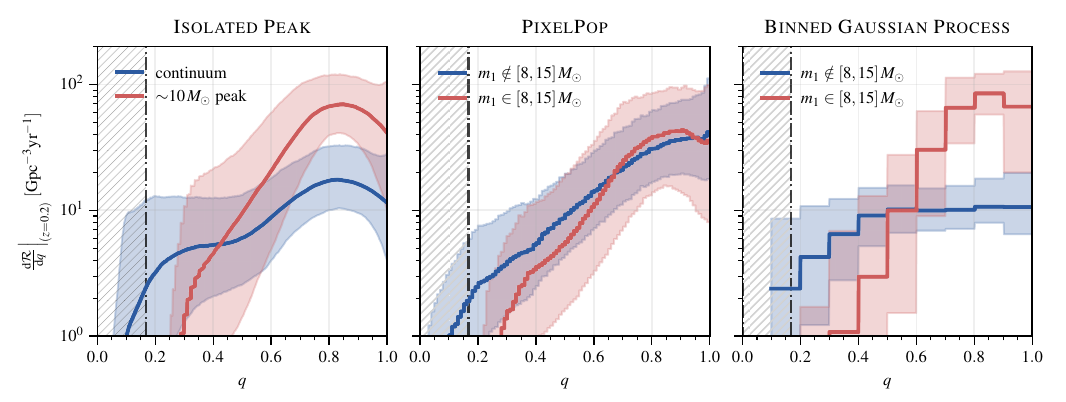}
	\par\medskip
	\includegraphics[width=\textwidth]{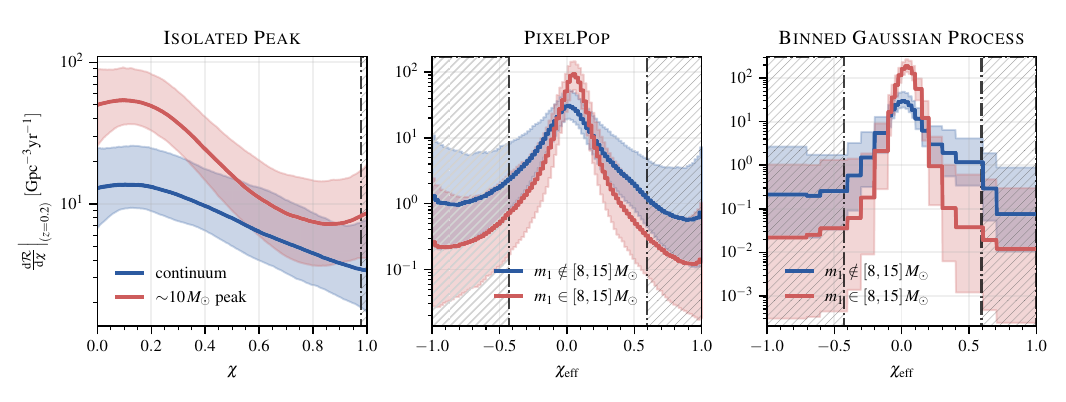}
	\caption{Mass-ratio (top row) and spin (bottom row) distributions for the ${\sim}10\,\mathrm{M}_{\odot}$ peak (red) versus the rest of the population (blue), inferred with \isolatedpeak (left), \pixelpop (middle), and \bgp (right).
    \pixelpop results come from a $(m_1, q)$ run in the top row, and an $(m_1, \chieff)$ run in the bottom row.
    \isolatedpeak natively models the spin magnitude $\chi$ (bottom left panel), while \pixelpop and \bgp model the effective inspiral spin \chieff (bottom middle and right panels). All models favor $q > 0.5$ for the peak. {The \bgp merger-rate-densities at $z=0.2$ are calculated assuming a powerlaw index of $2.7$ (see Tab.~\ref{tab:summary_of_models}).}}
\label{Fig:cosmic_cousins}
\end{figure*}

\textbf{With \gwtcfive, we are not able to confidently confirm a preference for asymmetric masses for systems in the ${\sim}10\, \Msun$ peak.}
While the new dataset could {allow for} an overabundance of $q \approx 0.8$ systems in the peak, such a feature is not strictly required by the data. 
{The degree of support for the $q \sim 0.8$ peak at ${\sim}10\,\Msun$ depends on modeling choices, as evidenced by our comparison to} the dedicated \isolatedpeak mode~\citep{Godfrey:2023oxb} to the more flexible \pixelpop and \bgp models (Fig.~\ref{Fig:cosmic_cousins}). The \isolatedpeak analysis models the ${\sim}10\,\Msun$ peak and the underlying continuum with distinct mass-ratio and spin distributions.
We compare its inferences for the peak (continuum) to those of the weakly-parametrized models within (outside) the $[8, 15] \, \Msun$ primary-mass range.

The \isolatedpeak model finds some preference for mergers in the ${\sim}10\,\Msun$ peak having unequal masses, measuring $q = \TenMsunIsoPeakQSummary$ based on posterior-predictive from the Gaussian peak component (red distribution in the top-left panel of Fig.~\ref{Fig:cosmic_cousins}).
However, unlike in \gwtcfour, this preference does not markedly distinguish the peak from the rest of the population: posterior-predictive draws from the continuum component give $q = \TenMsunIsoPeakContinuumQSummary$ (blue distribution in the top-left panel of Fig.~\ref{Fig:cosmic_cousins}), a broader range than for the peak but with a similar degree of support for $q=0.8$ versus $q=1$.
Even though systems in the peak favor $q=0.8$ on average, the posterior-predictive bands in the top-left panel of Fig.~\ref{Fig:cosmic_cousins} show that both the peak and the continuum can accommodate similar merger-rate densities for $q=1$ as $q=0.8$ within 90\% credibility.
In fact, the merger-rate density at $q=0.8$ is found to be $\TenMsunIsoPeakRateRatioQEightToQTenCredibleSummary$ times the value at $q=1$ for the peak and $\TenMsunIsoPeakContinuumRateRatioQEightToQTenCredibleSummary$ for the continuum, which are both consistent with unity.
In other words, based on the upper and lower bounds of these 90\%-credible ranges, \isolatedpeak does not find that the ${\sim}10\,\Msun$ peak significantly favors $q=0.8$ over $q=1$, although that cannot be ruled out either: the $q=0.8$ versus $q=1$ merger-rate density ratio could be as high as ${\sim}8:1$ and as low as ${\sim}1:2$.

The potential preference for $q=0.8$ in the ${\sim}10\,\Msun$ peak becomes even more ambiguous when we consider other models.
As seen in the top-middle panel of Fig.~\ref{Fig:cosmic_cousins}, \pixelpop does not find a marked preference for unequal masses for binaries in this peak {(red)}. On the contrary, the $(m_1,q)$-\pixelpop analysis suggests that systems in the $[8, 15]\, \Msun$ range are more likely on average to have equal rather than unequal masses.
This is also evident in the right panel of Fig.~\ref{fig:m1m2_fullspectrum} from the $(m_1, m_2)$-\pixelpop run, where the excess at ${\sim}10\, \Msun$ is consistent with the diagonal line $m_1 = m_2$.
\pixelpop finds the merger-rate density at $q=0.8$ to be $\TenMsunPixelPopInsideRateRatioQEightToQTenCredibleSummary$ times the rate at $q=1$ for the peak and $\TenMsunPixelPopOutsideRateRatioQEightToQTenCredibleSummary$ for the continuum.
This means that, while \pixelpop allows the ${\sim}10\,\Msun$ peak to favor $q=0.8$ over $q=1$ by as much as ${\sim}7:1$, the ratio could be as low as ${\sim}1:5$, so a preference for uneven masses in the peak is not required.

The other flexible model, \bgp, finds a similar result (top-right panel of Fig.~\ref{Fig:cosmic_cousins}).
The merger-rate density at $q=0.8$ is found to be $\TenMsunBGPInsideRateRatioQEightToQTenCredibleSummary$ times the value at $q=1$ for the peak {(red)} and $\TenMsunBGPOutsideRateRatioQEightToQTenCredibleSummary$ for the continuum {(blue)}.
At 90\% credibility, thus, the preference for $q=0.8$ over $q=1$ could be as high as ${\sim}4:1$ and as low as ${\sim}1:1.4$ for systems in the peak.
Unlike \pixelpop, \bgp applies a flexible model in the three-dimensional space of $(m_1, q, \chieff)$ but fixes the distribution in redshift and other spin degrees of freedom.

All in all, there is inconclusive evidence for distinctly unequal masses in the ${\sim}10\,\Msun$ peak: all three models considered here allow for such an asymmetry to exist preferentially {in} the peak more so than {in} the rest of the population, but none of the models strictly require it.
In this respect, the data remain consistent with the ${\sim}10\,\Msun$ peak originating in a specific formation channel that yields uneven masses, {such as} stable mass transfer, but the mass-ratio distribution alone is not currently sufficient to confidently ascertain this.
Nevertheless, all models favor $q > 0.5$ for the peak.

If systems in the ${\sim}10\,\Msun$ peak do originate in a specific formation channel, they may also have distinct spin properties.
The \isolatedpeak model, which explicitly assigns an independent spin-magnitude distribution to the peak, finds that \textbf{systems in the peak have a broad distribution of spin magnitudes that is consistent with the rest of the population, although with slightly more support for low magnitudes} (bottom-left panel of Fig.~\ref{Fig:cosmic_cousins}).
Concretely, the 90\%-credible upper limit on the spin magnitude is $\chi_{90\%} = \TenMsunIsoPeakContinuumChiNinetySummary$ for the continuum {(blue)} while only $\chi_{90\%} = \TenMsunIsoPeakChiNinetySummary$ within the peak {(red)}. {We performed an additional check in} the $(m_1, \chieff)$-\pixelpop run { (bottom-middle panel in Fig.~\ref{Fig:cosmic_cousins}), which} similarly finds that systems in the $[8, 15]\, \Msun$ range are more tightly constrained to have low effective inspiral spins than the rest of the population; the median of the \ac{PPD} implies $\chieff = \TenMsunPixelPopContinuumChiEffSummary$ for systems outside that range but $\chieff = \TenMsunPixelPopPeakChiEffSummary$ within it.
The $(m_1, q, \chieff)$ \bgp analysis {(bottom-right panel of Fig.~\ref{Fig:cosmic_cousins})} shows a qualitatively comparable effective-inspiral-spin structure, with a broader $\chieff$ distribution for systems in the $[8, 15]\, \Msun$ range than for the rest of the population; the median \ac{PPD} curves give $\chieff = \TenMsunBGPContinuumChiEffSummary$ outside the $[8, 15]\, \Msun$ range {(blue)} while $\chieff = \TenMsunBGPPeakChiEffSummary$ within it {(red)}.
{Both \pixelpop and \bgp find a slightly more pronounced preference for positive $\chieff$ for the peak.}

Finally, the $(m_1, \chieff)$-\pixelpop run suggests that \textbf{systems in the $[8, 15]\, \Msun$ range may display a correlation between effective inspiral spin and primary mass} that is not present in the rest of the population. The Spearman correlation coefficient \citep{ca468a70-0be4-389a-b0b9-5dd1ff52b33f} between $\log m_1$ and \chieff is found to be $\rho = \TenMsunPixelPopMOneChiEffSpearmanInsideSummary$ for systems in that range while $\rho = \TenMsunPixelPopMOneChiEffSpearmanOutsideSummary$ for the rest of the population ($|\chieff| \lesssim 0.25$ in both cases). {While not yet statistically conclusive, this correlation could be hinting at signatures of binary processes such as accretion-induced spin-up of the first born \ac{BH}~\citep{1999MNRAS.305..654K}}.

\section{Properties of the ${\sim} 35\,\Msun$ feature}
\label{sec:35_msun}

\begin{figure}[tb]
       \centering
        \includegraphics[width=0.47\textwidth]{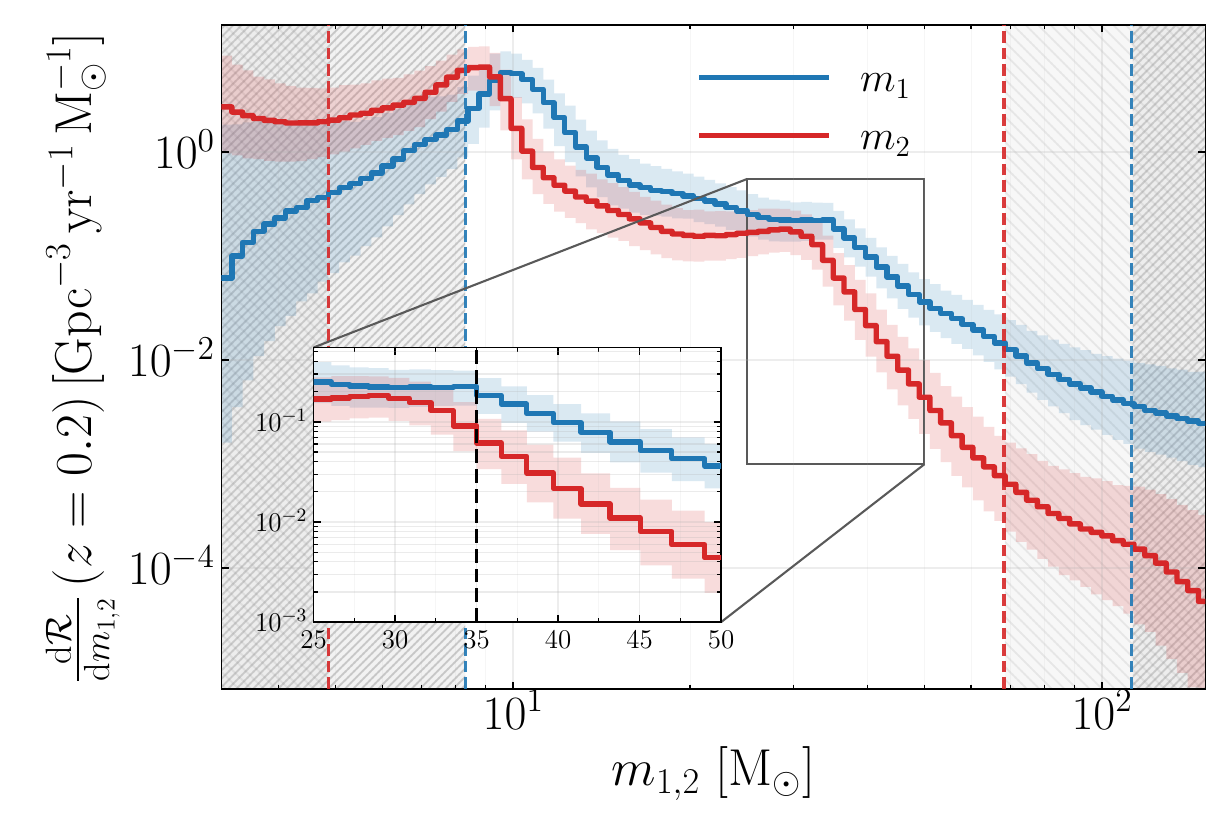}
        \caption{Marginal primary ($m_1$) and secondary ($m_2$) mass distributions reconstructed using $(m_1, m_2)$-\pixelpop model. The inset shows the merger-rate density in the mass range $25$--$50\,\Msun$. }
\label{Fig:35_m1m2}
\end{figure}

\begin{figure*}[tb]
        \centering
        \includegraphics[width=\textwidth]{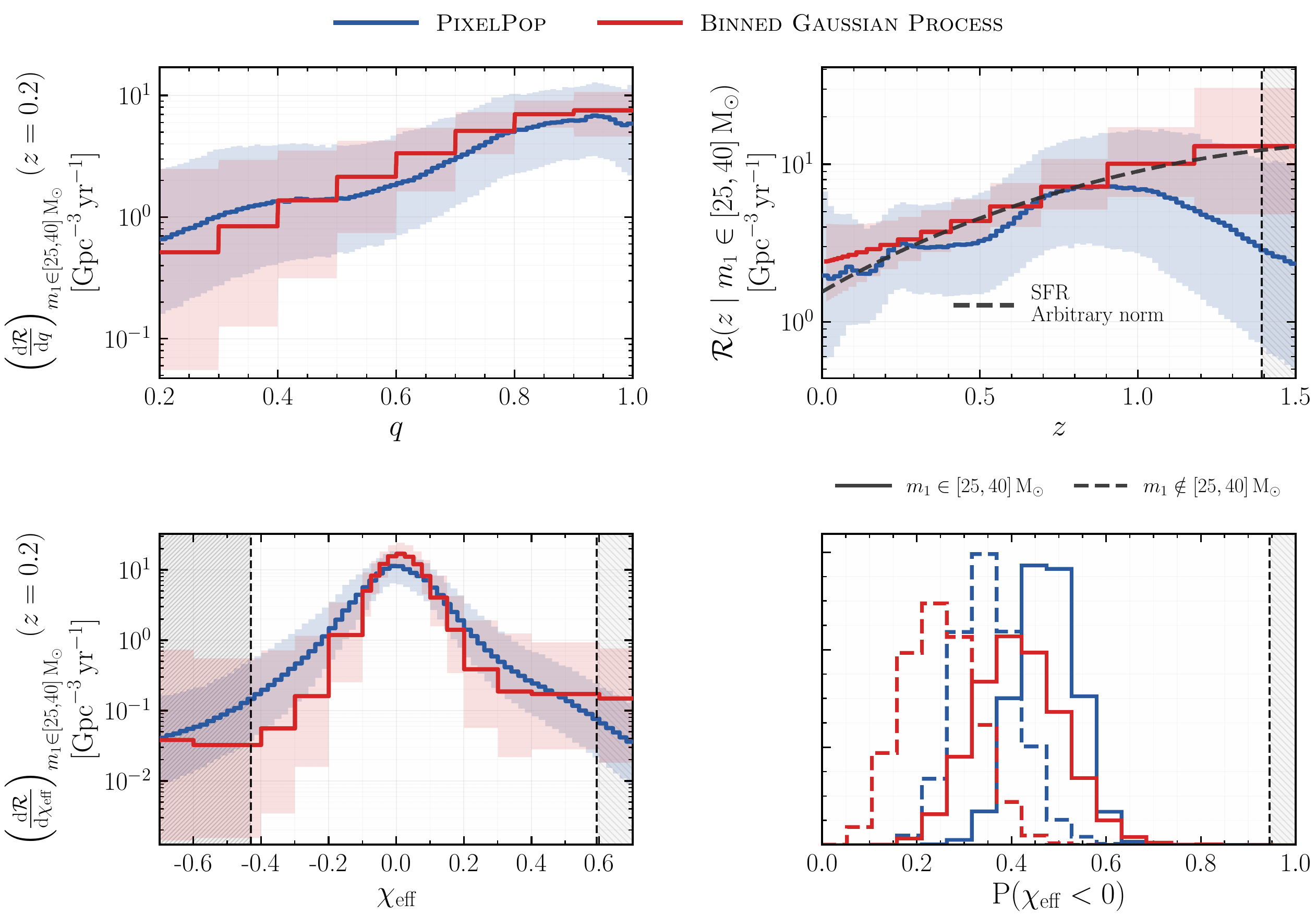}

        \caption{{BBH merger rate for $m_1 \in [25, 40]\,\Msun$, inferred using different variants of the \pixelpop model: $(m_1, q)$ (top left), $(m_1, z)$ (top right), and $(m_1, \chieff)$ (bottom left panel). The bottom right panel shows the inferred fraction of mergers with negative \chieff. These are overplotted with different variants of the  \bgp model as a consistency check: $(m_1, q, \chieff)$ (top left), $(m_1 ,m_2 ,z)$ (top right), and $(m_1, m_2, \chieff)$ (bottom panels). The agreement between \pixelpop and the \bgp demonstrates the robustness of our results in this mass range. In this mass range, the mass-ratio distribution likely peaks at $q = 1$, indicating a preference for equal-mass mergers in the $\sim 35\,\Msun$ population. The redshift distribution is broadly consistent with the star-formation rate and remains robust up to $z \sim 1.2$, beyond which modeling uncertainties dominate. {The effective inspiral spin distribution for the $\sim 35\,\Msun$ population is broadly consistent with that of the full population, with both exhibiting significant support at negative \chieff{}. But the $\sim 35\,\Msun$ population has a stronger preference for a symmetric $\chieff{}$ distribution about zero, as seen by the histograms of the fraction of binaries with $\chieff{} < 0$ (bottom right).}}. {The \bgp merger-rate-densities at $z=0.2$ are calculates assuming a powerlaw index of $2.7$ (see Tab.~\ref{tab:summary_of_models}).}}
\label{Fig:35_qzchieff}
\end{figure*}
\noindent
In previous catalogs, {we identified a potential local maxima in the BBH merger-rate density near} $m_1 \sim 35\,\Msun$ \citep{2021ApJ...913L...7A, 2023PhRvX..13a1048A, 2025arXiv250818083T}. While earlier work (e.g.,~\citealt{2023PhRvX..13a1048A}) identified a peak there with high significance, the analysis with \gwtcfour painted a more complex picture of the shape of the mass distribution, finding that the feature can either be a peak or a {turnover} marking the onset of a decline in merger-rate density~\citep{2025arXiv250818083T, 2026arXiv260300239C}. Such an ambiguity has also been noted in analyses that use more complex models for the mass-ratio distribution of binaries (e.g.,~\citealt{Farah:2023swu,Banagiri:2025dmy}). Here, we revisit the properties of the \acp{BH} that make up the $m_1 \sim 35\,\Msun$ feature with \gwtcfive using the weakly-parameterized models, \pixelpop and \bgp.

Figure~\ref{Fig:35_m1m2} shows the marginal primary- and secondary-mass distributions inferred using the bivariate $(m_1, m_2)$-\pixelpop analysis. For a comparison of the \pixelpop and \bgp $m_1$--$m_2$ distributions, see Fig.~\ref{Fig:m1m2_Appendix}. {The corresponding two-dimensional $(m_1,m_2)$ merger-rate density does not reveal additional structure beyond that seen in the one-dimensional $m_1$ and $m_2$ marginals, so we focus on the marginal distributions.} {In the mass range $[25 \text{--} 40]\,\Msun$, the reconstructed $m_1$ and $m_2$ distributions are largest at $m_1 = \CIPlusMinus{\thirtyfive[m1_peak_25_40]}\,\Msun$ and $m_2 = \CIPlusMinus{\thirtyfive[m2_peak_25_40]} \,\Msun$, respectively.} To quantify the shape of the two distributions, we compare the merger-rate density at several points in that range. For the primary mass, the {marginal merger-rate density at $m_1 = 30\,\Msun$ is a factor of $\CIPlusMinus{\thirtyfive[m1_rate_ratio_30_over_25]}$ times the merger-rate density at $m_1 = 25\,\Msun$, while it is $\CIPlusMinus{\thirtyfive[m1_rate_ratio_30_over_40]}$ times larger than the merger-rate density at $m_1 = 40\,\Msun$. For the secondary mass, the marginal merger-rate density at $m_2 = 30\,\Msun$ is a factor of $\CIPlusMinus{\thirtyfive[m2_rate_ratio_30_over_25]}$ times the merger-rate density at $m_2 = 25\,\Msun$, but $\CIPlusMinus{\thirtyfive[m2_rate_ratio_30_over_40]}$ times larger than the merger-rate density at $m_2 = 40\,\Msun$.}
{In other words, for both the primary- and secondary-mass distributions, we find that the merger-rate density at $25\,\Msun$ is consistent with the value at $30\,\Msun$, whereas the value at $40\,\Msun$ is clearly lower than at $30\,\Msun$}. This shows that, within current levels of observational uncertainty, \textbf{the feature in the mass spectrum marks a change in slope in both the primary- and secondary-mass distributions, with the secondary turning over much more rapidly than the primary}.

{Using both \pixelpop and \bgp, we now turn to other properties of the \ac{BBH} mergers comprising the ${\sim} 35\,\Msun$ feature. We start with their mass ratios in the top left of Fig.~\ref{Fig:35_qzchieff}. We find that the }marginal mass-ratio distribution in this mass range peaks at $q = 1$ in both reconstructions. Using $(m_1, q)$-\pixelpop, we find that the merger-rate density at $q = 1$ is $ \CIPlusMinus{\thirtyfive[pp_q_rate_ratio_1_over_0p7]}$ times larger than at $q = 0.7$. {This means that, while \pixelpop allows for a merger-rate density at $q=1$ that is a factor of $\approx 4.7$ higher than at $q=0.7$, it could also be lower by a factor of $\approx 0.5$; so a preference for equal masses, while favored, is not yet strongly required by the data.} This level of preference for equal-mass mergers is consistent with findings from both \gwtcthree and \gwtcfour for the ${\sim} 35\,\Msun$ population~\citep{2022ApJ...933L..14L, 2024ApJ...977...67L, 2025CQGra..42v5008K, 2025arXiv250818083T, Banagiri:2025dmy}. 

{The redshift evolution of the merger rate (top right of Fig.~\ref{Fig:35_qzchieff}) of the ${\sim} 35\,\Msun$ population}---inferred with both $(m_1, z)$-\pixelpop and \bgp---is broadly consistent with  \ac{SFR}. We observe mild deviations beyond $z \sim 1.2$. However, the inference at such high redshifts can be particularly sensitive to modeling and prior choice as it lies in a data-poor part of parameter space. {We find no evidence that the primary-mass distribution of this population evolves with redshift.}

The $(m_1, \chieff)$-\pixelpop analysis finds that the $\chieff$ distribution for the ${\sim} 35\,\Msun$ population is similar to that of the full population, with both distributions symmetric around zero. This corresponds to roughly $50\%$ of systems having negative, albeit small, effective inspiral spin. This result is consistent with some previous studies using \gwtcthree and \gwtcfour~\citep{2025ApJ...991...17R, 2025arXiv250818083T}. However, other studies report a preference for {${\sim}30\,\Msun$ \acp{BH} to have a \chieff distribution peaking at positive values}~\citep{2025PhRvD.112l3054S, Roy:2025ktr}. As shown in Figs.~\ref{Fig:35_qzchieff} and \ref{Fig:m1m2_Appendix}, we find that although \bgp finds a smaller fraction of mergers at negative \chieff, its results are statistically consistent with \pixelpop within $90\%$ credibility. {We also find no evidence for a correlation between $m_1$ and \chieff in the primary-mass range $[25\text{--}40]\,\Msun$.}

{The ${\sim}35\,\Msun$ feature was originally discussed in the context of pair-instability physics. This is because \acp{PPISN} are expected to compress a broad range of carbon-oxygen core masses into a narrow range of remnant masses, thereby producing a pile up in the \ac{BH} mass spectrum that could have potentially explained this feature~\citep{2016A&A...594A..97B, 2017ApJ...836..244W, 2017MNRAS.470.4739S,2018ApJ...856..173T,2019ApJ...882..121S,2019ApJ...882...36M,2023MNRAS.523.4539K,2025MNRAS.540...90W}}. However, subsequent studies have argued that producing a feature at such low masses through \acp{PPISN}  requires pushing the ${}^{12}\mathrm{C}(\alpha,\gamma){}^{16}\mathrm{O}$ reaction rate above its laboratory-measured value, together with additional assumptions about mixing, winds, envelope stripping, and rotation. Therefore, the observed break near ${\sim} 30\,\Msun$ is unlikely to be a direct signature of \acp{PPISN} alone~\citep{2022ApJ...937..112F, 2023MNRAS.526.4130H,2021MNRAS.501.4514C,2021MNRAS.504..146V,2025PhRvD.112f3053C,2024ApJ...976..121G,Roy:2025ktr,Antonini:2025ilj}.

This motivates the question of what astrophysical channel could instead explain the feature and its properties. Several possibilities have been proposed: chemically homogeneous evolution can form massive, nearly equal-mass binaries \citep{2016MNRAS.460.3545D, 2016MNRAS.458.2634M, 2016A&A...588A..50M}; stable mass transfer with quasi-homogeneous evolution can generate structure near the same mass scale \citep{2023MNRAS.520.5724B}, as is also the case for Population~III binaries \citep{2014MNRAS.442.2963K, 2021MNRAS.504L..28K}, dynamical formation in dense stellar environments, and hierarchical mergers \citep{2018PhRvL.120o1101R, 2023MNRAS.522..466A, 2025ApJ...991...17R}. However it remains unclear if any of these formation mechanisms can explain {\it all} of the properties of the \acp{BH} at these masses explored above~\citep{Roy:2025ktr, Banagiri:2025dmy}. The precise astrophysical origin of the ${\sim} 35 \, \Msun$ feature therefore remains open.

\section{High-Mass Features}\label{sec:high_mass}

\begin{figure}[tb]
    \centering
        \includegraphics[width=1.0\linewidth]{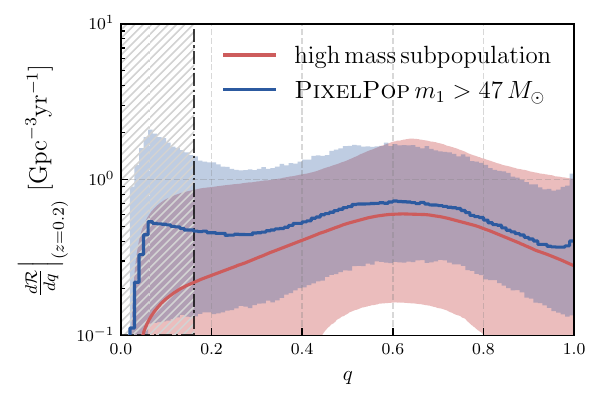}
        \caption{The mass-ratio distribution of the high-mass subpopulation with {$m_1 \geq \CIPlusMinus{\HighMassEstimates[threesubpop_onset_of_the_third_subpop]}\,\Msun$} using the \textsc{three-subpopulation} model (in red, see App.~\ref{sec:banagiri_three_subpop}) and the mass-ratio distribution of the high-mass mergers with $m_1>47\,\Msun$ using $(m_1, q)$-\pixelpop (blue). The spread in the inferred merger rate by \pixelpop at $q < 0.2$ is driven by its agnostic priors.}
        \label{fig:highmass_mass_ratio_distribution}
\end{figure}

We now investigate mergers at higher masses, $m \gtrsim 40\,\Msun$. 
In this region, the \ac{BBH} mass spectrum drops rapidly, consistent with a power law (Sec.~\ref{sec:pm1_bbh});
previous studies have reported a {potential} rise of the mass distribution at $m_1 \sim 60\text{--}70\,\Msun$~\citep{MaganaHernandez:2024qkz, 2025arXiv250818083T, 2026arXiv260103257P}, but we do not find a significant overdensity around this mass scale in \gwtcfive using \pixelpop{} {(Fig.~\ref{fig:m1m2_fullspectrum})}.
{Here we study mergers with $m_1 \gtrsim 40\,\Msun$ in search for signatures in the mass and mass-ratio distributions that differ from those at lower masses, potentially indicating that the high-mass \acp{BH} represent a distinct subpopulation.} \

As discussed in Sec.~\ref{sec:35_msun}, the secondary-mass distribution exhibits a more pronounced decline at higher masses compared to the primary (Fig.~\ref{Fig:35_m1m2}). We quantify the difference using the slopes, i.e., logarithmic probability density versus logarithmic mass, of the marginal $m_1$ and $m_2$ distributions, averaged over two different mass intervals using $(m_1, m_2)$-\pixelpop.
In the range $40\text{--}60\,\Msun$, the slope of the $m_2$ distribution is constrained to be {$\CIPlusMinus{\HighMassEstimates[m2_40_60Msun_slope]}$}, {significantly steeper at ${>}99.9\%$ credibility than in the lower-mass interval $13\text{--}25\,\Msun$ where the $m_2$ slope is $\CIPlusMinus{\HighMassEstimates[m2_13_25Msun_slope]}$}. The $m_1$ distribution also exhibits a steeper decline at higher masses, consistent with a broken power-law structure with a steeper spectral index above ${\sim} 35\,\Msun$ (Sec.~\ref{sec:pm1_bbh}). However, the $m_1$ slope in the $40\text{--}60\,\Msun$ range is constrained to be {$\CIPlusMinus{\HighMassEstimates[m1_40_60Msun_slope]}$} and is still shallower than the corresponding $m_2$ slope with $\HighMassEstimates[40_60Msun_slope_comparison]\%$ credibility.

{Using a model that explicitly allows for a sharp gap, \citet{Tong:2025wpz} reported one in the $m_2$ distribution beginning at  ${\sim} 45 \, \Msun$ with \gwtcfour data~(see also \citealp{Mould:2026sww}}). However, other {strongly-parameterized} models that allow for a more flexible gap shape do not clearly find the sharp onset of a gap~\citep{2026ApJ...998L..20R}. In \gwtcfive, \textbf{while {\pixelpop does not resolve a sharp gap in $m_2$ (Fig.~\ref{fig:m1m2_fullspectrum})}, we consistently find a clear suppression of systems with high $m_2$}. More conservatively, the cumulative distribution of $m_2$ also indicates only a small fraction of mergers above ${\sim} 40\,\Msun$ (Fig.~\ref{fig:cdf_m1_m2} in App.~\ref{sec:m1m2_cdf}).

{The relative scarcity of high-mass secondaries is consistent with the inferred pairing distribution (Fig.~\ref{fig:highmass_mass_ratio_distribution}) of \ac{BBH} mergers at high masses}. We find that \textbf{binaries with $m_1 \gtrsim 40\,\Msun$ can exhibit low mass ratios}, consistent with the properties of a high-mass subpopulation reported in, e.g., \cite{Banagiri:2025dmy, 2026arXiv260317987R}. This tendency for {unequal masses} can also be qualitatively observed in several high-mass binaries discovered in O4b such as GW241127\_061008, GW241102\_144729, GW241230\_084504, GW250109\_074552 etc~\citep{GWTC:Results}. Using \threeSubpopulation, which treats these high-mass mergers as a subpopulation with separate spin and mass-ratio properties (\citealp{Banagiri:2025dmy} and App.~\ref{sec:banagiri_three_subpop}), we constrain the onset of this subpopulation to {$m_1 = \CIPlusMinus{\HighMassEstimates[threesubpop_onset_of_the_third_subpop]}\,\Msun$}{. We find a similar transition in \chieff, Sec.~\ref{sec:HS_results}.}

In this model, the mass-ratio distribution for this high-mass subpopulation is described by a Gaussian and we infer its mean to be at {$q = \CIPlusMinus{\HighMassEstimates[mean_q_of_the_third_subpop]}$} (Fig.~\ref{fig:highmass_mass_ratio_distribution}).
Results for mergers with $m_1 > 47\,\Msun$ from $(m_1, q)$-\pixelpop are overplotted for comparison; they similarly indicate that high-mass mergers preferentially follow a mass-ratio distribution peaked away from unity, supporting the tendency toward unequal-mass systems in this regime.
Together with the mass-ratio distributions at lower $m_1$, this indicates that \ac{BBH} pairing properties evolve with primary mass (cf.~top row of Fig.~\ref{Fig:cosmic_cousins} for mass ratios at $m_1 \sim 10\,\Msun$ and Fig.~\ref{Fig:35_qzchieff} for mass ratios at $m_1 \sim 35\,\Msun$). This could be a signature of underlying differences in the physics of \ac{BBH} formation.

{ One of the predictions of stellar evolution theory is a gap in \ac{BH} masses arising from very massive stars that undergo \acp{PISN} and \acp{PPISN}}~\citep{1964ApJS....9..201F,1967PhRvL..18..379B,Rakavy:1967rgs,Heger:2002by,Woosley:2007qp,Woosley:2016hmi,2017MNRAS.470.4739S,Marchant:2018kun,2019ApJ...887...53F}. Although the precise locations of the lower and upper edges of this pair-instability mass gap depend sensitively on uncertain physical assumptions~\citep{Marchant:2018kun,2019ApJ...887...53F, 2021MNRAS.501.4514C,Renzo:2024jhc}, the presence of such a feature is a robust prediction of most stellar evolution models. 

\textbf{The primary-mass distribution in \gwtcfive, however, shows no clear evidence for such a \ac{PISN} gap}: although it steepens at high mass, it extends smoothly beyond $100\,\Msun$. {This suggests that, if there is a pair-instability gap, it should be polluted by some other formation pathways, e.g., hierarchical mergers from dense stellar clusters}. The physics of several of these alternative formation channels, which are generally expected to be rare to begin with, can make it less likely that they contribute simultaneously to both components of a binary. Binaries with only one component from such a channel will generally be dominant. For example, the rate of mergers involving one second-generation (2G) \ac{BH}---formed from the merger of first-generation (1G) \acp{BH} rather than from stellar collapse---in dense stellar clusters could be more than an order of magnitude higher than that for mergers where both components are 2G \acp{BH}~\citep{Rodriguez:2019huv}. {Motivated by this, previous studies have tested whether the $m_2$ distribution may provide a cleaner view of a putative \ac{PISN} gap~\citep{Tong:2025wpz,Ray:2025xti,Mould:2026sww}}.

The increased support for low $q$ at higher masses in Fig.~\ref{fig:highmass_mass_ratio_distribution} {tentatively} supports the picture of increasing contamination from non-standard formation pathways. In particular, this trend is consistent with 2G+1G hierarchical mergers populating a putative pair-instability gap in the primary. While under the simplest assumption for dynamically assembled 2G+1G mergers one would expect a characteristic mass ratio near $q \sim 0.5$ \citep{Rodriguez:2019huv, 2020ApJ...900..177K, 2021ApJ...915L..35K, 2025ApJ...987..146B}, our inferred $p(q)$ for $m_1 \gtrsim 47\,\Msun$ with \gwtcfive (Fig.~\ref{fig:highmass_mass_ratio_distribution}) is comparatively broad and, within uncertainties, compatible with a nearly flat distribution. {It is possible that the lack of a sharper peak in $p(q)$ is merely an observational limitation from the paucity of high-mass mergers. However theoretical expectations for $p(q)$ in dense clusters also depend sensitively on the \ac{BH} natal spin and pairing prescriptions \citep{Torniamenti:2024uxl, Sedda:2026xqr}, and the limited number of available simulations may not yet provide an exhaustive sampling of the relevant parameter space for such systems \citep{Wang:2016sag,Kremer:2019iul,Rodriguez:2019huv,Hong:2020dsl,Rantala:2025rva}}. 
Thus, given both the current observational uncertainties and the model dependence of these predictions, the mass-ratio distribution alone does not provide conclusive evidence for or against 1G+2G mergers in dense star clusters, nor uniquely identifies the dominant formation channel of high-mass \acp{BH}. 

In the next section, we turn to the spin distribution of \acp{BH} to identify signatures of alternate formation channels such as hierarchical mergers.

\begin{figure}[t!]
	\centering
	\includegraphics[width=1.05\linewidth]{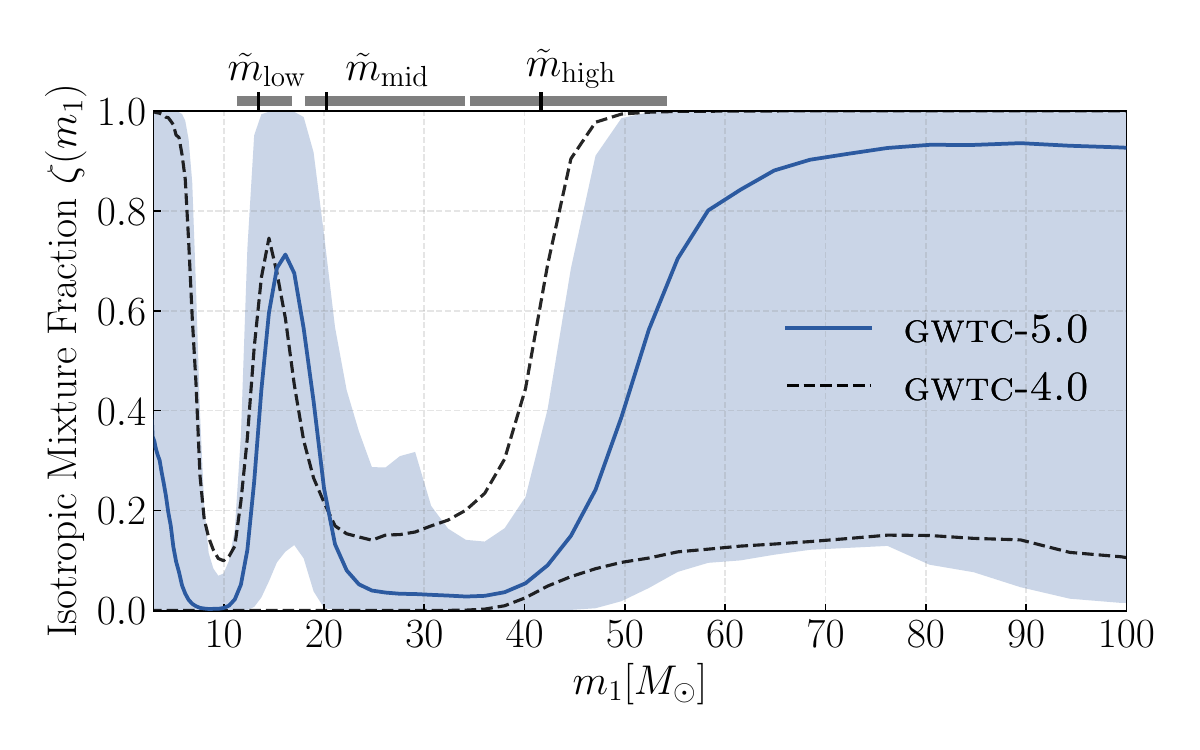}
	\caption{Mass-dependent mixture fraction $\zeta(m_1)$ between the Gaussian and uniform \chieff components as a function of primary \ac{BH} mass obtained with the \chieffmixture{} model. {Here $\zeta$ denotes the fraction of systems in the uniform component}. The solid line shows the posterior median inferred from \gwtcfive{} and the shaded band the corresponding 90\% credible interval. The dashed lines indicate the 90\% credible bounds from \gwtcfour{}. The bars and markers at the top are the credible intervals of the transition masses (\mlow, \mmid, \mhigh) inferred with the \chieffthree{} model.}
	\label{fig:mass_dep_mix}
\end{figure} 

\section{Signatures of hierarchical merger subpopulations}
\label{sec:hierarchical}

As discussed in Sec.~\ref{Sec:Overview_spins}, the observed GW population includes \acp{BH} with a broad distribution of spin magnitudes and tilts, with evidence for multiple subpopulations, including systems {with} at least one rapidly-spinning \ac{BH}. While isolated binary evolution can account for some of these features, dynamical assembly in dense environments, such as AGN disks and star clusters, provides a natural pathway to the observed \ac{BBH} diversity via hierarchical mergers, in which merger remnants participate in subsequent mergers (e.g., \citealt{Gerosa:2021mno}). Such channels can naturally produce larger \ac{BH} masses, high spin magnitudes and large tilt angles, even if the 1G progenitors rotated slowly. {The \ac{LVK} recently reported the detection of two \ac{BBH} mergers with rapid primary spins and unequal mass ratios~\citep{2025ApJ...993L..21A}---potential signatures of 1G+2G mergers}.

If \acp{BH} above the \ac{PISN} mass gap are predominantly hierarchical, \ac{BH} spin properties may change across the \ac{PISN} mass scale. Second-generation \acp{BH} are expected to have spin magnitudes clustering around $\chi \sim 0.7$ \citep{2005PhRvL..95l1101P,2008PhRvD..77b6004B,2016ApJ...825L..19H,2008ApJ...684..822B}. If such remnants are retained in dense environments and subsequently merge with 1G \acp{BH} with low, isotropic spins, the \chieff{} distribution would be approximately symmetric around zero with a half-width of ${\sim} 0.47$ \citep{Antonini:2024het,Rodriguez:2019huv,Baibhav:2020xdf,2022ApJ...935L..26F}.
In contrast, alternative channels generally do not make robust predictions for \chieff, as the resulting spins depend more sensitively on uncertain assumptions about the progenitors.
The characteristic \chieff morphology expected from hierarchical mergers thus provides a falsifiable fingerprint of this scenario, which can be tested by searching for mass-dependent structure in the \chieff distribution.

\subsection{Results}
\label{sec:HS_results}

We look for transitions in the \chieff distribution at different masses using \gwtcfive{} data.
Support for such transitions has been reported using by analyses {previous datasets}~\citep{2022ApJ...928...75H, Antonini:2024het,Antonini:2025ilj,Antonini:2025zzw,Tong:2025wpz,Tong:2025xir,Plunkett:2026pxt, Banagiri:2025dmy, Ray:2025xti,2026ApJ..1001L..40F,2026ApJ...999L..30V}. {Meanwhile, some analyses also found evidence that the spin-magnitude distribution correlates with \ac{BH} primary mass \citep{2024arXiv240601679P,2026PhRvD.113d3048B}.}

We adopt the \chieffmixture model and infer $\zeta(m_1)$, the mass-dependent fraction of systems assigned to a uniformly-distributed \chieff subpopulation over a narrower Gaussian component (App.~\ref{app:zeta_model}).
Figure~\ref{fig:mass_dep_mix} shows the \gwtcfive result, which displays nontrivial {peak}: a pronounced low-mass feature at $10\text{-–}20\,\Msun$, where the median $\zeta(m_1)$ rises to ${\sim} 70\%$, followed by a dip at intermediate masses, and a second increase above ${\sim} 40\,\Msun$. \textbf{The data therefore support the presence of multiple mass scales at which a broader, approximately uniform $\chi_{\rm eff}$ component becomes relevant}. 
For the uniform \chieff component we find a mean $\mu = \CIPlusMinus{\HierarchicalMacros[uniform_mean]}$ with corresponding half-width $\sigma = \CIPlusMinus{\HierarchicalMacros[uniform_halfwidth]}$. The narrow truncated Gaussian component has a location and a width parameter given by $\mu = \CIPlusMinus{\HierarchicalMacros[gaussian_mean]}$ and $\log_{10}\sigma = \CIPlusMinus{\HierarchicalMacros[gaussian_logstd]}$, respectively. The mixture fraction $\zeta(m_1)$ remains largely unconstrained below ${\sim} 10\,\Msun$. 

To further localize these features in mass, we use the \chieffthree model. In contrast to the \chieffmixture model, which allows for a smooth variation of the mixing fraction across the entire mass range, the \chieffthree model parametrizes the population using discrete transitions between spin regimes at transition masses \mlow{}, \mmid{}, and \mhigh{} (App.~\ref{app:three_transition_model}). We find support for a multi-transition description, with all three transition masses well constrained by the data (Fig.~\ref{fig:three_transition_post} in App.~\ref{app:three_transition_model}). We infer transitions at { $\mlow =  \CIPlusMinus{\HierarchicalMacros[triple_transition_m_t_low]} \,\Msun$, $\mmid = \CIPlusMinus{\HierarchicalMacros[triple_transition_m_t_mid]} \,\Msun$, and $\mhigh = \CIPlusMinus{\HierarchicalMacros[triple_transition_m_t_high]} \,\Msun$}, qualitatively consistent with \chieffmixture (top axis in Fig.~\ref{fig:mass_dep_mix}). {The \chieffthree model does not assume, a priori, that the low-mass and high-mass uniform \chieff{} components have the same \chieff{} distribution. It finds the mean and the half-width of the Uniform component between $\mlow$ and $\mmid$ to be $\CIPlusMinus{\HierarchicalMacros[triple_transistion_uniform_mean_low]}$ and $\CIPlusMinus{\HierarchicalMacros[triple_transistion_uniform_halfwidth_low]}$, respectively. Similarly, the mean and the half-width of the Uniform component over masses of $\mhigh$ is $\CIPlusMinus{\HierarchicalMacros[triple_transistion_uniform_mean_high]}$ and $\CIPlusMinus{\HierarchicalMacros[triple_transistion_uniform_halfwidth_high]}$ , respectively. These are both consistent with the expected signatures for 1G + 2G mergers.}
These results also broadly agree with the \tgmm finding from Sec.~\ref{Sec:Overview_spins} that the \ac{BBH} population has a dominant low-spin component and a subdominant component with at least one rapidly-spinning \ac{BH} and median magnitude of $\chi \sim 0.7$ (App.~\ref{sec:TGIMM}).

The \chieffmixture model infers that the volumetric merger rate associated with the uniform and approximately-symmetric \chieff subpopulation is $R_{\rm U}^{\zeta} = \CIPlusMinus{\HierarchicalMacros[rate_uniform]}\,\perGpcyr$ at $z=0.2$. This is consistent with the \chieffthree model, which yields  $R_{\rm U, high\,mass}= \CIPlusMinus{\HierarchicalMacros[rates_2g1g_high_mass]} \,\perGpcyr$  for $m_1 > \mhigh{}$ and $R_{\rm U, low\,mass}= \CIPlusMinus{\HierarchicalMacros[rates_2g1g_low_mass]}\,\perGpcyr$  for $\mlow{}< m_1 < \mmid{}$, corresponding to a total rate  $R_{\rm U} = \CIPlusMinus{\HierarchicalMacros[rates_2g1g_total]}\,\perGpcyr$. The uniform component thus corresponds to ${\sim} 1$--$10\%$ {of the overall \ac{BBH} merger rate}.

\subsection{Astrophysical implications}
\label{subsec:HS_astrointerpretation}

\subsubsection{The high-mass sector}
The transition in spin properties above \mhigh $\approx 40\,\Msun$ suggests a change in the channels contributing to the observed mergers. The inferred width and mean of the uniform \chieff component is consistent with expectations from the standard 1G+2G merger scenario at high masses~\citep{Antonini:2024het,Baibhav:2020xdf,Gerosa:2017kvu,Rodriguez:2019huv,2017ApJ...840L..24F},
but many factors could lead to confusion from other channels.
Furthermore, the observational uncertainty on the mixture fraction is high and we cannot rule out low occupancy of the uniform component at high masses (Fig.~\ref{fig:mass_dep_mix}), and neither can we rule out additional formation channels. 

 Mergers in AGN disks~\citep{Vaccaro:2023cwr,2019PhRvL.123r1101Y,Gerosa:2017kvu,2020MNRAS.494.1203M,2018ApJ...866...66M,2010CQGra..27k4007M,2020A&A...636A.104B,2009A&A...497..243D,Song:2016mme,Yoon:2005tv}, chemically homogeneous evolution in tidally-locked massive binaries~\citep{2016MNRAS.460.3545D,2016A&A...588A..50M}, and primordial \acp{BH}~\citep{Bardeen:1972fi,DeLuca:2020sae,DeLuca:2025fln,Blinnikov:2016bxu} have all been suggested as mechanisms to produce either high-mass or high-spin systems or both. {Uncertainties in stellar evolution  \citep[e.g.,][]{Spruit:1999cc,Spruit:2001tz,Heger:2004qp,Heger:1999ax,2018A&A...616A..28Q,Cantiello:2014uja,Denissenkov:2006tk,Zahn:2007uk,2019ApJ...881L...1F} may also enable the formation of rapidly rotating 1G \acp{BH} with masses above ${\sim} 40\,\Msun$~\citep{Shibata:2025lde,Uchida:2018ago,Croon:2025gol,Gottlieb:2025ugy,Spera:2018wnw,DiCarlo:2019pmf,Kremer:2020wtp,2024MNRAS.534.1868G,2024MNRAS.528.5140A,2025ApJ...994L..37K}}

The inferred transition mass \mhigh{} lies suggestively close to common estimates for the lower edge of the \ac{PISN} mass gap.
However, the mapping between progenitor and final \ac{BH} properties depends sensitively on multiple layers of uncertain stellar physics that can shift the effective onset of the gap~\citep{2021MNRAS.501.4514C,Ugolini:2025lzo,2021MNRAS.504..146V,Renzo:2024jhc,Mapelli:2019ipt,Fernandez:2017pcd,Luo:2025njy}, leading to 1G \acp{BH} with masses above $40\,\Msun$.
This makes a direct association between the observed transition and the underlying physics difficult.

\subsubsection{The low-mass sector} 

We observe a pronounced increase in the mixing fraction $\zeta(m_1)$ for $m_1 \sim 10$--$20\,\Msun$, indicating that the uniform \chieff component becomes more prominent in this interval.
Hierarchical mergers formed in dense stellar environments are predicted to span a broad mass range, making their presence at low primary masses possible (e.g., \citealt{2021ApJ...913L..19T,2022MNRAS.511.5797M,Santoliquido:2020bry,Sedda:2023qlx,Ye:2025ano}). However, the apparent enhancement of hierarchical-like signatures in the ${\sim} 10$--$20\,\Msun$ interval, followed by a suppression at intermediate masses (${\sim} 20$--$40\,\Msun$), is difficult to explain. 
One possible explanation lies in the structure of the 1G \ac{BH} mass function, which depends sensitively on supernova physics and stellar evolution (e.g., \citealt{Spera:2022byb,Mapelli:2021taw}). Several models predict a depletion or partial gap in the first-generation \ac{BH} mass spectrum between ${\sim} 11\, \Msun$ and $16\,\Msun$ from binary stripped stars \citep{Woosley:2020mze,Patton:2020tiy,2020ApJ...890...51E,Schneider:2023mxe,Maltsev:2025bgs,Galaudage:2024meo,Willcox:2025poh}, which could enhance the relative visibility of hierarchical mergers in this mass range. At the same time, the location and depth of such a feature remain model dependent and impacted by uncertainties in stellar evolution and cluster dynamics (e.g., \citealt{Schneider:2023mxe, Chieffi:2020aml,2024ApJ...975..253A,2025ApJ...995..177T,2025PhRvD.111l3046G}), complicating the interpretation.

\newcommand{\Linqchieffdmu}{-0.03^{+0.15}_{-0.15}}
\newcommand{\Linqchieffdlnsig}{-2.20^{+0.69}_{-0.59}}
\newcommand{\LinqchieffdlnsigcredNeg}{99}
\newcommand{\Splineqchieffdmu}{-0.05^{+0.23}_{-0.19}}
\newcommand{\Splineqchieffdlnsig}{-1.91^{+0.80}_{-0.73}}
\newcommand{\CopulaetaqchiNeg}{63}
\newcommand{\Linzchieffdmu}{-0.03^{+0.08}_{-0.08}}
\newcommand{\Linzchieffdlnsig}{0.48^{+0.31}_{-0.42}}
\newcommand{\Splinezchieffdmu}{-0.06^{+0.11}_{-0.13}}
\newcommand{\Splinechieffdlnsig}{0.22^{+0.47}_{-0.54}}

\newcommand{\PixelPopqchieffdmu}{0.17^{+0.59}_{-0.55}}
\newcommand{\PixelPopqchieffdlnsig}{-2.05^{+2.49}_{-2.40}}
\newcommand{\PixelPopqchieffSpearman}{26}
\newcommand{\PixelPopqchieffSpearmanReduced}{27}
\newcommand{\Copulaetazchi}{3.25^{+1.35}_{-1.53}}
\newcommand{\CopulaetazchiNeg}{99.8}
\newcommand{\CopulazchiSpearmanReduced}{0.10}
\newcommand{\PixelPopzchieffSpearmanReduced}{-0.03^{+0.16}_{-0.16}}
\newcommand{\LinSkewzchieffdlnsig}{0.53^{+0.51}_{-0.50}}
\newcommand{\Copulaetamz}{0.46^{+1.84}_{-1.95}}
\newcommand{\CopulaetamzNeg}{66}

\section{Broad Population Correlations}
\label{Sec:Correlations}

Here we present an overview of the broad correlations between pairs of parameters in the \ac{BBH} populations~\citep{2023PhRvX..13a1048A, 2025arXiv250818083T}. 
We discuss evidence for correlations between the distributions of mass ratio and effective inspiral spin (Sec.~\ref{Sec:qchieff}), and between effective inspiral spin and redshift (Sec.~\ref{Sec:zchieff}). 
Model variations and inconclusive results are shown in Apps.~\ref{app:qchieffvarients} and \ref{app:zchieffvarients}. 
These results further suggest contributions to the \ac{BBH} population from multiple formation pathways.

\subsection{Mass ratio and spin correlations}
\label{Sec:qchieff}

\begin{figure}[tb]
\includegraphics[scale=1]{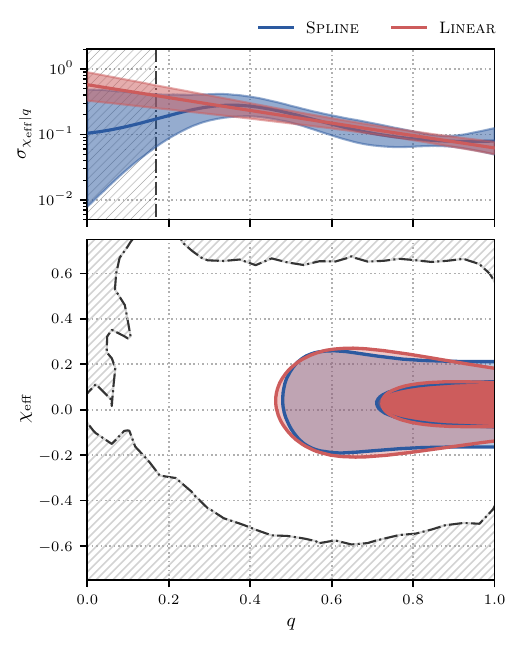} 
\caption{Top: the inferred standard deviation of the \chieff distribution as a function of mass ratio, for the \lincorr (red) and \splcorr (blue) models. 
The shaded bands show the $90\%$ credible interval, indicating that the standard deviation {tends to decrease} with increasing $q$. 
Bottom: the posterior predictive distribution of the joint $(\chieff\text{--}q)$ distribution, for the \lincorr and \splcorr models, where the outlines contain $90\%$ and $50\%$ of the posterior volume.
The hatching in both panels shows the areas where there are $<1\%$ of parameter estimation samples for \gwtcfive observations.
}
\label{fig:chieffq}
\end{figure}

Correlations between $\chieff$ and $q$ have been {reported} since \gwtctwo~\citep{2021ApJ...922L...5C, 2023PhRvX..13a1048A, 2022MNRAS.517.3928A, 2023ApJ...958...13A,  2025arXiv250818083T}.
{While \gwtcfour data found a correlation between $q$ and \chieff, we could not determine whether the \chieff shifts in mean, broadens, or both, with $q$.}
\textbf{In \gwtcfive we find decreased evidence that the mean of \chieff evolves with $q$ and increased evidence that its width does, compared to \gwtcfour.}
Figure~\ref{fig:chieffq} shows the joint distribution of $\chieff$ and $q$ for the \lincorr and \splcorr models.
The top panel shows the evolution of the width of the $\chieff$ distribution with $q$, quantified by the standard deviation $\sigma_{\chi_\mathrm{eff}|q}$. 

The \lincorr model relates both the mean and width of the \chieff distribution through two independent linear functions, directly parametrizing the gradient of the mean \deltamuchiq and the gradient of the natural log of the width \deltasigmachiq.
This model finds that the peak of the \chieff distribution does not shift with $q$: we measure $\deltamuchiq=\Linqchieffdmu$, consistent with no evolution of the mean. 
Conversely, the model supports a more pronounced evolution of the width of the distribution with $q$, with more unequal-mass binaries exhibiting a wider distribution of \chieff values. Specifically, we obtain $\deltasigmachiq=\Linqchieffdlnsig$ which rules out no evolution of the width with $q$ at ${>}{\LinqchieffdlnsigcredNeg}\%$ credibility and implies that more unequal-mass binaries have a larger spread in \chieff.
This trend persists when we leave out the recently-discovered unequal-mass, highly-spinning systems GW241011 and GW241110, meaning that these two events do not entirely drive the correlation (see App.~\ref{app:qchieffvarients}). 

The \splcorr model relates the mean and width of the \chieff distribution to $q$ through two independent cubic splines, allowing for a more flexible correlation.  For this model, we calculate average slopes for \deltamuchiq and \deltasigmachiq in the range $q>1/3$, since for smaller mass ratios the width is informed by the spline knot at $q=0$ where we have no observations. Therefore, our calculation of \deltasigmachiq captures the \splcorr result's overall evolution of $\sigma_{\chi_\mathrm{eff}|q}$, avoiding the impact of modeling choices on local gradients. We find $\deltamuchiq=\Splineqchieffdmu$ and $\deltasigmachiq=\Splineqchieffdlnsig$. 
This result agrees with the \lincorr model, showing an evolution of the width of \chieff with $q$ without evolution of the mean, although uncertainties are larger due to the increased model flexibility. 

We additionally test the presence of a correlation in $(q,\chieff)$ using a \copcorr model, which relates the two marginal distributions using a Frank copula via a correlation coefficient $\eta$~\citep{2022MNRAS.517.3928A}. 
We find $\etachiq<0$ at $\CopulaetaqchiNeg\%$ credibility, compared to $90\%$ credibility with \gwtcfour. 
However, unlike the \lincorr and \splcorr models, the \copcorr model cannot account for a broadening of the \chieff distribution with mass ratio.
The decrease in $\etachiq$ compared to \gwtcfour is consistent with the diminishing evidence for a direct anticorrelation between the mean of \chieff and $q$.

The wider range of \chieff observed at small $q$ could be explained by a single $q$-dependent channel, or a mix of channels that add up to broaden \chieff at small $q$.
A significant fraction of dynamically assembled \acp{BH} at lower mass ratios may widen \chieff \citep{Mould:2022ccw}. 
Hierarchical mergers of 1G+2G \acp{BH} are expected to have $q\approx0.5$ and $\spinone\approx0.7$, resulting in a broadening of \chieff with decreasing $q$.
The presence of low-mass-ratio, highly-spinning binaries qualitatively agrees with the hierarchical-merger-like signatures presented in Secs.~\ref{Sec:Overview_spins}, \ref{sec:high_mass} and \ref{sec:hierarchical}. 
With a strong model, \cite{Vijaykumar:2026zjy} reported evidence for a hierarchical subpopulation of mergers in \gwtcfour giving rise to a similar broadening. Other mechanisms could also explain the observed trend: isolated binaries undergoing tidal spin-up, stable mass transfer, and mass ratio reversal may all result in low-$q$, high-spin binaries~\citep{2022ApJ...938...45B, 2022ApJ...933...86Z, 2024A&A...689A.305O}.
However, the reduced evidence for a shift in the mean suggests that the \chieff distribution at unequal mass ratios is more isotropic than previously reported. 
This agrees with previous works that found that the aligned-spin components do not exhibit a correlation with mass ratio \citep{Tiwari:2021yvr, Tiwari:2025oah,Vijaykumar:2026zjy}. 
Such an isotropic distribution of \chieff is difficult to produce with most isolated formation channels although this can depend on the complex interplay between spins, natal kicks and other factors (e.g.,~\citealp{Baibhav:2024rkn}).

\subsection{Redshift and spin correlations}
\label{Sec:zchieff}

\begin{figure}[tb]
\includegraphics[scale=1]{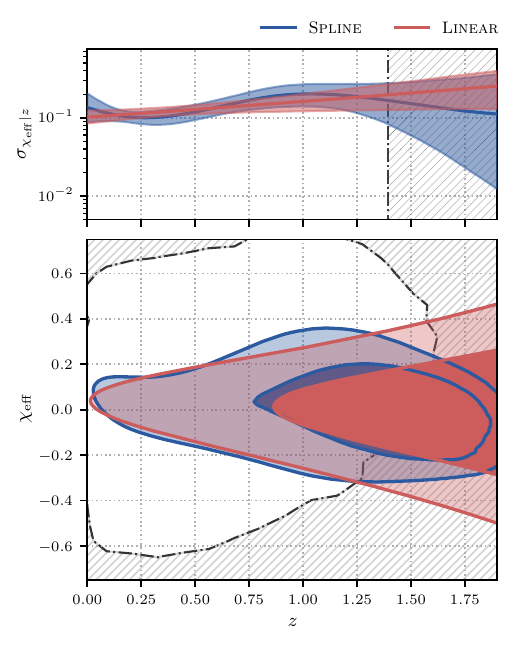} 
\caption{Top: The inferred standard deviation of the \chieff distribution as a function of redshift, for the \lincorr (red) and \splcorr (blue) models. The shaded bands show the $90\%$ credible interval. Bottom: the posterior predictive distribution of the joint $(\chieff\text{--}z)$ distribution, for the \lincorr and \splcorr models. The outlines contain $90\%$ and $50\%$ of the volume. The hatched area in both plots indicates the region where there are $<1\%$ parameter estimation samples for \gwtcfive observations. 
}
\label{fig:chieffz}
\end{figure}

Evidence of a broadening of \chieff with increasing redshift has been observed since \gwtcthree \citep{Biscoveanu:2022qac, 2025arXiv250818083T}. 
Figure \ref{fig:chieffz} shows the joint $(\chieff\text{--}z)$ distribution and the width of the \chieff distribution as a function of redshift ($\sigma_{\chi_\mathrm{eff}|z}$) for the \lincorr and \splcorr models. 
{With the \lincorr model we find $\deltamuchiz=\Linzchieffdmu$ and $\deltasigmachiz=\Linzchieffdlnsig$, 
which indicates a preference for broadening of the width of the distribution with $z$ at $93\%$ credibility.} 
The joint distribution from the \splcorr model qualitatively supports broadening, even if the evolution of $\sigma_{\mathrm{eff}|z}$ indicates some non-linear structure in the evolution of the width. 
With the \splcorr model we find $\deltamuchiz=\Splinezchieffdmu$ and $\deltasigmachiz=\Splinechieffdlnsig$, where we evaluate the gradients over the range where we have ${>}1\%$ of parameter estimation samples. 
This is consistent with no evolution of \chieff across the broad range of redshifts where we have observations.

In summary, \textbf{we find model-dependent evidence for \chieff broadening with redshift} with the \lincorr and \splcorr models. 
Evidence for this broadening is supported by additional models, which we include in App.~\ref{app:zchieffvarients}. 
The results in this section are caveated by the fact that correlations in $(\chieff,z)$ are degenerate with correlations between mass and redshift, or mass and \chieff \citep{Biscoveanu:2022qac}. 
We discussed evidence for correlations between primary mass and \chieff in Sec.~\ref{sec:hierarchical}.

A broadening in \chieff with redshift could be a signature of multiple formation channels. For example, \cite{Farah:2026jlc} find that hierarchical mergers can produce such a broadening, due to the steeper evolution of the hierarchical \ac{BBH} merger rate with redshift. Meanwhile, tidal spin up in isolated binaries can produce more \emph{positive} \chieff at larger redshift due to interplay between metallicity and orbital separation, with low-metallicity environments at high redshift resulting in tighter binaries that are both tidally spun up and short-lived~\citep{2022A&A...665A..59B}. However, this mechanism cannot easily explain a roughly symmetric broadening under common supernova kicks~\citep{Baibhav:2024rkn}.

\section{Conclusion}\label{Sec:conclusion}

We have presented analyses of the compact objects in the \gwtcfive{} catalog. The \gwtcfivendet candidate events in the dataset have further enhanced our understanding of the astrophysical properties of stellar-mass \acp{BH}. Among the most exciting findings is the growing evidence that multiple subpopulations of \acp{BH}, each with distinct properties, are producing the observed sources. Based on the number of public alerts sent about significant \ac{BBH} candidates, we expect the forthcoming GWTC-6.0 catalog to yield another $\gtrsim 68$ sources~\citep{GraceDB_O4}, further improving our understanding of the number, rates and properties of these populations. 

\section{Data availability}\label{Sec.DataRelease}

The analysis results and figure generation scripts used for this paper are released as \citep{GWTC5_pop}, while the data products for estimating sensitivity can be found in \citep{ligo_scientific_collaboration_2026_19500064,ligo_scientific_collaboration_2026_19500052}.

\section*{Acknowledgements}
This material is based upon work supported by NSF's LIGO Laboratory, which is a
major facility fully funded by the National Science Foundation.
The authors also gratefully acknowledge the support of
the Science and Technology Facilities Council (STFC) of the
United Kingdom, the Max-Planck-Society (MPS), and the State of
Niedersachsen/Germany for support of the construction of Advanced LIGO 
and construction and operation of the GEO\,600 detector. 
Additional support for Advanced LIGO was provided by the Australian Research Council.
The authors gratefully acknowledge the Italian Istituto Nazionale di Fisica Nucleare (INFN),  
the French Centre National de la Recherche Scientifique (CNRS) and
the Netherlands Organization for Scientific Research (NWO)
for the construction and operation of the Virgo detector
and the creation and support  of the EGO consortium. 
The authors also gratefully acknowledge research support from these agencies as well as by 
the Council of Scientific and Industrial Research of India, 
the Department of Science and Technology, India,
the Science \& Engineering Research Board (SERB), India,
the Ministry of Human Resource Development, India,
the Spanish Agencia Estatal de Investigaci\'on (AEI),
the Spanish Ministerio de Ciencia, Innovaci\'on y Universidades,
the European Union NextGenerationEU/PRTR (PRTR-C17.I1),
the ICSC - CentroNazionale di Ricerca in High Performance Computing, Big Data
and Quantum Computing, funded by the European Union NextGenerationEU,
the Comunitat Auton\`oma de les Illes Balears through the Conselleria d'Educaci\'o i Universitats,
the Conselleria d'Innovaci\'o, Universitats, Ci\`encia i Societat Digital de la Generalitat Valenciana and
the CERCA Programme Generalitat de Catalunya, Spain,
the Polish National Agency for Academic Exchange,
the National Science Centre of Poland and the European Union - European Regional
Development Fund;
the Foundation for Polish Science (FNP),
the Polish Ministry of Science and Higher Education,
the Swiss National Science Foundation (SNSF),
the Russian Science Foundation,
the European Commission,
the European Social Funds (ESF),
the European Regional Development Funds (ERDF),
the Royal Society, 
the Scottish Funding Council, 
the Scottish Universities Physics Alliance, 
the Hungarian Scientific Research Fund (OTKA),
the French Lyon Institute of Origins (LIO),
the Belgian Fonds de la Recherche Scientifique (FRS-FNRS), 
Actions de Recherche Concert\'ees (ARC) and
Fonds Wetenschappelijk Onderzoek - Vlaanderen (FWO), Belgium,
the Paris \^{I}le-de-France Region, 
the National Research, Development and Innovation Office of Hungary (NKFIH), 
the National Research Foundation of Korea,
the Natural Sciences and Engineering Research Council of Canada (NSERC),
the Canadian Foundation for Innovation (CFI),
the Brazilian Ministry of Science, Technology, and Innovations,
the International Center for Theoretical Physics South American Institute for Fundamental Research (ICTP-SAIFR), 
the Research Grants Council of Hong Kong,
the National Natural Science Foundation of China (NSFC),
the Israel Science Foundation (ISF),
the US-Israel Binational Science Fund (BSF),
the Leverhulme Trust, 
the Research Corporation,
the National Science and Technology Council (NSTC), Taiwan,
the United States Department of Energy,
and
the Kavli Foundation.
The authors gratefully acknowledge the support of the NSF, STFC, INFN and CNRS for provision of computational resources.

This work was supported by MEXT,
the JSPS Leading-edge Research Infrastructure Program,
JSPS Grant-in-Aid for Specially Promoted Research 26000005,
JSPS Grant-in-Aid for Scientific Research on Innovative Areas 2402: 24103006,
24103005, and 2905: JP17H06358, JP17H06361 and JP17H06364,
JSPS Core-to-Core Program A.\ Advanced Research Networks,
JSPS Grants-in-Aid for Scientific Research (S) 17H06133 and 20H05639,
JSPS Grant-in-Aid for Transformative Research Areas (A) 20A203: JP20H05854,
the joint research program of the Institute for Cosmic Ray Research,
University of Tokyo,
the National Research Foundation (NRF),
the Computing Infrastructure Project of the Global Science experimental Data hub
Center (GSDC) at KISTI,
the Korea Astronomy and Space Science Institute (KASI),
the Ministry of Science and ICT (MSIT) in Korea,
Academia Sinica (AS),
the AS Grid Center (ASGC) and the National Science and Technology Council (NSTC)
in Taiwan under grants including the Science Vanguard Research Program,
the Advanced Technology Center (ATC) of NAOJ, 
the Mechanical Engineering Center of KEK
and Vietnam National Foundation for Science and Technology Development 
(NAFOSTED) 103.01-2025.147.

Additional acknowledgements for support of individual authors may be found in the following document: \\
\texttt{https://dcc.ligo.org/LIGO-M2300033/public}.
For the purpose of open access, the authors have applied a Creative Commons Attribution (CC BY)
license to any Author Accepted Manuscript version arising.
We request that citations to this article use 'A. G. Abac {\it et al.} (LIGO-Virgo-KAGRA Collaboration), ...' or similar phrasing, depending on journal convention.

\bibliography{}

\appendix
\section{Dataset and Methods}
\label{App.Methods}

\subsection{Waveform model choices}
\label{App.waveforms}

For newly identified \ac{CBC} candidates in \ac{O4b}, we use posterior samples generated with the \BILBY package~\citep{2019ApJS..241...27A,2020MNRAS.499.3295R}.

For \ac{BBH} candidates from \ac{O1} - \ac{O3}, all analyses in this paper made the same waveform choices as detailed in \citet{2025arXiv250818082T}. For events in \ac{O4}, our analyses broadly made two different choices of waveforms. All analyses used posterior samples generated with \SURSEVENDQFOUR{}~\citep{2019PhRvR...1c3015V} when available. When samples with \SURSEVENDQFOUR{} were not available, e.g. when the signal duration is too long to be analyzed with it, \pixelpop, \fullpop,  \lincorr and the \splcorr models used posterior draws generated with the \IMRPhenomXPHMST{} waveform~\citep{2021PhRvD.103j4056P, 2025PhRvD.111j4019C}. The \defbbh model, the \threeSubpopulation model, \bgp, \isolatedpeak, \tgmm, \chieffthree{}, \bivskewchichi, \chieffmixture{}, and the copula analyses used \textsc{Mixed} samples -- a mixture of posterior draws from the \IMRPhenomXPHMST{} and \SEOBNRFIVEPHM{} waveforms analyses~\citep{2023PhRvD.108l4037R, 2023PhRvD.108l4035P} -- for \ac{O4a} and samples from \IMRPhenomXPHMST{} for \ac{O4b}~\footnote{Note, however, that the \defbbh model, \threeSubpopulation model, \chieffthree{} and the copula analyses used \textsc{Mixed} samples for a small handful of events from \ac{O4b}, namely GW241111\_111552, GW241110\_124123, GW240925\_005809 and GW240910\_103535}.

Our full-spectrum models also made different waveform choices for \ac{NS}-containing events. For the \acp{BNS} mergers, GW170817 and GW190425, both \fullpop and \pixelpop used \IMRPhenomPTWONRTidal{}. For GW200105 and GW200115 \pixelpop used \textsc{Mixed} samples while \fullpop used \IMRPhenomNSBH{}. Finally for GW230529\_181500, \pixelpop used draws from \IMRPhenomNSBH{} while \fullpop used \textsc{Mixed} samples.

We note that we do not expect our population inferences to be sensitive to the choices between these waveforms. 

\subsection{Estimation of the detector network sensitivity}
\label{App.Selection}

We numerically estimate the sensitivity of the detector network (i.e., measure $\xi(\Lambda)$) during \ac{O3}, \ac{O4a} and \ac{O4b} by re-weighting a large set of simulated signals (injections) added to real data and searched by the same pipelines used to produce the actual catalog~\citep{2025arXiv250818083T,2025PhRvD.112j2001E}. This yields a list of \AtOnePerYear (\AtOnePerFourYears) injections above the detection threshold of FAR$ <1$ per year ($ <0.25$ per year). For \ac{O1} and \ac{O2} only, a different procedure is used, which thresholds on \ac{SNR} instead of \ac{FAR}~\citep{2023PhRvX..13a1048A}, yielding \AtSNRTen injections above \ac{SNR}$=10$. \color{black}{We note that while \ac{O4b} officially started on \OfourBStartDate{} \OfourBStartTime{}, the sensitivity file only considered injections from 2024 April 12 \OfourBStartTime{}}. This is a small difference of two days and we do not expect this to have any statistically meaningful impact on the results. Injections in other observing chunks were not impacted. 
\section{Overview Models}\label{Sec.AppendixOverview}

\subsection{\pixelpop}
\label{Sec:pixel_pop_appendix}

{The \pixelpop analysis employs a high-resolution binned model that is able to infer joint population distribution of multiple \ac{CBC} parameters and their correlations with minimal assumptions~\citep{Heinzel:2023hlb}. To accomplish this, the joint space of \ac{CBC} parameters is finely binned and the comoving merger rate density in each bin is inferred as an independent parameter using hierarchical Bayesian inference. To enforce a notion of smoothness, the rate within each bin $\beta$ is coupled to its nearest neighbors using a log-normal \ac{ICAR}},

\begin{equation}
	p \left(\ln \mathcal{R}_{\beta} \,\middle|\, \{ \ln \mathcal{R}_{b \neq \beta} \}, \sigma \right) \propto \exp \left[ -\sum_b \frac{(\ln \mathcal{R}_{\beta} - \ln \mathcal{R}_{b})^2}{2 \sigma^2} \right].
\end{equation}

{where $b$ is an index over the nearest neighbors of bin $\beta$ and the global parameter $\sigma$ defines the smoothness of the \ac{ICAR}. In practice, the joint prior density over all bins can be computed simultaneously as the matrix equation shown in \cite{Heinzel:2023hlb}. Here, we take the intrinsic CAR limit to improve computational efficiency without making sacrifices on the inference fidelity. See the discussion in App.~B of \cite{Alvarez-Lopez:2025ltt}. 
}

{In this paper, we generally use variants of \pixelpop that model the joint population distributions of one or two \ac{CBC} parameters, with 100 bins per dimension. Therefore, our one- and two-dimensional \pixelpop models have $10^2$ and $10^4$ rate parameters in total, respectively. Table~\ref{Tab:pixelpop_variants} shows the different variants used in this papers along with the relevant sections. For \ac{CBC} parameters that are not binned, we fall back on the strongly-parameterized models described in Tab.~\ref{tab:summary_of_models}. We use a uniform prior on $\ln \sigma \in [-3, 5]$. We direct the reader to \cite{Heinzel:2023hlb, Heinzel:2024hva} and \cite{Alvarez-Lopez:2025ltt} for more details on \pixelpop.}

\begin{table*}[h]
	\begin{tabular*}{0.95\textwidth}{@{\extracolsep{\fill}} lcc @{}}
\hline
\hline
\pixelpop variant & Dimensionality & Sections used\\
\hline
$(m_1, m_2)$-\pixelpop & 2 & Secs.~\ref{sec:full_spectrum}, \ref{sec:full_spectrum_rates}, \ref{sec:pm1_bbh}, \ref{sec:high_mass}\\
$(\chi, \cos\theta)$-\pixelpop & 2 & Sec.~\ref{Sec:Overview_spins}\\
$(m_1, q)$-\pixelpop & 2 & Secs.~\ref{sec:pq_bbh}, \ref{sec:35_msun}, \ref{sec:high_mass} \\
$(m_1, \chieff)$-\pixelpop  & 2 & Secs.~\ref{Sec:10msun}, \ref{sec:35_msun}\\
$(m_1, z)$-\pixelpop & 2 & Sec.~\ref{sec:35_msun}\\ 
$(\chieff, \chip)$-\pixelpop & 2 & App.~\ref{Sec.InconclusiveCorr}\\
$(\chieff, q)$-\pixelpop & 2 & Sec.~\ref{Sec:qchieff} \\
$(\chieff, z)$-\pixelpop & 2 & Sec.~\ref{Sec:zchieff} \\
$z$-\pixelpop & 1 & Sec.~\ref{sec:bbh_z} \\
\hline
\hline
\end{tabular*}
\caption{\label{Tab:pixelpop_variants} Variants of \pixelpop used in this paper}
\end{table*}

\subsection{\fullpop}
\label{sec:fulllpop}

\begin{figure}[h]
\centering
    \includegraphics[width=0.85\textwidth]{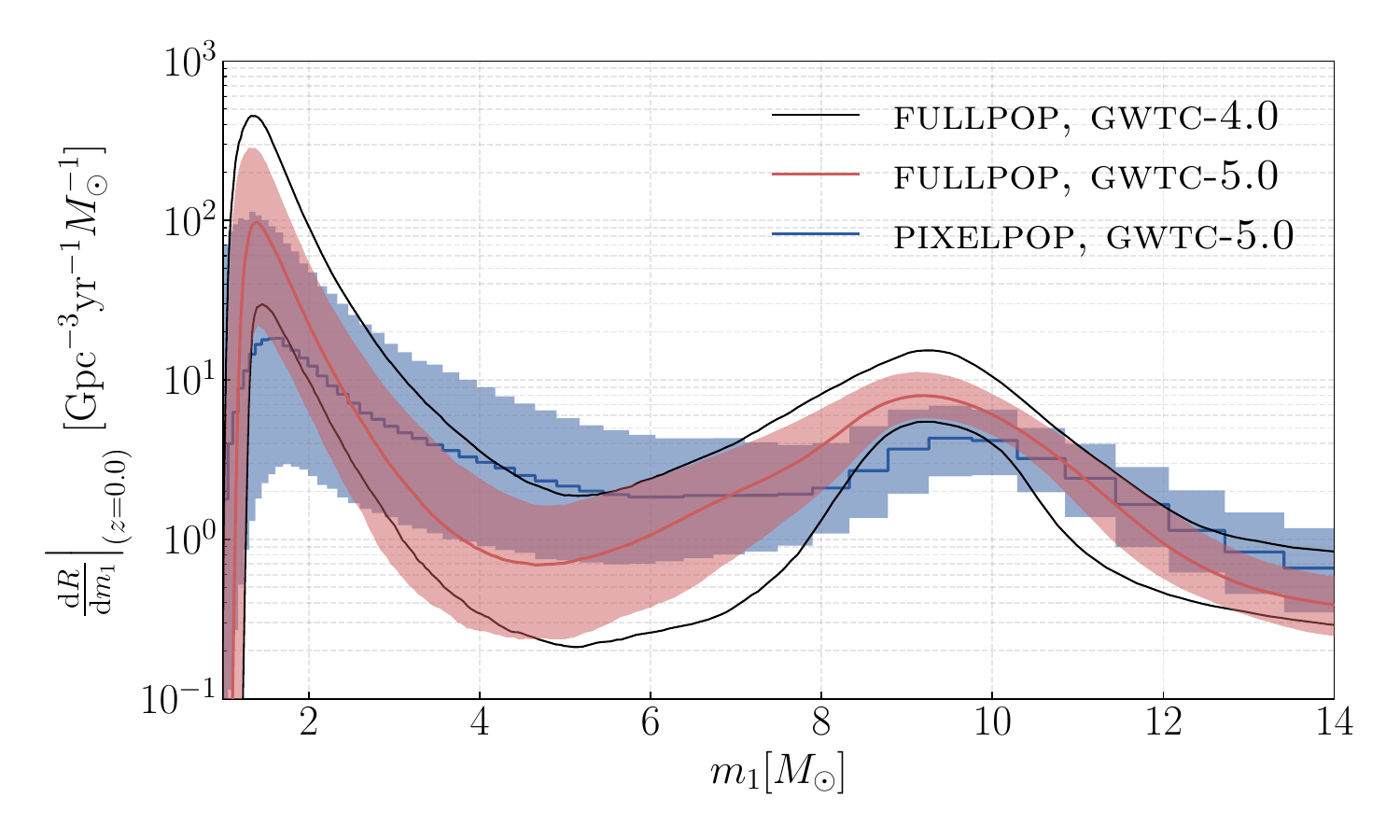}
    \caption{\label{fig:lowmass_fullspectrum}The primary mass distribution below $14\,\Msun$. The solid lines show the median distribution for each model, with the shaded bands depicting the $90\%$ credible regions. With \gwtcfive, we find that our models effectively rule out a completely empty lower mass gap. The \fullpop model parameterizes the depth of the gap with the hyperparameter $A$, with $A=1$ indicating a completely empty gap between $3 - 5\,\Msun$. With \gwtcfive we find $A \leq \OverviewEstimates[fullpop_gap_depth][99th percentile]$ at $99\%$ credibility.}
\end{figure}

The strongly-parameterized \fullpop~\citep{2025arXiv250818083T, Farah:2021qom, Mali:2024wpq}, models the complete \ac{CBC} distribution. The model parameterizes both the primary and secondary masses as drawn from a common underlying population, $\pi_m (m \mid \Lambda)$, with a pairing function, $f (m_1, m_2)$,

\begin{equation}
    \pi(m_1, m_2\mid\Lambda) \propto \pi_m(m_1\mid\Lambda)\pi_m(m_2\mid\Lambda)f(m_1, m_2)\,\Theta(m_2 < m_1),
\label{eq: pdb joint mass model}
\end{equation}
where $\Theta$ is the Heaviside function. The mass model, $\pi_m (m \mid \Lambda)$, is parameterized as
\begin{align}
\pi_m(m\mid\Lambda) &=& [1 + c_1 \mathcal{N}_{[m_{\text{min}}, m_{\text{max}}]}(m\mid\mu_1, \sigma_1) + c_2 \mathcal{N}_{[m_{\text{min}}, m_{\text{max}}]}(m\mid\mu_2, \sigma_2)]
n_1(m \mid m_{\text{NSmax}}, m_{\text{BHmin}}, \eta_{\text{NSmax}}, \eta_{\text{BHmin}}, A) \nonumber \\
&& \times n_2(m \mid m_{\text{UMGmin}}, m_{\text{UMGmax}}, \eta_{\text{UMGmin}}, \eta_{\text{UMGmax}}, A_2) h(m \mid m_{\text{NSmin}}, \eta_{\text{NSmin}}) l(m \mid m_{\text{BHmax}}, \eta_{\text{BHmax}}) \nonumber \\
&& \times
\begin{cases}
m^{\alpha_1} & \text{if } m < m_{\text{NSmax}} \\
m^{\alpha_{\text{dip}}} m_{\text{NSmax}}^{\alpha_1 - \alpha_{\text{dip}}} & \text{if } m_{\text{NSmax}} \leq m < m_{\text{BHmin}} \\
m^{\alpha_2} m_{\text{NSmax}}^{\alpha_1 - \alpha_{\text{dip}}} m_{\text{BHmin}}^{\alpha_{\text{dip}} - \alpha_2} & \text{if } m \geq m_{\text{BHmin}}.
\end{cases}.
\label{eq: pdb mass model}
\end{align}

This $\pi_m(m\mid\Lambda)$ represents a universal mass function to describe the primary- and secondary-mass distributions. Note that the marginal mass distribution is different from the universal mass distribution due to the pairing formalism.
$\mathcal{N}_{[a,b]}(\mu, \sigma)$ represents a truncated normal distribution over $[a,b]$ with location and width parameters $\mu$ and $\sigma$. 
The high-pass, low-pass and notch functions are defined as:

\begin{align}
    l(m\mid m_{\text{BHmax}}, \eta_{\text{BHmax}}) &= \left[1+\left(\frac{m}{m_{\text{BHmax}}}\right)^{\eta_{\text{BHmax}}}\right]^{-1},\\
    h(m\mid m_{\text{NSmin}}, \eta_{\text{NSmin}}) &= 1 - l(m\mid m_{\text{NSmin}}, \eta_{\text{NSmin}}),\\
    n_1(m\mid m_{\text{NSmax}}, m_{\text{BHmin}}, \eta_{\text{NSmax}}, \eta_{\text{BHmin}}, A) &= 1 - A l(m\mid m_{\text{NSmax}}, \eta_{\text{NSmax}}) h(m\mid m_{\text{BHmin}}, \eta_{\text{BHmin}}),\\
	n_2(m\mid m_{\text{UMGmin}}, m_{\text{UMGmax}}, \eta_{\text{UMGmin}}, \eta_{\text{UMGmax}}, A_2) &= 1 - A_2 l(m\mid m_{\text{UMGmin}}, \eta_{\text{UMGmin}}) h(m\mid m_{\text{UMGmax}}, \eta_{\text{UMGmax}}).
\end{align}

The pairing function
\begin{equation}
f(m_1,m_2 \mid \beta_{\rm BH}, \beta_{\rm NS}) = \begin{cases}
    \left(\dfrac{m_2}{m_1}\right)^{\beta_{1}} & {\rm if} \; m_2 < 5\,\Msun \\
    \left(\dfrac{m_2}{m_1}\right)^{\beta_{2}} & {\rm if} \; m_2 > 5\,\Msun \\
\end{cases}
\end{equation}
controls how much merging binaries favor/disfavor equal masses. We allow for alternative pairing for binaries with low-mass secondaries (\acp{NSBH} or \ac{BBH} with the secondary component in the lower-mass gap, e.g., GW190814, \citealt{2020ApJ...896L..44A}).
We show the priors and describe the parameters of the \fullpop model in Tab.~\ref{tab:parameters_pdb}. {The primary-mass distribution below $15\,\Msun$ is shown in Fig.~\ref{fig:lowmass_fullspectrum}.}

\begin{deluxetable}{ccccc}
\tablecaption{Summary of \fullpop model parameters and priors. \label{tab:parameters_pdb}}
\tablehead{
\colhead{Category} & \colhead{Parameter} & \colhead{Unit} & \colhead{Description} & \colhead{Prior}
}
\startdata
Pairing Function & $\beta_1$ & -- & {Power-law index} below $5 \, \Msun$ & U($-2, 3$) \\
& $\beta_2$ & -- & {Power-law index} above $5\, \Msun$ & U($-2, 7$) \\
\tableline
Broken Power-Law & $\alpha_1$ & -- & {Power-law index} below $m_{\text{NS max}}$ & U($-10, 2$) \\
& $\alpha_{\text{dip}}$ & -- & {Power-law index} between $m_{\text{NS max}}$ and $m_{\text{BH min}}$ & U($-3, m2$) \\
& $\alpha_2$ & -- & {Power-law index} above $m_{\text{BH min}}$ & U($-3, 2$) \\
& $m_{\text{brk}}$ & $\Msun$ & Break point between $\alpha_1$ and $\alpha_2$ & $5$ \\
\tableline
Highpass Filter & $m_{\text{NS min}}$ & $\Msun$ & Low-mass roll-off & U($1, 1.4$) \\
& $\eta_{\text{min}}$ & -- & Sharpness at $m_{\text{NS min}}$ & $50$ \\
\tableline
Lowpass Filter & $m_{\text{BHmax}}$ & $\Msun$ & High-mass roll-off & U($60, 200$) \\
& $\eta_{\text{max}}$ & -- & Sharpness at $m_{\text{BHmax}}$ & U($-4, 12$) \\
\tableline
Low-Mass Notch & $m_{\text{NS max}}$ & $\Msun$ & Lower notch edge & U($1.4, 5$) \\
& $\eta^{\text{low}}_1$ & -- & Sharpness at $m_{\text{NS max}}$ & $50$ \\
& $m_{\text{BH min}}$ & $\Msun$ & Upper notch edge & U($5, 9$) \\
& $\eta^{\text{high}}_1$ & -- & Sharpness at $m_{\text{BH min}}$ & $50$ \\
& $A_1$ & -- & Notch depth & $0$ \\
\tableline
High-Mass Notch & $m_{\text{UMGmin}}$ & $\Msun$ & Lower notch edge & U($30, 90$) \\
& $\eta^{\text{low}}_2$ & -- & Sharpness at $m_{\text{UMGmin}}$ & $30$ \\
& $m_{\text{UMGmax}}$ & $\Msun$ & Upper notch edge & U($60, 150$) \\
& $\eta^{\text{high}}_2$ & -- & Sharpness at $m_{\text{UMGmax}}$ & $30$ \\
& $A_2$ & -- & Depth of high-mass notch & U($0, 1$) \\
\tableline
Low-Mass Peak & $\mu^{\text{peak}}_2$ & $\Msun$ & Peak location & U($6, 12$) \\
& $\sigma^{\text{peak}}_2$ & $\Msun$ & Peak width & U($0, 5$) \\
& $c_2$ & -- & Peak height & U($0, 500$) \\
\tableline
High-Mass Peak & $\mu^{\text{peak}}_1$ & $\Msun$ & Peak location & U($17, 50$) \\
& $\sigma^{\text{peak}}_1$ & $\Msun$ & Peak width & U($4, 20$) \\
& $c_1$ & -- & Peak height & U($0, 1000$) \\
\enddata
\tablecomments{$U(a,b)$ denotes a uniform distribution in the range $[a,b]$.}
\end{deluxetable}

\subsection{\tgmm}
\label{sec:TGIMM}

In the \tgmm analysis, we {write} the joint component spin magnitude as a sum of two truncated bivariate Gaussian mixtures~\citep{2026ApJ...996...71H}. 

\begin{equation}
    p(\boldsymbol{\chi}) = \eta_a \mathcal{N}(\boldsymbol{\chi} \mid \boldsymbol{\mu}^a, \boldsymbol{\Sigma}^a) + \eta_b \mathcal{N}(\boldsymbol{\chi} \mid \boldsymbol{\mu}^b, \boldsymbol{\Sigma}^b), 
\end{equation} 
where $\boldsymbol{\chi} = (\chi_1, \chi_2)$, and $\boldsymbol{\mu}$ and $\boldsymbol{\Sigma}$ are the mean vectors and the covariance matrices of the two-dimensional distribution. The indices $a$ and $b$ refer to the dominant and the subdominant modes respectively. The fraction of mergers in each mode is parameterized by $\eta_a$ and $\eta_b$, with $\eta_a + \eta_b = 1$.

Unlike the component spin models highlighted in Sec.~\ref{Sec:Overview_spins}, this model fits the two components independently allowing us to separately infer the spins of each components and any possible correlations. Moreover by using more than one mode, the model is able to find substructure in this joint distribution that is not clearly distinguishable in IID models.

\begin{figure}
    \centering
    \includegraphics[width=0.99\textwidth]{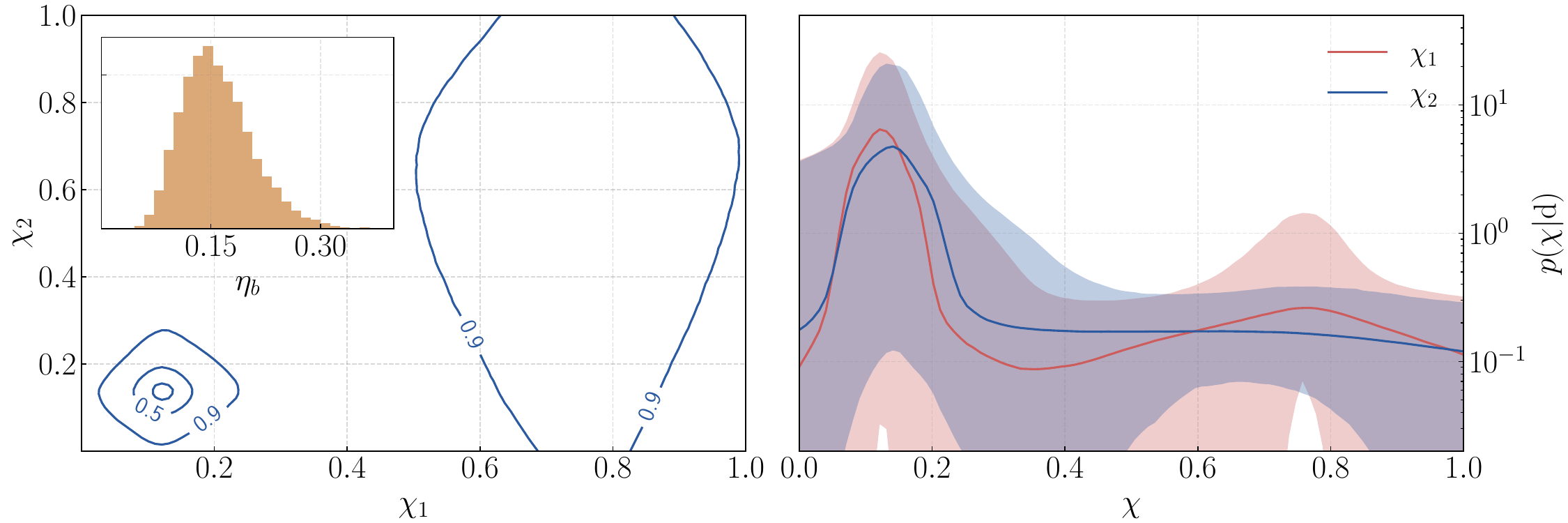}
    \caption{ \label{Fig:TGMM} A plot of the $10\%, 50\% \text{ and } 90\%$ HDI contours for the median of the joint $\chi_1 - \chi_2$ distribution on the left, as inferred with \tgmm. We find a dominant mode centered away from zero spins but below spin magnitudes of $0.2$ in both components. The subdominant mode is localized at $\chi_1 \sim 0.7$ but with a broad distribution for $\chi_2$. The analysis finds that the fraction of \ac{BBH} mergers in the subdominant mode, given by $\eta_b$, confidently excludes zero as shown in the inset. The marginal distributions of $\chi_1$ and $\chi_2$ are shown on the right. The dominant modes are again well recovered in both components. The subdominant mode is consistent with either a broad $\chi_2$ distribution or a potential peak at $\chi_2 \sim 0.7$.}
\end{figure}

\subsection{\defbbh Model}
\label{Sec:default_bbh_corners}
\noindent

This section presents details the default strongly-parameterized models used in this paper (see Tab.~\ref{tab:summary_of_models}). These are fully consistent with the fiducial models used by the \ac{LVK} in our analysis of \gwtcfour data in \cite{2025arXiv250818083T}.

\bpltp: The fiducial \ac{BBH} mass model is a mixture between a broken power law and two left-truncated Gaussian peaks, with low-mass tapering applied to the full distribution. The broken power law is given by

\begin{eqnarray}
	p_\text{BP}(m_1 \mid \alpha_1, \alpha_2, m_\text{break}, m_\text{1,low}, m_\text{high}) = \frac{1}{N}
	\begin{cases}
		  \left(\dfrac{m_1}{m_\text{break}}\right)^{-\alpha_1} & m_\text{1,low} \leq m_1  < m_\text{break} \\
		  \left(\dfrac{m_1}{m_\text{break}}\right)^{-\alpha_2} & m_\text{break} \leq m_1  < m_\text{high},
	\end{cases}
\end{eqnarray}

\noindent where $\alpha_1$ and $\alpha_2$ are the power-law indices, the transition between the low-mass and high-mass power law occurs at $m_\text{break}$, and the normalization constant is

\begin{equation}
	N = m_{\rm break} \left[
    \frac{1 - \left({m_{1,\rm low}}/{m_{\rm break}}\right)^{1-\alpha_1}}{1-\alpha_1}
    + \frac{\left({m_{1,\rm high}}/{m_{\rm break}}\right)^{1-\alpha_2} - 1}{1-\alpha_2}\right].
\end{equation}

\noindent The full mixture distribution is

\begin{align}
 \pi(m_1 )\propto & \Biggl[ \lambda_0 p_{\rm BP}(m_1\mid\alpha_1, \alpha_2, m_{\rm break}, m_{1,{\rm low}}, m_{\rm high}) + \lambda_1 \mathcal{N}_{lt}(m_1 \mid \mu_1, \sigma_1, \text{low} = m_{1,{\rm low}}) \\ \nonumber
& + (1 - \lambda_0 - \lambda_1) \mathcal{N}_{lt}(m_1\mid\mu_2, \sigma_2, \text{low} = m_{1,{\rm low}}) \Biggr] S(m_1 \mid m_{1,{\rm low}}, \delta_{m,1}),
\end{align}

\noindent where $\mathcal{N}_{lt}$ is a left-truncated normal distribution. The Planck tapering function $S$ ensures a smooth turn-on of the distribution in the range $(m_{\rm 1,low}, m_{\rm 1,low} + \delta_{m,1}]$ and is given by

\begin{eqnarray}
S(m\mid m_{\rm low}, \delta_m) =  \begin{cases}
0 & m < m_{\rm low}, \\\
[1 + f(m - m_{\rm low}, \delta_m)]^{-1}  & m_{\rm low} \leq m < m_{\rm low} + \delta_m, \\
1 & m_{\rm low} + \delta_m \leq m,
\end{cases}
\end{eqnarray}

\noindent with
\begin{equation}
f(m', \delta_m) = \text{exp} \Biggl( \frac{\delta_m}{m'} + \frac{\delta_m}{m' - \delta_m} \Biggr).
\end{equation}

\begin{figure}
    \centering
    \includegraphics[width=1.0\textwidth]{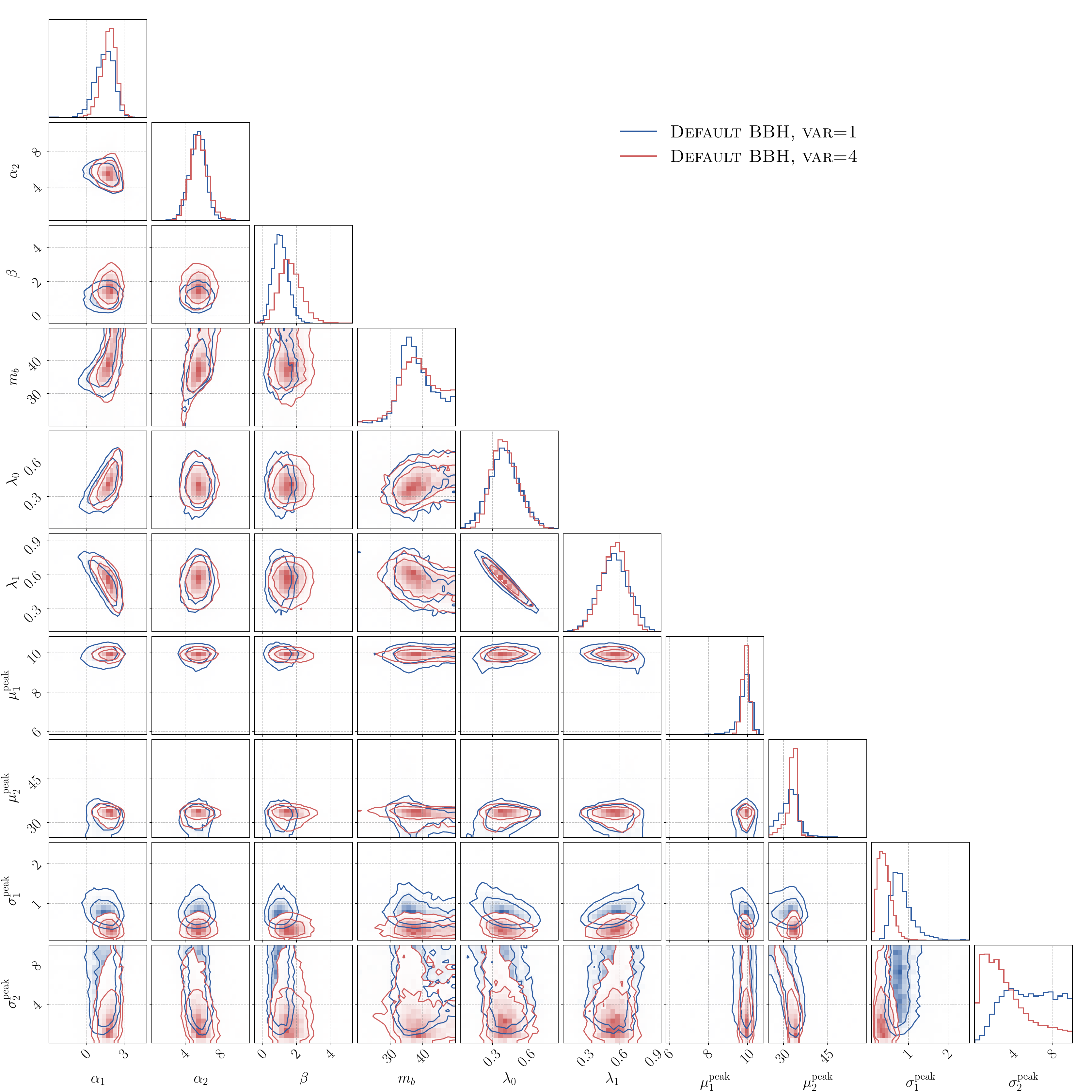}
    	    \caption{\label{fig:bpl2pk_corners} The posterior distribution of the parameters (see Tab.~\ref{tab:bpl2pk_parameters} for definition) of the \defbbh mass model. The two colors corresponds to inferences with two different variance cuts (see Sec.~\ref{Sec:methods}). Other than expected differences in the width of the Gaussian peaks, other parameter estimates are consistent between both runs.}
\end{figure}

We model the mass ratio as a power law with index $\beta_q$ and low-mass tapering applied to secondary mass $m_2$, conditioned on primary mass $m_1$,

Table~\ref{tab:bpl2pk_parameters} shows the parameter descriptions and prior ranges. Figure~\ref{fig:bpl2pk_corners} shows the posterior distributions of the hyperparameters for two different variance cuts.

\begin{deluxetable}{ccc}
\tablecaption{Summary of \bpltp model parameters and priors.}
\tablehead{
\colhead{Parameter} & \colhead{Description} & \colhead{Prior}
}
\label{tab:bpl2pk_parameters}
\startdata
$\alpha_1$ & {Power-law index} of 1st primary-mass power law & U$(-4, 12)$ \\
$\alpha_2$ & {Power-law index} of 2nd primary-mass power law & U$(-4, 12)$ \\
$m_{\rm break}$ & {Power-law break} location & U$(20, 50)$ \\
$\mu_1$ & Location of the first peak & U$(5, 20)$ \\
$\sigma_1$ & Width of the first peak & U$(0,10)$ \\
$\mu_2$ & Location of the second peak & U$(25, 60)$ \\
$\sigma_2$ & Width of the second peak & U$(0,10)$ \\
$m_{\rm 1, low}$ & Lower edge of taper function & U$(3, 10)$ \\
$\delta_{\rm m,1}$ & Mass range of low mass tapering & U$(0,10)$ \\
$\lambda_0, \lambda_1$ & Mixing fractions between power law and peaks & ${\rm Dir}(\mathbf{\alpha}=(1,1,1))$ \\
$m_{\rm high}$ & Maximum mass for distribution, which is pinned to $m_{\rm high} = 300\,\Msun$ by default & $\delta(m_{\rm high} - 300)$ \\
\tableline
$\beta_q$ & Power-law index of mass-ratio power law & U$(-2,7)$ \\
$m_{\rm 2, low}$ & Lower edge of taper function in $m_2$ & U$(3, m_{\rm 1, low})$ \\
$\delta_{\rm m,2}$ & Mass range of low-mass tapering in $m_2$ & U$(0,10)$ \\
\enddata
\end{deluxetable}

\defbbh component spin model: We model the spin magnitudes ($\chi_i$) as a truncated Gaussian distribution between $0$ and $1$, assuming they are identically and independently distributed: 
\begin{equation}
    \pi(\chi_i \mid \mu_\chi, \sigma_\chi) = \mathcal{N}_{[0,1]}(\chi_1 \mid \mu_\chi, \sigma_\chi)\mathcal{N}_{[0,1]}(\chi_2 \mid \mu_\chi, \sigma_\chi)\,.
\label{eqn:spin_mag_model}
\end{equation}

\begin{figure}
    \centering
    \includegraphics[width=0.8\textwidth]{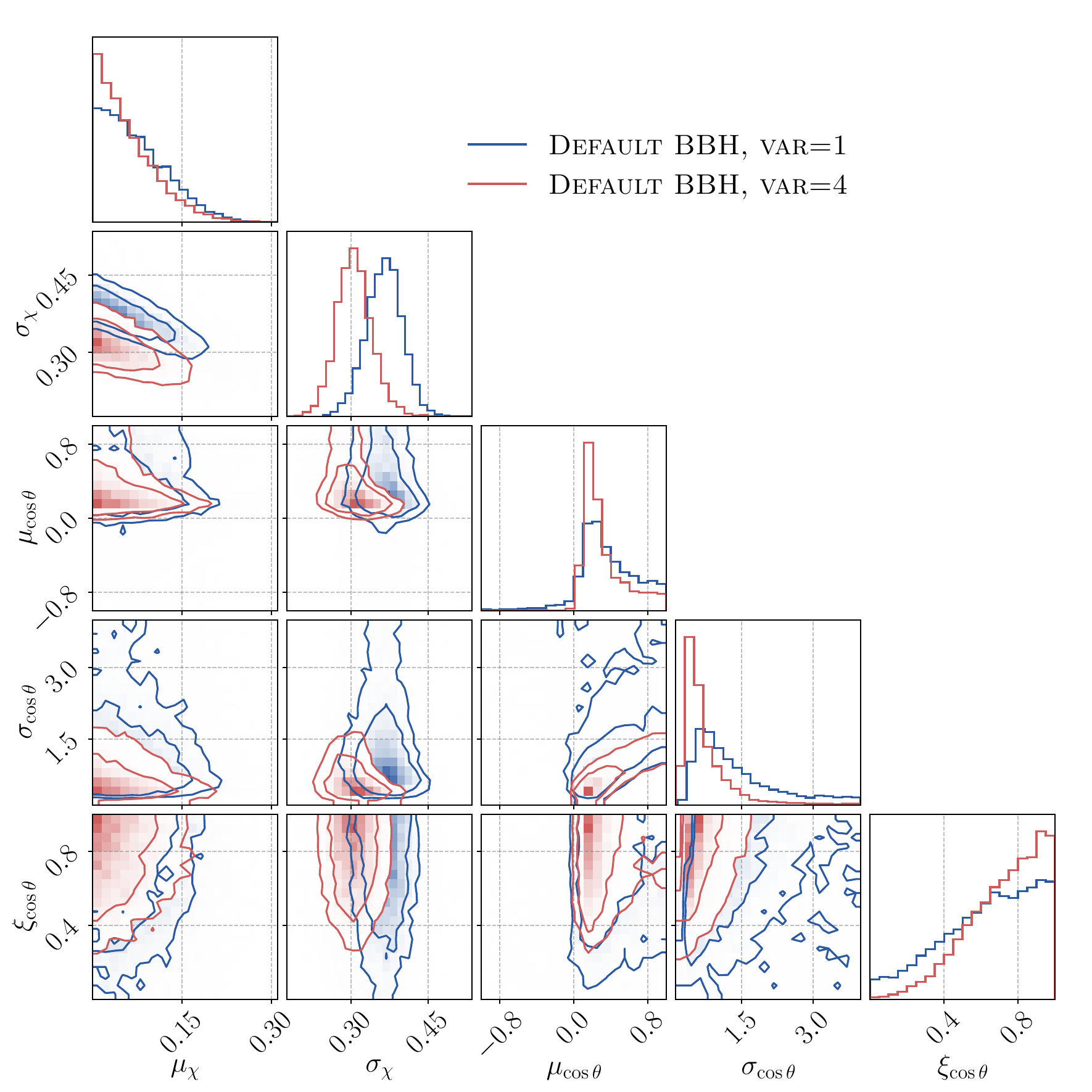}
    	    \caption{\label{fig:def_spin_corners} The posterior distribution of the parameters (see Tab.~\ref{tab:def_bbh_spin_priors} for definition) of the \defbbh spin model. The two colors corresponds to inferences with two different variance cuts (see Sec.~\ref{Sec:methods}). Other than expected differences in the width of the Gaussians, other parameter estimates are consistent between both runs.}
\end{figure}

We model the distribution of the cosine spin tilt angle ($\cos\theta_i$) as a mixture between a Gaussian distribution truncated on $-1$ to $1$ and an isotropic distribution, assuming they are identically but \textit{not} independently distributed:
\begin{equation}
    \pi(\cos\theta_i \mid \mu_t, \sigma_t, \zeta ) = \zeta \,\mathcal{N}_{[-1,1]}(\cos\theta_1 \mid \mu_t, \sigma_t)\mathcal{N}_{[-1,1]}(\cos\theta_2 \mid \mu_t, \sigma_t) + \frac{1-\zeta}{4}\,.
\label{eqn:spin_tilt_model} 
\end{equation}
We here allow for the location of the Gaussian subpopulation to vary, following~\cite{Vitale:2022dpa}. Priors on the hyperparameters are given in Tab.~\ref{tab:def_bbh_spin_priors}. Posteriors of the model parameters are shown in Fig.~\ref{fig:def_spin_corners} with two different variance cuts.

\begin{deluxetable}{ccc}
\tablecaption{Summary of \defbbh model parameters for spin magnitudes (Eq.~\ref{eqn:spin_mag_model}) and tilt angles (Eq.~\ref{eqn:spin_tilt_model}).}
\tablehead{
\colhead{Parameter} & \colhead{Description} & \colhead{Prior}
}
\label{tab:def_bbh_spin_priors}
\startdata
$\mu_\chi$ & Location of the $\chi$ distribution & U($0$, $1$) \\
$\sigma_\chi$ & Width of the $\chi$ distribution & U($0.005$, $1$) \\
$\mu_t$ & Location of the Gaussian component of the $\cos\theta$ distribution & U($-1$, $1$) \\
$\sigma_t$ & Width of the Gaussian component of the $\cos\theta$ distribution & U($0.01$, $4$) \\
$\zeta$ & Fraction in the Gaussian component of the $\cos\theta$ distribution & U($0$, $1$) \\
\tableline
$t_{\rm min}$ & Minimum of the $\cos\theta$ distribution & U($-1$, $1$) \\
\enddata
\end{deluxetable}

\begin{figure}
    \centering
    \includegraphics[width=0.5\textwidth]{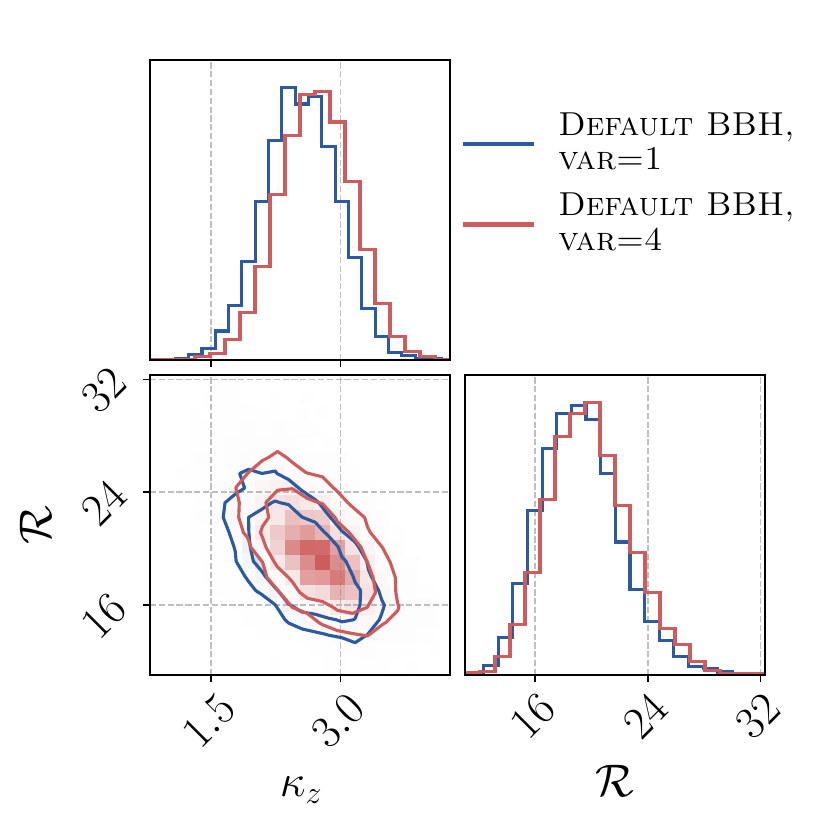}
    	    \caption{\label{fig:def_z_corners} The posterior distribution of the parameters (see Tab.~\ref{tab:redshift_model_priors} for definition) of the \defbbh redshift model. The two colors corresponds to inferences performed using two different convergence cuts.}
\end{figure}

\defbbh redshift model: We model redshift evolution by the comoving merger rate density. In particular, we use the model
\begin{equation}
\pi(z\mid\kappa) \propto \frac{1}{1+z}\frac{\mathrm{d} V_c}{\mathrm{d} z}(1+z)^\kappa
\label{eq:powerlawredshift}
\end{equation}
where the prefactor converts from a rate density in comoving volume and source-frame time to detector-frame time and redshift. 
In other words, the comoving rate density scales as $\mathcal{R} \propto (1+z)^\kappa$.
We use a prior on $\kappa$ as specified in Tab.~\ref{tab:redshift_model_priors}. The posterior inferences of the parameters are shown in Fig.~\ref{fig:def_z_corners}.

\begin{deluxetable}{ccc}
\tablecaption{Summary of \defbbh redshift model parameter (top) and priors and Madau-Dickinson model parameters (bottom), and the corresponding priors.}
\tablehead{
\colhead{Parameter} & \colhead{Description} & \colhead{Prior}
}
\label{tab:redshift_model_priors}
\startdata
$\kappa_z$ & Power-law index on comoving merger rate evolution & U($-10, 10$) \\
\tableline
$\gamma$ & The low redshift index of the Madau-Dickinson model & U($-10, 10$) \\
$z_{\rm peak}$ & The peak redshift parameter of the Madau-Dickinson model & U($0, 2.5$)\\
$\alpha$ & The high redshift index of the Madau-Dickinson model & U($-10, 10$) \\
\enddata
\end{deluxetable}

\subsection{Madau--Dickinson Model}
\label{Sec:md_model}

We test an alternative redshift model based on the parametric form of the Madau--Dickinson Star Formation Rate. 

\begin{equation}
    \psi(z \mid \gamma, \alpha, z_p) \propto \frac{(1 + z)^\gamma}{1 + ((1 + z)/(1 + z_p))^\alpha}.
    \label{Eq:MD}
\end{equation}

At $z \ll z_p$, the spectral index of the model is given by 
\begin{equation}
    \kappa_{\rm eff} = \gamma - \frac{\alpha}{1 + (1 + z_p)^\alpha}
    \label{Eq:lowz_MD}
\end{equation}

\begin{figure}
    \centering
    \includegraphics[width=0.95\textwidth]{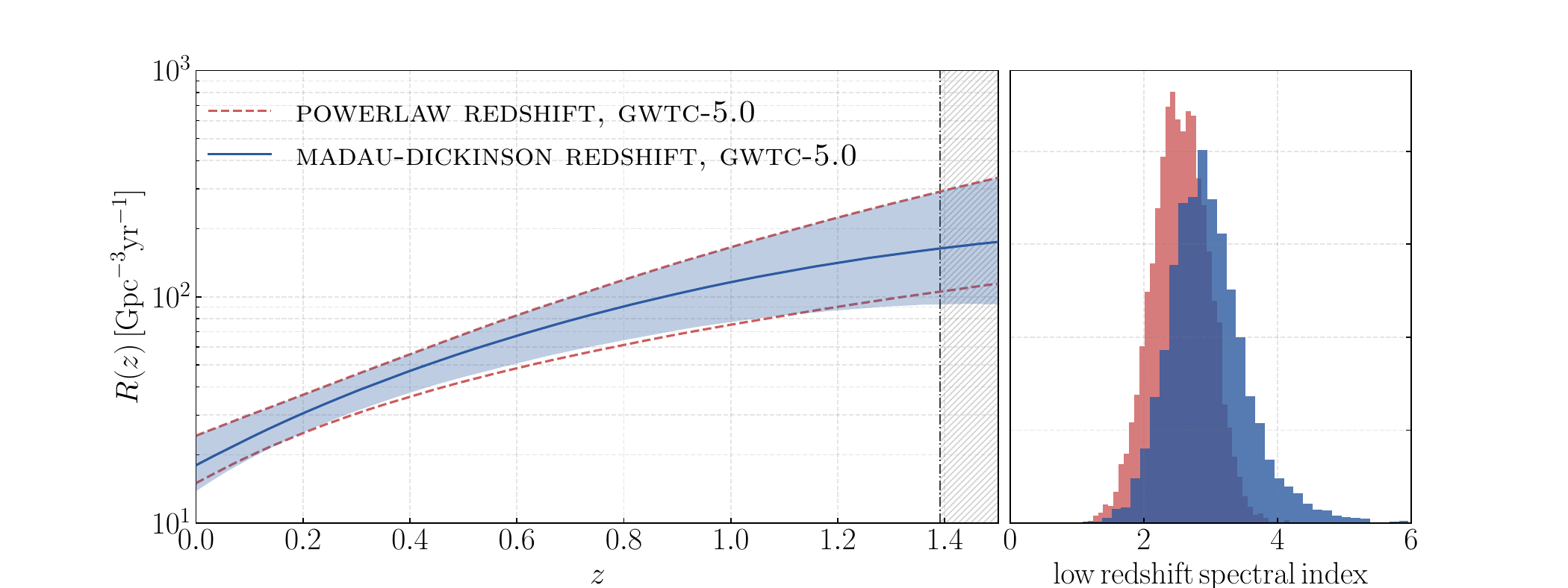}
    	    \caption{\label{fig:md_redshift} The redshift distribution of the \ac{BBH} population under the Madau--Dickinson model (Eq.~\ref{Eq:MD}) is shown on the left (with the \defbbh redshift model shown in comparison). In the right, the inferred value of the low redshift spectral index $\kappa_z$ assuming $p(z) \propto (1 + z)^{\kappa_z}$, and for the Madau--Dickinson model~(Eq.~\ref{Eq:lowz_MD}).}
\end{figure}

Model priors are shown in Tab.~\ref{tab:redshift_model_priors} and the inferred redshift distribution in Fig.~\ref{fig:md_redshift}. No major deviations are noticeable between the inferred redshift distribution of the Madau--Dickinson model and the \defbbh redshift model.

\subsection{\bivskewchichi}
\label{sec:bivariate_skewnormal}

The \bivskewchichi uses a truncated version of the multivariate skewnormal model~\citep{Azzalini:1996}. This model combines the bivariate Gaussian~\citep{2020ApJ...895..128M} and the skewnormal models~\citep{Banagiri:2025dmy} used to fit effective spins in previous \ac{LVK} analyses~\citep{2025arXiv250818083T, 2023PhRvX..13a1048A}. This model therefore allows both the \chieff and \chip distributions to be skewed and be covariant with each other.

The model is formally defined as,

\begin{equation}
	p(\chieff, \chip) \propto
  \begin{cases}
     2 \mathcal{N} (\vec{\chi} \mid \vec{\mu}, \Sigma) \, \Phi (\vec{\chi} - \vec{\mu} \cdot \vec{\alpha})  & \text{if } \chieff \in [-1, 1] \text{ and } \chip \in [0, 1], \\
   0 & \text{otherwise}\\
  \end{cases}.
\end{equation}

\noindent Here, $\vec{\chi} = (\chieff, \chip)$ is the vector of effective spins. The parameters $\vec{\mu} = (\mu_{\rm eff}, \mu_{\rm p})$ represent the mean vector, and $\Sigma$, the covariance matrix with normalized covariance $\rho$:

\begin{equation}
	\Sigma = \begin{bmatrix}
\sigma^2_{\rm eff} & \rho \sigma_{\rm eff} \sigma_{\rm p}\\
\rho \sigma_{\rm eff} & \sigma_{\rm p}^2
	\end{bmatrix},
\end{equation}

\noindent and $\vec{\alpha} = (\alpha_{\rm eff}, \alpha_{\rm p})$ as the skewness matrix. While $\alpha_{\rm eff}$ and $\alpha_{\rm p}$ formally can take any real numbered value, for consistency with \cite{2025arXiv250818083T} and \cite{Banagiri:2025dmy}, we define $\epsilon_{\rm eff}$ and $\epsilon_{\rm p}$ such that,

\begin{equation}
	\alpha_{\rm eff (p)} = \epsilon_{\rm eff (p)} + \tan \frac{\pi \epsilon_{\rm eff (p)}}{2}.
\end{equation}

\begin{figure}
    \centering
    \includegraphics[width=1.0\textwidth]{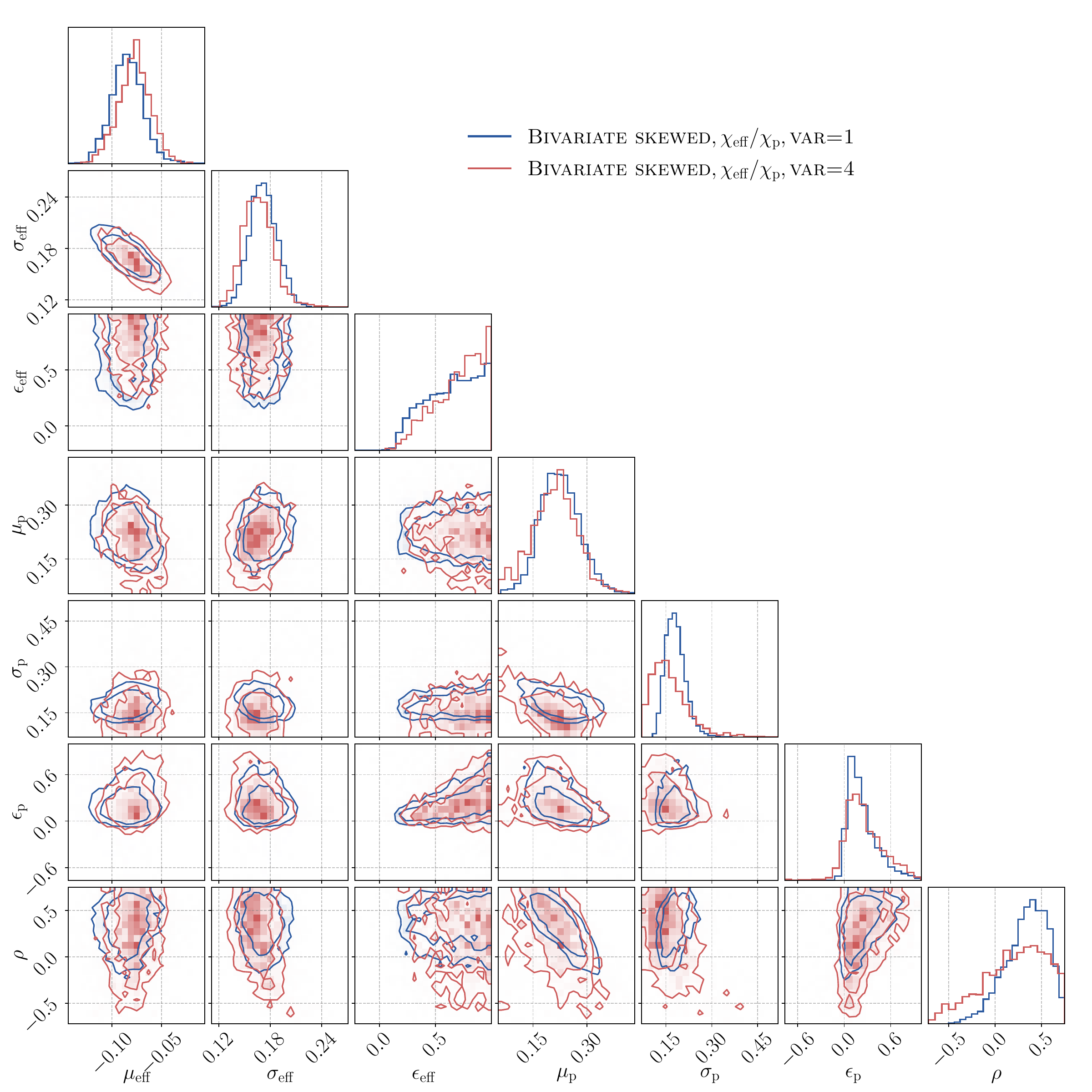}
    	    \caption{\label{fig:bsn_corners} The posterior distribution of the parameters for the \bivskewchichi model. {The two colors corresponds to inferences with two different variance cuts (see Sec.~\ref{Sec:methods}). We find that the run with higher variance cut infers a $\mu_{\rm p}$ value that is slightly lower but still statistically consistent with the default run with a variance cut of 1}. We see that \bivskewchichi prefers a positive skewness  ($\epsilon_{\rm eff} >0$ at $\OverviewEstimates[bsn_chieff_skew_cred]$ credibility) while for $\chi_{\rm p}$, the model prefers small skew. The correlation coefficient between \chieff and $\chi_{\rm p}$ is consistent with zero. }
\end{figure}

Table~\ref{tab:effectiveSpinPriors} shows the parameter priors and Fig.~\ref{fig:bsn_corners} shows the posteriors on the model parameters. We find that the \bivskewchichi model and the (\chieff{}, $\chi_p$)-\pixelpop model disagree on the degree of skewness of the \chieff{} distribution. Using the asymmetry about the mode of the distribution as a measure of skewness, we find that the former model infers skewness to be $\CIPlusMinus{\OverviewEstimates[bsn_chieff_skew]}$, while the latter infers a value of $\CIPlusMinus{\OverviewEstimates[pp_chieff_skew]}$ -- consistent with zero.

\begin{deluxetable}{ccc}
\tablecaption{Summary of \bivskewchichi model parameters}
\tablehead{
\colhead{Parameter} & \colhead{Description} & \colhead{Prior}
}
\label{tab:effectiveSpinPriors}
\startdata
$\mu_\mathrm{eff}$ & Location of the $\chieff$ distribution & U($-1$, $1$)  \\
$\sigma_\mathrm{eff}$ & Width of the $\chieff$ distribution & U($0.05$, $1$)  \\
$\mu_p$ & Location of the $\chi_\mathrm{p}$ distribution & U($0.05$, $1$)\\
$\sigma_p$ & Width of the $\chi_\mathrm{p}$ distribution & U($0.07$, $1$)\\
$\rho$ & Degree of correlation between $\chieff$ and $\chi_\mathrm{p}$ & U($-0.75$, $0.75$) \\
$\epsilon_{\rm eff}$ & \chieff{} component of the skew of the distribution & U($-1$, $1$) \\
$\epsilon_{\rm p}$ & $\chi_\mathrm{p}$ component of the skew of the distribution & U($-1$, $1$) \\
\enddata
\end{deluxetable}

\section{Mass Ratio and Spin Correlation Variants}
\label{app:qchieffvarients}

\begin{figure}
\centering
\includegraphics[scale=1]{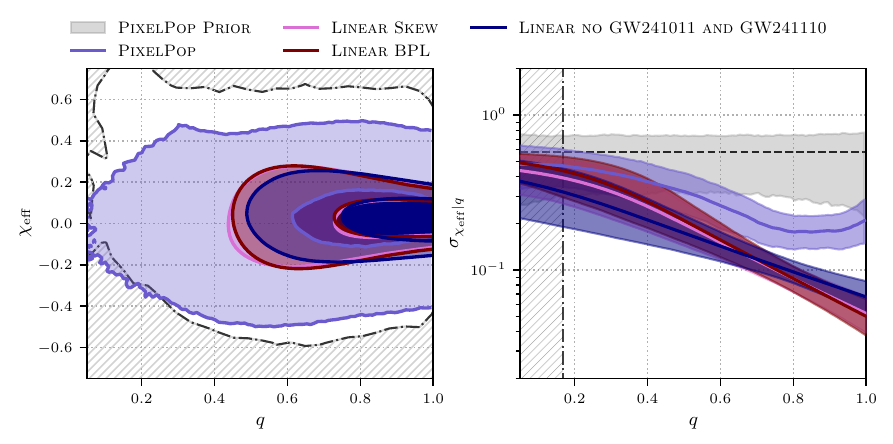} 
\caption{\label{fig:chieffqmusig_linearvars} Left: the joint $(q,\chieff)$ \ac{PPD} for the $(q,\chieff)$-\pixelpop model, for the \linbpl (pink) and \linske (dark red) variants of the default \lincorr model, and for the default \lincorr model omitting highly-spinning, unequal-mass-ratio events GW241011 and GW241110 (dark blue).
Right: the inferred standard deviation of the \chieff distribution as a function of mass ratio for the same models.
The shaded bands show the $90\%$ credible interval, where the grey band shows the $99\%$ credible interval of the $(q,\chieff)$-\pixelpop prior on $\sigma_{\chi_\mathrm{eff}}(q)$. 
At low $q$, the standard deviation reaches a maximum value at $\sim 1/\sqrt{3}$, equivalent to the standard deviation of a uniform distribution over 
$(-1,1)$, shown in the black dashed line. 
The standard deviation shows a preference for a narrowing \chieff with increasing $q$ for all \lincorr model variants. 
The $(q,\chieff)$-\pixelpop result also shows a preference for  narrowing \chieff with increasing $q$, though to a lesser degree due to modeling assumptions.
}
\end{figure}

We verify the correlation between mass ratio and effective spin (Sec.~\ref{Sec:qchieff}) with variations of the \lincorr model and the weakly-parametrized $\pixelpop$ model.
Namely, we test the \lincorr ($q,\chieff$) model with a broken power-law distribution in $q$ instead of a power-law distribution (\linbpl), and a skewnormal distribution in $\chieff$ instead of a truncated Gaussian (\linske).
The joint ($q,\chieff$) distribution and the width of $\chieff$ with $q$ for all model variants are shown in Fig.~\ref{fig:chieffqmusig_linearvars}. 
We find that the variants agree with the default model result: the position of the peak of the $\chieff$ distribution does not evolve with $q$, and that the width of the $\chieff$ distribution gets wider towards more unequal $q$. 
In addition, we test the impact of unequal-mass, high-spin, sources, such as GW241011 and GW241110, with a leave-one-out analysis using the default \lincorr model, finding that these observations do not impact the result.
This confirms that the evolution of the width of \chieff with $q$ is a robust feature of the data, and is not driven by particular choices of marginal distributions nor by individual events GW241011 and GW241110. 

The more flexible $\pixelpop$ is also in agreement with the lack of evolution of the mean of \chieff, with $\delta \mu_{\chi_\mathrm{eff}|q=0.7} = \PixelPopqchieffdmu$. 
$\pixelpop$ finds the Spearman correlation statistic is negative at $\PixelPopqchieffSpearman\%$ credibility over the entire space ($\PixelPopqchieffSpearmanReduced\%$ credibility over the data-dominated region $[q>0.5, -0.25<\chieff<0.25]$) indicating little evidence for an anti-correlation \citep{ca468a70-0be4-389a-b0b9-5dd1ff52b33f}. 
$\pixelpop$ also identifies broadening, though this is less prominent than the strongly-parametrized models. 
We find $\delta \mathrm{ln} \sigma_{\chi_\mathrm{eff}|q=0.7}= \PixelPopqchieffdlnsig$ at $q=0.7$, which is approximately where the gradients are the largest and uncertainties on the mean and width are smallest.
It is not clear whether this evolution is a transition between distinct features along the mass-ratio spectrum with different widths of the \chieff distribution, or a gradual transition along $q$ as is assumed by the \lincorr model. 

There may be several explanations for why the evolution of $\sigma_{\chi}(q)$ disagrees between the \lincorr and $\pixelpop$ models. 
The standard deviation has a mathematical limit, approximately equal to the standard deviation of a truncated uniform distribution over $(-1,1)$. 
The result for $\pixelpop$ reaches this limit at $q<0.3$, hence we caution against over interpreting $\sigma_{\chi}$ in this region.
We calculate all other results plotted in Fig.~\ref{fig:chieffqmusig_linearvars} consistently with the $\pixelpop$ result, hence these also reach this limit at low $q$. 
This limit is not fundamental to the population models.%
Towards more equal mass ratios ($q\gtrsim0.7$), $\pixelpop$ finds a broader distribution of \chieff than the \lincorr and \splcorr models. 
$\pixelpop$ a priori disfavours narrow distributions, as highlighted by the model's prior on the width of \chieff (grey interval in the right panel of Fig.~\ref{fig:chieffqmusig_linearvars}). 
We find that this is due to a combination of $\pixelpop$'s ICAR prior preference for smoother distributions, and that measuring the width of a narrow distribution requires strong constraints in the tails. 
While stronger approaches such as the \lincorr and \splcorr models extrapolate in the tails using their more rigid functional form, weaker models like $\pixelpop$ constrain data-poor regions using a combination of the prior (favoring smoothness) and the information due to non-detection (upper bounds).
The inflated uncertainty away from the bulk leads to a bias towards large standard deviations.
Compared to the prior, the data at $q\gtrsim 0.6$ are highly informative towards a narrower \chieff distribution, indicating a true preference for a narrower \chieff distribution at more equal mass ratios. 
The disagreement between the \lincorr and $\pixelpop$ models for $q>0.7$ could also come from the \lincorr model's assumption that the evolution of the width of \chieff is the same at all $q$. 
Instead, it may be that the width of the \chieff distribution is narrowest at $q<1$, which is not allowed within the modelling assumptions of the \lincorr model which requires monotonic evolution of the width of \chieff. 

\section{Redshift and Spin Correlation Variants}
\label{app:zchieffvarients}

\begin{figure}
\centering
\includegraphics[scale=1.]{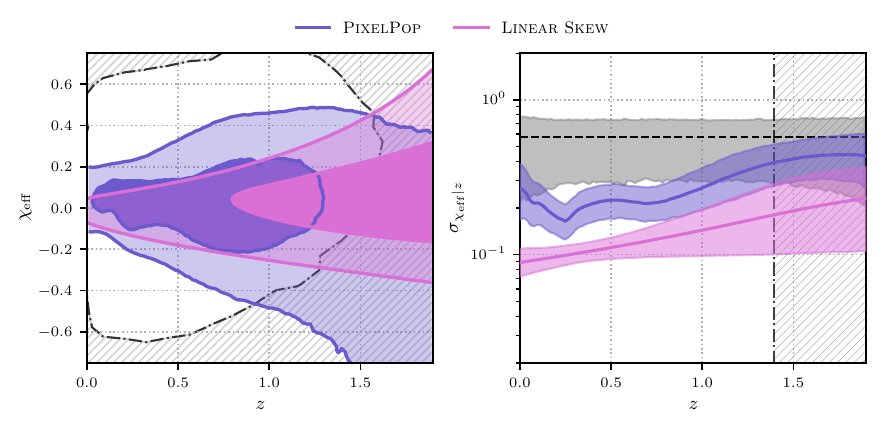} 
\label{fig:chieffzmusig_vars}
\caption{Left: the joint $(\chieff,z)$ distribution with the $\pixelpop$ model (purple) and the \linske model (pink). 
Right: the inferred standard deviation of the \chieff distribution as a function of redshift for the \lincorr model variant and the $(z,\chieff)$-\pixelpop model.
The shaded bands show the $90\%$ credible interval, where the grey band shows the $99\%$ credible interval of the $(z,\chieff)$-\pixelpop prior on $\sigma_{\chi_\mathrm{eff}}(z)$. 
At large redshifts, the standard deviation of the $\pixelpop$ model reaches a maximum value at $\sim 1/\sqrt{3}$, equivalent to the standard deviation of a uniform distribution over $(-1,1)$, shown in the black dashed line. 
The standard deviation shows a preference for a broadening of \chieff with increasing $z$ for the \lincorr model variant. 
The $(q,\chieff)$-\pixelpop result also shows a preference for  broadening of \chieff with increasing $z$, though to a lesser degree due to do modeling assumptions.}
\end{figure}

Figure \ref{fig:chieffzmusig_vars} shows the joint $(\chieff,z)$ distribution and the width of $\chieff$ with $z$ for two model variants: the $(z,\chieff)$-\pixelpop model, and a modified \lincorr model with a skewnormal marginal \chieff distribution (\linske).
The peak of the joint $(\chieff,z)$ distribution differs between both models due to modeling assumptions; the \lincorr model assumes a powerlaw in redshift which is strictly monotonic. 
The marginal redshift evolution is consistent within uncertainties for both $\pixelpop$ and a power-law model as seen in Fig.~\ref{fig:p_z_bbh}. 

We find that the \linske model prefers a slightly steeper evolution of the width of \chieff with $z$ than the default \lincorr model, though with larger uncertainties, with $\deltasigmachiz= \LinSkewzchieffdlnsig$. 
This is concurrent with an increasing skew of \chieff with larger redshifts. 

$\pixelpop$ also indicates some degree of broadening of \chieff with $z$.
Within $\chieff\in[-0.5,0.5]$, $z<1$---where we have the most information from observations---we find that the Spearman correlation coefficient for the broadening of the distribution for $\pixelpop$ is positive at $94.8\%$ credibility. 
This evidence for broadening can be seen from the increasing $\sigma_{\chi_\mathrm{eff}}(z)$ until $z\sim1.25$. 
Outside this range, this evidence diminishes and the evidence is consistent with no broadening, but this is likely due to the larger uncertainties. 
$\pixelpop$ finds that the mean of the \chieff distribution is consistent with no correlation for all values of $z$. 
Over the restricted domain of $\chieff\in[-0.5,0.5]$, $z<1$, the Spearman correlation coefficient is $\PixelPopzchieffSpearmanReduced$, indicating no correlation of the mean of \chieff with redshift, while outside this domain $\pixelpop$ is largely prior dominated due to a lack of observations.

With a \copcorr model, we find evidence for a positive correlation in $(\chieff,z)$. 
We measure the \copcorr model's correlation parameter $\etachiz=\Copulaetazchi$, or $\etachiz>0$ with $\CopulaetazchiNeg\%$ credibility --- up from $98\%$ for \gwtcfour. 
The correlation in the \copcorr model may be driven by the high-redshift region where we lack sensitivity, and hence the evidence for correlation could come from extrapolation.

Unlike $\pixelpop$, \lincorr, and \splcorr models, the \copcorr model lacks flexibility to fit broadening. 
If the \chieff distribution only broadens with redshift and there is no evolution in the mean, it might be expected that the \copcorr model finds no evidence for correlation, with \etachiz consistent with zero. 
One caveat is that the orbital hang-up effect leads to \acp{BBH} with negative \chieff being quieter than those with positive \chieff, all else held the same \citep{2006PhRvD..74d1501C}. 
This may cause the median of the $\pixelpop$ rate, as shown in Fig.~\ref{fig:chieffz}, to be wider towards negative \chieff as the model is less able to rule out contributions from negative \chieff \ac{BBH}.
Under $\pixelpop$ modelling assumptions, this manifests as broadening, and could obscure an underlying correlation. 
On the other hand, the correlation in the \copcorr model may be driven by the high-redshift region where we lack sensitivity, and hence the evidence for correlation could come from extrapolation. 
To compare to the $\pixelpop$ result, over the restricted domain of $\chieff\in[-0.5,0.5]$, $z<1$, the Spearman correlation coefficient for the \copcorr model is $\CopulazchiSpearmanReduced$, marginalised over the hyperposterior. 
This is in agreement with no correlation where we are most informed by observations, indicating that the \copcorr model is likely extrapolating to high redshifts.

\begin{deluxetable}{lll}
\tablecaption{Priors for the \lincorr model for the joint $q-\chieff$ distribution.
\label{tab:lincorr_qchieff_priors}}
\tablewidth{0pt}
\tablehead{
    \colhead{Parameter} & \colhead{Description} & \colhead{Prior}
}
\startdata
$\mu_{\chieff|q=0}$       & Mean of $\chieff$ distribution at $q=0$                           & $\mathrm{U}(-1,1)$  \\
$\mu_{\chieff|q=1}$       & Mean of $\chieff$ distribution at $q=1$                           & $\mathrm{U}(-2,2)$  \\
$\ln\sigma_{\chieff|q=0}$ & Natural log standard deviation of $\chieff$ distribution at $q=0$ & $\mathrm{U}(-5,0)$  \\
$\ln\sigma_{\chieff|q=1}$ & Natural log standard deviation of $\chieff$ distribution at $q=1$ & $\mathrm{U}(-12,4)$ \\
\enddata
\end{deluxetable}

\section{Inconclusive Correlation Results}\label{Sec.InconclusiveCorr}

Both our \gwtcfour analyses~\citep{2025arXiv250818083T} and follow-up work (e.g., \citealp{2025PhRvD.111l3046G,2025PhRvD.112h3028S,2025A&A...698A..85L}) failed to find model-independent evidence for an evolution of the primary-mass function with redshift. In \gwtcfive too we do not find that the primary mass evolves linearly with redshift.
We investigate this with a \copcorr model, which allows for a correlation between $m_1$ and $z$ with a variable strength $\eta_\mathrm{m_1,z}$. 
We find this correlation parameter to be $\eta_\mathrm{m_1,z}=\Copulaetamz$, consistent with zero. 
\pixelpop finds no strong evidence for evolution of the primary nor secondary mass distribution with redshift. 
However, particular features in the primary-mass spectrum may vary with redshift, as discussed in Sec.~\ref{sec:35_msun} of the main text.

We also explore the space of \chieff{} and \chip{} \citep[e.g.,][]{Plunkett:2026pxt}, but we do not find any evidence of correlations with any of the \copcorr, \bivskewchichi, or \pixelpop models.

\section{Correlation Models}

All strongly-parametrized correlation model parameters and priors are specified in tables below.

\begin{deluxetable}{lll}
\tablecaption{Priors for the \lincorr model for the joint $z-\chieff$ distribution. Note that the prior on the width of the peaks in $m_1$ differs from the \bpltp model priors to reduce Monte Carlo uncertainties.
\label{tab:lincorr_zchieff_priors}}
\tablewidth{0pt}
\tablehead{
    \colhead{Parameter} & \colhead{Description} & \colhead{Prior}
}
\startdata
$\sigma_1$ & Width of the first peak in $m_1$ & $\mathrm{U}(0.25,10)$ \\
$\sigma_2$ & Width of the second peak in $m_1$ & $\mathrm{U}(1,10)$ \\
\hline
$\mu_{\chieff|z=0}$       & Mean of $\chieff$ distribution at $z=0$                           & $\mathrm{U}(-1,1)$  \\
$\mu_{\chieff|z=0.5}$       & Mean of $\chieff$ distribution at $z=0.5$                           & $\mathrm{U}(-1,1)$  \\
$\ln\sigma_{\chieff|z=0}$ & Natural log standard deviation of $\chieff$ distribution at $z=0$ & $\mathrm{U}(-5,0)$  \\
$\ln\sigma_{\chieff|z=0.5}$ & Natural log standard deviation of $\chieff$ distribution at $z=0.5$ & $\mathrm{U}(-3,5)$ \\
\enddata
\end{deluxetable}

\begin{deluxetable}{lll}
\tablecaption{Priors for the \splcorr model for the joint $q-\chieff$ and joint $z-\chieff$ distributions.
\label{tab:splcorr_priors}}
\tablewidth{0pt}
\tablehead{
    \colhead{Parameter} & \colhead{Description} & \colhead{Prior}
}
\startdata
$\mu_{\chieff}$       & Mean of $\chieff$ distribution at \splcorr model knots                           & $\mathrm{U}(-1,1)$  \\
$\ln\sigma_{\chieff}$ & Natural log standard deviation of $\chieff$ distribution at \splcorr model knots & $\mathrm{U}(-5,0)$  \\
\enddata
\end{deluxetable}

\begin{deluxetable}{lll}
\tablecaption{Priors for the \lincorr \textsc{Skewnormal} and  \lincorr \textsc{Broken Power Law} models for the joint $q-\chieff$ distribution.
\label{tab:lincorrvars_qchieff_priors}}
\tablewidth{0pt}
\tablehead{
    \colhead{Parameter} & \colhead{Description} & \colhead{Prior}
}
\startdata
$\mu_{\chieff|q=0}$       & Mean of $\chieff$ distribution at $q=0$                           & $\mathrm{U}(-1,1)$  \\
$\mu_{\chieff|q=1}$       & Mean of $\chieff$ distribution at $q=1$                           & $\mathrm{U}(-1,1)$  \\
$\log_{10}\sigma_{\chieff|q=0}$ & $\log_{10}$ standard deviation of $\chieff$ distribution at $q=0$ & $\mathrm{U}(-1.5,0.3)$  \\
$\log_{10}\sigma_{\chieff|q=1}$ & $\log_{10}$ standard deviation of $\chieff$ distribution at $q=1$ & $\mathrm{U}(-1.5,0.3)$ \\ \hline
$\epsilon_{\chieff|q=0}$ & Skew of $\chieff$ distribution at $q=0$ for the \textsc{Skewnormal} model & $\mathrm{U}(-1,1)$  \\
$\epsilon_{\chieff|q=1}$ & Skew of $\chieff$ distribution at $q=1$ for the \textsc{Skewnormal} model & $\mathrm{U}(-1,1)$ \\ \hline
$\beta_1$ & Power law index of the mass ratio distribution when $q < q_b$ for the \textsc{Broken Power Law} model & $\mathrm{U}(-2,7)$  \\
$\beta_2$ & Power law index of the mass ratio distribution when $q > q_b$ for the \textsc{Broken Power Law} model & $\mathrm{U}(-2,7)$  \\
$q_b$ & Mass ratio at which the power-law index changes from $\beta_1$ to $\beta_2$ for the \textsc{Broken Power Law} model & $\mathrm{U}(0,1)$ \\ \hline
\enddata
\end{deluxetable}

\begin{deluxetable}{lll}
\tablecaption{Priors for the \lincorr \textsc{Skewnormal} model for the joint $z-\chieff$ distribution.
\label{tab:lincorrvars_zchieff_priors}}
\tablewidth{0pt}
\tablehead{
    \colhead{Parameter} & \colhead{Description} & \colhead{Prior}
}
\startdata
$\mu_{\chieff|z=0}$       & Mean of $\chieff$ distribution at $z=0$                           & $\mathrm{U}(-1,1)$  \\
$\mu_{\chieff|z=1.9}$       & Mean of $\chieff$ distribution at $z=1.9$                           & $\mathrm{U}(-1,1)$  \\
$\log_{10}\sigma_{\chieff|z=0}$ & $\log_{10}$ standard deviation of $\chieff$ distribution at $z=0$ & $\mathrm{U}(-1.5,0.3)$  \\
$\log_{10}\sigma_{\chieff|z=1.9}$ & $\log_{10}$ standard deviation of $\chieff$ distribution at $z=1.9$ & $\mathrm{U}(-1.5,0.3)$ \\ \hline
$\epsilon_{\chieff|z=0}$ & Skew of $\chieff$ distribution at $z=0$ for the \textsc{Skewnormal} model & $\mathrm{U}(-1,1)$  \\
$\epsilon_{\chieff|z=1.9}$ & Skew of $\chieff$ distribution at $z=1.9$ for the \textsc{Skewnormal} model & $\mathrm{U}(-1,1)$ \\ \hline
\enddata
\end{deluxetable}

\section{Models for rapidly-spinning subpopulations}

In Sec.~\ref{sec:hierarchical}, we model the mass dependence of the \chieff distribution using both weakly- and strongly-parametrized descriptions. Here, we give more details on the models.

\begin{figure}[h]
	\centering
	\includegraphics[width=0.7\textwidth]{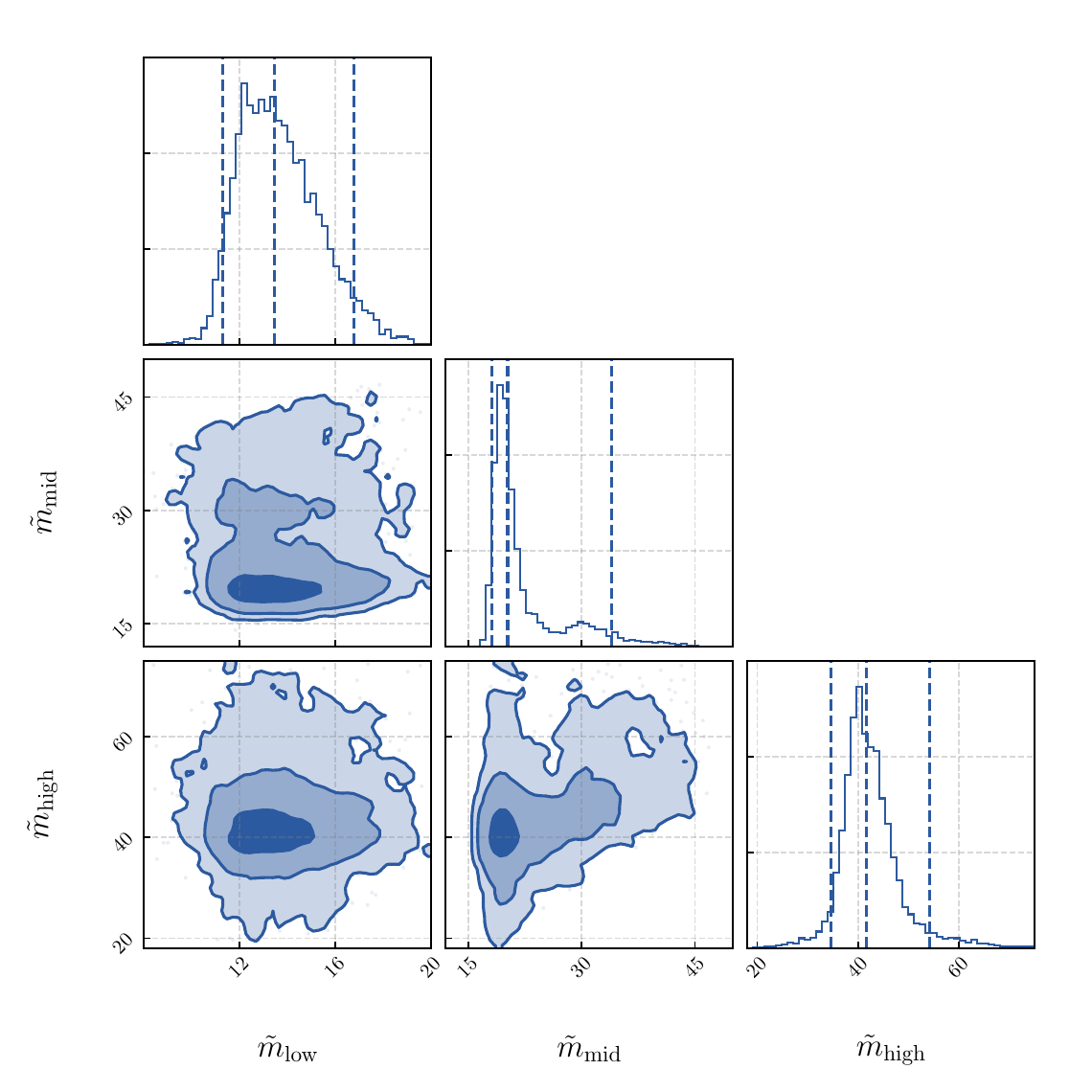}
	\caption{Posterior distributions of the transition masses \mlow{}, \mmid{}, and \mhigh{} inferred with the three-transition model. One-dimensional marginals are shown on the diagonal, with dashed lines marking the median and 90\% credible interval; two-dimensional credible regions are shown off-diagonal.}
	\label{fig:three_transition_post}
\end{figure}

\subsection{\chieffmixture}
\label{app:zeta_model}

\chieffmixture is a weakly-parametrized model in which the \chieff distribution at fixed $m_1$ is described as a mixture of a truncated Gaussian and a uniform component \citep{Antonini:2025ilj, Antonini:2025zzw},
\begin{equation}
	\pi(\chieff\mid m_1,\Lambda)=
	\zeta(m_1)\,\mathcal{U}(\chieff\mid \chi_{\min},\chi_{\max})
	+\bigl(1-\zeta(m_1)\bigr)\,\mathcal{N}_{\rm trunc}(\chieff\mid \mu,\sigma).
\end{equation}
Here, $\mu$ and $\sigma$ describe the Gaussian component, while $\chi_{\min}$ and $\chi_{\max}$ define the support of the uniform component. The mass dependence is encoded in the mixture fraction $\zeta(m_1)$, which is modeled non-parametrically as
\begin{equation}
	\zeta(m_1)=S\!\bigl(\Psi(\ln m_1)\bigr),
\end{equation}
where $S(x)=(1+e^{-x})^{-1}$ ensures $0\leq\zeta(m_1)\leq1$. The latent function $\Psi$ is drawn from a Gaussian process,
\begin{equation}
	\Psi(\ln m_1)\sim \mathcal{GP}\bigl(0,k(x,x';a_\zeta,\ell_\zeta)\bigr),
\end{equation}
with squared-exponential kernel $k$. The hyperparameters $a_\zeta$ and $\ell_\zeta$ control the amplitude and correlation length of the process, respectively. {The Gaussian process is evaluated on a grid uniform in $\ln m_1$, and $\zeta(m_1)$ is obtained via interpolation between grid points. This model allows the relative contribution of the broad spin component to vary smoothly with primary mass. The list of model parameters and their priors are shown in Tab.~\ref{tab:chieffMixturePriors}.}

\begin{deluxetable}{ccc}
\tablecaption{Summary of \chieffmixture model parameters and priors (top) and the Gaussian process hyperparameters (bottom).}
\tablehead{
\colhead{Parameter} & \colhead{Description} & \colhead{Prior}
}
\label{tab:chieffMixturePriors}
\startdata
$\mu$ & Location of the Gaussian component & U($-1$, $1$)  \\
$\log \sigma$ & Scale of the Gaussian component & U($-1.5$, $0$)  \\
$\chi_{\rm max}$ & Upper bound of the uniform component & U($0.1$, $1$)  \\
$\chi_{\rm min, unscaled}$ & Unscaled lower bound of the uniform component & U($0$, $1$)  \\
\tableline
$a_\zeta$ & Gaussian process amplitude parameter & $\mathcal{HN}(5)$ \\
$\ell_\zeta$ & Gaussian process correlation length parameter & $\mathcal{N}(-0.5, 1)$ \\
\enddata
\tablecomments{The true lower bound, $\chi_{\rm min}$ is obtained by the transformation $\chi_{\rm min} = \chi_{\rm min, unscaled} (1 + \chi_{\rm max}) - 1$. $\mathcal{HN}(a)$ denotes a half-normal distribution with scale parameter $a$.}
\end{deluxetable}
\subsection{\chieffthree}
\label{app:three_transition_model}

As a more structured alternative, we consider a strongly-parametrized model in which the \chieff distribution is specified across four primary-mass intervals separated by three transition masses, $\mlow$, $\mmid$, and \mhigh{} \citep{Tong:2025xir}. In all mass intervals, the baseline component is a truncated Gaussian ($\mathcal{N}_{\rm trunc}(\chieff\mid \mu,\sigma)$) with shared hyperparameters $\mu$ and $\sigma$. In the intervals $\mlow<m_1<\mmid$ and $m_1>\mhigh$, this Gaussian component is mixed with a uniform component. The support of the uniform distribution is allowed to differ between the two mass ranges, and the corresponding mixture fractions are described by $\xi_\text{low-mass}$ and $\xi_\text{high-mass}$. The full model is
\begin{equation}
	\pi(\chieff\mid m_1,\Lambda)=
	\begin{cases}
		\mathcal{N}_{\rm trunc}(\chieff\mid \mu,\sigma), & m_1 < \mlow,\\
		\xi_\text{low-mass}\,\mathcal{U}(\chieff\mid \chi_\text{min}^\text{low-mass},\chi_\text{max}^\text{low-mass})
		+ (1-\xi_\text{low-mass})\,\mathcal{N}_{\rm trunc}(\chieff\mid \mu,\sigma),
		& \mlow < m_1 < \mmid,\\
		\mathcal{N}_{\rm trunc}(\chieff\mid \mu,\sigma), & \mmid < m_1 < \mhigh,\\
		\xi_\text{high-mass}\,\mathcal{U}(\chieff\mid \chi_\text{min}^\text{high-mass},\chi_\text{max}^\text{high-mass})
		+ (1-\xi_\text{high-mass})\,\mathcal{N}_{\rm trunc}(\chieff\mid \mu,\sigma),
		& m_1 > \mhigh.
	\end{cases}
\end{equation}

This parameterization introduces discrete transition masses and allows the broad spin component to appear in two distinct mass ranges. The priors for the parameters of the \chieffthree model are summarized in Tab.~\ref{tab:three_transition_priors}.
	
\begin{deluxetable}{lll}
\tablecaption{Priors for the \chieffthree model.
\label{tab:three_transition_priors}}
\tablewidth{0pt}
\tablehead{
    \colhead{Parameter} & \colhead{Description} & \colhead{Prior}
}
\startdata
$\mu$ & Mean of the Gaussian component & U(-1,1)\\
$\log_{10}\sigma$ & Logarithm base 10 of the standard deviation of the Gaussian component & U$(-2, 1)$ \\
\mlow & the transition mass of the low-mass spinning subpopulation & U(5, 30)\\
\mmid & the interval of the mass of the low-mass spinning subpopulation & U(\mlow, \mlow+30) \\
\mhigh & the interval of the mass of the low-mass spinning subpopulation & U(\mmid, \mmid+80)\\ 
$\xi_\text{low-mass}$ & Mixture fraction of the low-mass spinning subpopulation & U$(0, 1)$ \\
$\xi_\text{high-mass}$ & Mixture fraction of the high-mass spinning subpopulation & U$(0, 1)$ \\
$\chi_\text{min}^\text{low-mass}$ & Minimum \chieff of the low-mass spinning subpopulation  & U$(-1, 1)$ \\
$\chi_\text{max}^\text{low-mass}$ & Maximum \chieff of the low-mass spinning subpopulation & U$(-1, 1)$ \\
$\chi_\text{min}^\text{high-mass}$ & Minimum \chieff of the high-mass spinning subpopulation & U$(-1,1)$ \\
$\chi_\text{max}^\text{high-mass}$ & Maximum \chieff of the high-mass spinning subpopulation & U$(-1,1)$ \\
\enddata
\end{deluxetable}

\section{Cumulative distribution of primary and secondary masses}
\label{sec:m1m2_cdf}

\begin{figure}[ht]
	\centering
		\includegraphics[width=0.5\textwidth]{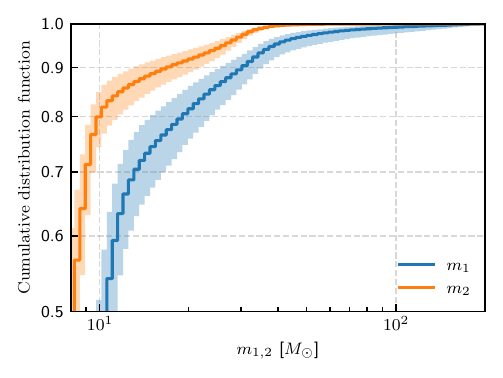}
		\caption{The cumulative distribution function of $m_1$ distribution and $m_2$ distribution using the $(m_1,m_2)$-\pixelpop model, focusing on the high-mass range.}
		\label{fig:cdf_m1_m2}
	\end{figure}

In Fig.~\ref{fig:cdf_m1_m2}, we show the cumulative distribution functions of the $m_1$ and $m_2$ distributions inferred using the $(m_1, m_2)$-\pixelpop model, with a focus on the high-mass range. 
The plot reveals that only a small fraction of mergers have $m_2\gtrsim 40\,\Msun$, whereas a non-negligible fraction of \acp{BBH} have $m_1 \in [50\text{--}100] \,\Msun$. 
In particular, we find that 99.9\% of \acp{BBH} have {$m_2 \lesssim \CIPlusMinus{\HighMassEstimates[99.9th_percentile_m2_dist]}\,\Msun$}. Although the uncertainty is relatively large due to the highly flexible assumptions of the \pixelpop model, this finding is consistent with the \gwtcfour results~\citep{Tong:2025wpz}.

\section{Three-subpopulation model}
\label{sec:banagiri_three_subpop}

We also explore a model that allows for three subpopulations at different $m_1$ ranges, as proposed in~\cite{Banagiri:2025dmy}.
The three subpopulations differ in either their mass-ratio distributions, their spin-magnitude distributions or both. We refer to this model as \textsc{three-subpopulation} model. 
The mass-ratio distribution of \textsc{three-subpopulation} model is modeled as:
\begin{equation}
    p(q) \propto 
\begin{cases}
    q^{\beta_0}, & \text{if } m_1 \leq m^t_{\mathcal{AB}},\\
    q^{\beta_1}, & \text{if } m^t_{\mathcal{AB}} \leq m_1 \leq m^t_{\mathcal{BC}},\\
    \mathcal{N}(q \mid \mu_q, \sigma_q),              &\text{if } m_1 > m^t_{\mathcal{BC}},\\
\end{cases}
\label{Eq:Multipop_pq}
\end{equation}
and the spin-magnitude distribution as:
\begin{equation}
    p(\chi) \propto 
\begin{cases}
    \mathcal{N}(\chi \mid \mu_{\chi_0}, \sigma_{\chi_0}),& \text{if } m_1 \leq m^t_{\mathcal{BC}},\\
    \mathcal{N}(\chi \mid \mu_{\chi_1}, \sigma_{\chi_1}),              &\text{if } m_1 > m^t_{\mathcal{BC}}.\\
\end{cases}
\label{Eq:Multipop_pchi}
\end{equation}
The priors for the parameters of the \textsc{three-subpopulation} model are summarized in Tab.~\ref{tab:three_subpopulation_priors}.
We employ the same spin tilt and redshift distributions as the basic mass model for all three subpopulations.

    \begin{deluxetable}{lll}
\tablecaption{Priors for the \textsc{three-subpopulation} model.
\label{tab:three_subpopulation_priors}}
\tablewidth{0pt}
\tablehead{
    \colhead{Parameter} & \colhead{Description} & \colhead{Prior}
}
\startdata
$m^t_{\mathcal{ZA}}$ & First potential transition mass & U$(5, 20) \, M_{\odot}$ \\
$m^t_{\mathcal{AB}}$ & Second potential transition mass & U$(20, 65)\, M_{\odot}$ \\
$m^t_{\mathcal{BC}}$ & Third potential transition mass & U$(30, 75)\, M_{\odot}$ \\
$\beta_q^i$ & Mass-ratio power-law index of the $i$-th subpopulation & U$(-6, 11)$ \\
$\mu_q$ & Mean of the mass ratio distribution of the subpopulation & U$(0.01, 1)$ \\
$\sigma_q$ & Width of the mass ratio distribution of the subpopulation & U$(0.01, 1)$ \\
$\mu^i_{\chi}$ & Mean of the spin-magnitude distribution of the $i$-th subpopulation & U$(0, 1)$ \\
$\sigma^i_{\chi}$ & Width of the spin-magnitude distribution of the $i$-th subpopulation & U$(0.005, 2)$ \\
$\sigma_1$ & Width of the first peak & U$(0, 5)$ \\
$\kappa_z$ & Power-law index on comoving merger rate evolution & U($-8, 8$) \\
$m_{\rm high}$ & Maximum mass for distribution, which is pinned to $m_{\rm high} = 200\,\Msun$ for this model & $\delta(m_{\rm high} - 200)$ \\
\enddata
\end{deluxetable}

{Consistent with \gwtcfour results in \citet{Banagiri:2025dmy}, we find that the high-mass subpopulation has significantly flatter mass-ratio distribution (Fig.~\ref{fig:highmass_mass_ratio_distribution})}.

\section{Primary and Secondary Mass Marginals: \pixelpop and \bgp Comparison}
\label{app:full_mass}

\noindent
\begin{figure*}[ht]
        \centering
        \includegraphics[width=0.9\textwidth]{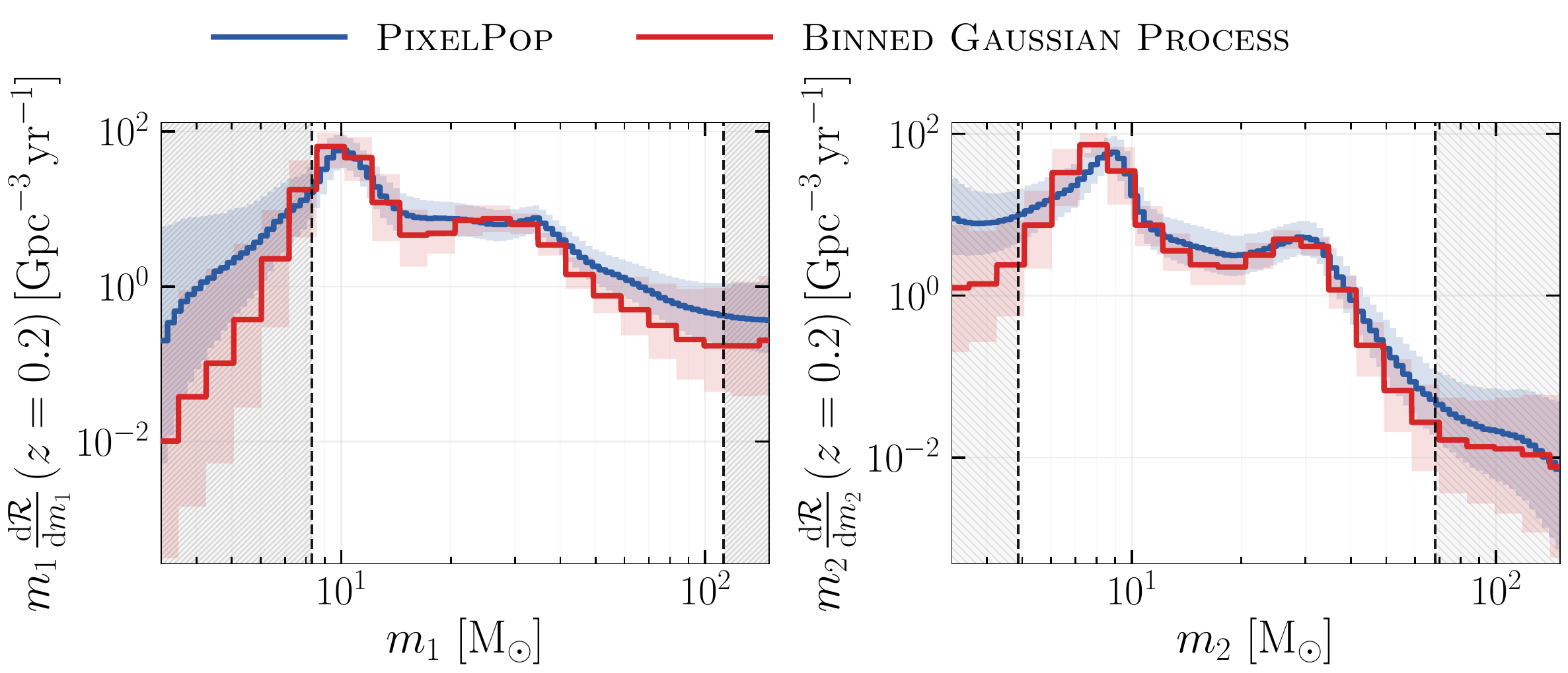}

        \caption{We compare the primary- and secondary-mass marginals from $(m_1,m_2)$-\pixelpop and the $(m_1, m_2, \chieff)$ variant of \bgp, marginalizing over all other parameters and evaluating at $z=0$. The two approaches are in good agreement, showing the robustness of our results.}
\label{Fig:m1m2_Appendix}
\end{figure*}
In this appendix, we compare the primary- and secondary-mass marginals inferred using the $(m_1, m_2)$ variant of \pixelpop and the $(m_1, m_2, \chieff)$ variant of \bgp. While the two approaches differ in their dimensionality and smoothing assumptions, they provide an important cross-check of the inferred mass distribution. As shown in Fig.~\ref{Fig:m1m2_Appendix}, the resulting $m_1$ and $m_2$ marginals are broadly consistent across the two methods.

\ifprintauthors
\allauthors
\fi

\end{document}